\documentclass[12pt]{article}
\usepackage{epsf,amssymb,amsmath,psfrag}
\usepackage[dvips]{graphicx}

\catcode`\@=11
\def \thesection {\arabic{section}.}

\def \be  {\begin{equation}}
\def \ee  {\end{equation}}
\def \ba  {\begin{eqnarray}}
\def \ea  {\end{eqnarray}}
\def \baa {\begin{eqnarray*}}
\def \eaa {\end{eqnarray*}}
\def \bb  {\begin {thebibliography} }
\def \eb  {\end{thebibliography}}
\def \lab #1 {\label{#1}}

\newcommand\re[1]{(\ref{#1})}
\def \qqquad {\qquad\quad}
\def \qqqquad {\qquad\qquad}
\def \matrix #1 {\left(\begin{array}{cc} #1 \end{array}\right)}

\def \tr {\mathop{\rm tr}\nolimits}
\def \str {\mathop{\rm str}\nolimits}

\def \e  {\mathop{\rm e}\nolimits}
\def \Span {\mathop{\rm span}\nolimits}

\newcommand\lr[1]{{\left({#1}\right)}}
\newcommand \widebar [1] {\overline{#1}}

\newcommand \vev [1] {\langle{#1}\rangle}

\newcommand \ket [1] {|{#1}\rangle}

\newcommand{\as}{\ifmmode\alpha_{\rm s}\else{$\alpha_{\rm s}$}\fi}
\newcommand{\asbar}{\ifmmode\bar{\alpha}_{\rm s}\else{$\bar{\alpha}_{\rm
s}$}\fi}

\newcommand{\ft}[2]{{\textstyle\frac{#1}{#2}}}

\def \jj {j}

\font\cmss=cmss12 
\def\inbar{\,\vrule height1.5ex width.4pt depth0pt}
\def\IC{\relax\hbox{$\inbar\kern-.3em{\rm C}$}}
\def\IZ{\relax{\hbox{\cmss Z\kern-.4em Z}}}
\def\IR{{\hbox{{\rm I}\kern-.2em\hbox{\rm R}}}}

\def\IP{{\hbox{{\rm I}\kern-.2em\hbox{\rm P}}}}
\def\II{\hbox{{1}\kern-.25em\hbox{l}}}

\newcommand{\lsemisum}{{\mathinner{\ni\mkern-12.3mu\rule{0.15mm}
   {2.0mm}\mkern10mu}}}

\def\numberbysection{\@addtoreset{equation}{section}
                     \def\theequation{\thesection\arabic{equation}}}
\numberbysection

\newcommand \Mybf[1] {\mbox{\boldmath$ {#1} $}}

\newbox\lett\newdimen\lheight\newdimen\lwidth
\def\ontop#1#2{\setbox\lett=\hbox{#2}\lheight\ht\lett
\multiply\lheight by 12 \divide\lheight by 10\relax%
\lwidth\wd\lett \multiply\lwidth by 8 \divide\lwidth by 10\relax #2\kern-
\lwidth%
\raise\lheight\hbox{{$\scriptstyle #1$}}\kern.1ex}

\newcommand{\insertfig}[2]{\mbox{\epsfxsize=#1cm \epsfbox{#2.eps}}}

\begin{document}

\begin{titlepage}
\begin{flushright}
\begin{tabular}{l}
LPT--Orsay--06--70 \\
hep-th/0610332
\end{tabular}
\end{flushright}

\vskip1cm

\centerline{\large \bf Baxter $Q-$operator for graded $SL(2|1)$ spin chain}

\vspace{1cm}

\centerline{\sc A.V. Belitsky$^a$, S.\'E. Derkachov$^{b,a}$,
                G.P. Korchemsky$^c$, A.N. Manashov$^{d,e}$}

\vspace{10mm}

\centerline{\it $^a$Department of Physics, Arizona State
University} \centerline{\it Tempe, AZ 85287-1504, USA}

\vspace{3mm}

\centerline{\it  $^b$St.\ Petersburg Department of Steklov Mathematical Institute,
} \centerline{\it Russian Academy of Sciences, 191023 St.-Petersburg, Russia}

\vspace{3mm}

\centerline{\it $^c$Laboratoire de Physique Th\'eorique\footnote{Unit\'e
                    Mixte de Recherche du CNRS (UMR 8627).},
                    Universit\'e de Paris XI}
\centerline{\it 91405 Orsay C\'edex, France}

\vspace{3mm}

\centerline{\it $^d$Department of Theoretical Physics,  St.-Petersburg State
University}

\centerline{\it 199034, St.-Petersburg, Russia}

\vspace{3mm}

\centerline{\it $^e $Institut f{\"u}r Theoretische Physik, Universit{\"a}t
Regensburg} \centerline{\it D-93040 Regensburg, Germany}

\def\thefootnote{\fnsymbol{footnote}}%
\vspace{1cm}

\centerline{\bf Abstract}

\vspace{5mm}

We study an integrable noncompact superspin chain model that emerged in recent
studies of the dilatation operator in the $\mathcal{N} = 1$ super-Yang-Mills
theory. It was found that the latter can be mapped into a homogeneous Heisenberg
magnet with the quantum space in all sites corresponding to infinite-dimensional
representations of the $SL(2|1)$ group. We extend the method of the Baxter
$Q-$operator to spin chains with supergroup symmetry and apply it to determine
the eigenspectrum of the model. Our analysis relies on a factorization property
of the $\mathcal{R}-$operators acting on the tensor product of two generic
infinite-dimensional $SL(2|1)$ representations. It allows us to factorize an
arbitrary transfer matrix into a product of three `elementary' transfer matrices
which we identify as Baxter $Q-$operators. We establish functional relations
between transfer matrices and use them to derive the TQ-relations for the
$Q-$operators. The proposed construction can be generalized to integrable models
based on supergroups of higher rank and, in distinction to the Bethe Ansatz, it
is not sensitive to the existence of the pseudovacuum state in the quantum space
of the model.

\end{titlepage}

\setcounter{footnote} 0

{\small \tableofcontents}

\newpage

\section{Introduction}

Integrable lattice spin chain models with supergroup symmetries play an
important r\^ole in various areas of theoretical physics ranging from condensed
matter to supersymmetric gauge theories. In particular, these models arose in
studies of strongly correlated electronic systems in relation with high $T_c$
superconductivity, in the quantum Hall effect and recently made their appearance
on both sides of the gauge/string correspondence.

In condensed matter physics, the interest in the one-dimensional supersymmetric
$t-J$ model has been renewed by Anderson's suggestion that two-dimensional systems
may share common features with one-dimensional systems \cite{And90}.  This model
describes electrons on a one-dimensional lattice with a Hamiltonian that
includes nearest-neighbor hopping ($t$) and nearest-neighbor spin exchange and charge
interactions ($J$). The Hilbert space of the model is constrained to exclude
double occupancy so that at a given lattice site there are only three possible
electronic states: the Fock vacuum $\ket{0}$, spin-up $\ket{\uparrow}$ and
spin-down $\ket{\downarrow}$ states. For special values of the couplings, $J=2t$,
the model can be mapped into an integrable Heisenberg magnet with the spin
operators in each site being generators of the three-dimensional atypical representation
of the $SL(2|1)$ group \cite{FoeKar92,EssKor92}. Its exact eigenspectrum can be found
within the nested Bethe ansatz approach.

In supersymmetric Yang-Mills (SYM) theories, integrable lattice spin chain models
appeared in studies of the scale dependence of composite light-cone single-trace
operators
\be\label{O}
\mathbb{O}(z_1,\theta_1,\ldots,z_N,\theta_N) = \tr \left[\Phi(z_1
n,\theta_1)\ldots \Phi(z_N n,\theta_N) \right]\,,
\ee
built from chiral superfields $\Phi(z n_\mu,\theta)$ ``living'' on the light ray
defined by a light-like vector, $n_\mu^2=0$. The expansion of $\Phi(z n_\mu,\theta)$
in powers of $\theta$ produces bosonic, $\phi$, and fermionic, $\chi$, fields%
\footnote{In $\mathcal{N}=1$ SYM theory, the fields $\chi(z n_\mu)$ and $\phi(z
n_\mu)$ can be identified as a helicity $(-1/2)$ component of the gaugino field
and a helicity $(-1)$ component of the gauge field strength.} which are assumed
to be holomorphic functions of $z$ at the origin
\be\label{Phi-dec}
\Phi_{\mathcal{N}=1}(z n_\mu,\theta) = \chi(z n_\mu) + \theta\phi(z
n_\mu)=\sum_{k\ge 0} z^k\cdot
\chi_k + \theta z^k\cdot \phi_k\,,
\ee
with $\phi_k=(n\cdot\partial)^k  \phi(0)$ and $\chi_k=(n\cdot\partial)^k
\chi(0)$.
In gauge theories with $\mathcal{N}>1$ supercharges, the superfield depends on
$\mathcal{N}$ Grassmann variables $\theta^A$ (with $A=1,\ldots,\mathcal{N}$) and
its expansion involves more terms. The scale dependence of the operators \re{O}
is driven by the dilatation operator which can be calculated in gauge theory as
a series in the coupling constant. To one-loop order and in the multi-color
limit, the dilatation operator in super-Yang-Mills theories with $\mathcal{N}-$supercharges
takes the form
\be\label{DO}
\mathbb{H}_N=H_{12}  + \ldots+ H_{N-1,N} + H_{N1}\,.
\ee
The two-particle Hamiltonian, say $H_{12}$, acts only on the $1^{\rm st}$ and
$2^{\rm nd}$ superfields inside the trace \re{O} and is given by the following
integral operator \cite{BelDerKorMan04}
\begin{eqnarray}\label{V-super}
\lefteqn{H_{12} \, {\Phi}(Z_1) {\Phi}(Z_2) = \int_0^1 \frac{d\alpha}{\alpha} \,
\bigg\{ 2 {\Phi}(Z_1) {\Phi}(Z_2)} &&
\\
&& - (1-\alpha)^{j - 1} \left[{\Phi} ((1-\alpha) Z_1 + \alpha Z_2) {\Phi}(Z_2)
+{\Phi} (Z_1) {\Phi}((1-\alpha)Z_2 + \alpha Z_1)\right] \bigg\} \, . \nonumber
\end{eqnarray}
Here $j=(3-\mathcal{N})$ is twice the superconformal spin of the superfields
$\Phi(Z_k)\equiv \Phi(z_k n,\theta_k)$ (with $k=1,2$) and a short-hand notation
is used for the sum of vectors in the $(\mathcal{N}+1)-$dimensional superspace
$\beta Z_1 + \alpha Z_2\equiv (\beta z_1 + \alpha z_2,\beta\theta_1^A + \alpha
\theta_2^A)$. The dilatation operator $\mathbb{H}_N$ defined in this way can be
mapped into a Hamiltonian of a (graded) integrable Heisenberg $SL(2|\mathcal{N})$
magnet. The length of the spin chain equals the number of superfields entering
\re{O} and its eigenspectrum determines the spectrum of anomalous dimensions of
the operators \re{O} to one-loop accuracy.

The appearance of the global $SL(2|\mathcal{N})$ symmetry in gauge theory is not
of course accidental and has a clear physical origin. The Lagrangian of SYM
theory with $\mathcal{N}$ supercharges is invariant under the
$SU(2,2|\mathcal{N})$ group of superconformal transformations \cite{Soh85}. The
$SL(2|\mathcal{N})$ symmetry of the one-loop dilatation operator arises as a
reduction of this symmetry for the light-like operators \re{O}. The superfield
$\Phi(z_k n_\mu,\theta_k)$ belongs to an irreducible chiral representation of the
$SL(2|\mathcal{N})$ group labeled by its superconformal spin $j$. The
corresponding graded vector space $\mathbb{V}_j$ defines the quantum space in
$k^{\rm th}$ site of the lattice model so that the Hilbert space for the
Hamiltonian \re{DO} is given by the tensor product of $N$ copies of this space,
${\mathbb{V}_j}^{\otimes N}$. According to \re{Phi-dec}, for $\mathcal{N}=1$ the
linear vector space $\mathbb{V}_j$ is spanned by the monomials
$\mathbb{V}_j=\Span\{z^k,\theta z^k\,|\,k\in \mathbb{N}\}$ and, therefore, it is
necessarily infinite-dimensional. This should be compared to the supersymmetric
$t-J$ model in which case the corresponding $SL(2|1)$ representation is
three-dimensional. Still, one can associate with each site of the $t-J$ model a
superfield given by a linear combination of three states
\be
\Phi_{tJ}(z,\theta) =1\cdot \ket{\uparrow} + z\cdot \ket{\downarrow}
+\theta\cdot
\ket{0}
\, ,
\ee
where $\{1,\theta,z\}$ define the basis of the graded linear space $\mathrm{v}_1$
corresponding to the atypical fundamental representation of the $SL(2|1)$. The
Hamiltonian of the $t-J$ model can be realized as an operator acting on the
product of superfields $\Phi_{tJ}(z_1,\theta_1) \ldots \Phi_{tJ}(z_N,\theta_N)\in
{\mathrm{v}_1}^{\otimes N}$. One of the advantages of dealing with superfields is
that $\Phi_{tJ}(z_1,\theta_1)$ can be realized as an invariant component of {\sl
reducible} (but indecomposable) $SL(2|1)$ representation $\mathbb{V}_{j=-1}$.
This allows one to treat in a unifying manner both compact and noncompact graded
spin chain models. In both cases, the Hilbert space of the model contains a
pseudovacuum state and this opens up a possibility to construct the nested Bethe
ansatz solution. For noncompact super-spin chains, the number of eigenstates is
infinite for a finite length of the spin chain $N$ and completeness of the Bethe
ansatz proves to be an extremely nontrivial issue. This calls up for an
alternative approach which does not rely on the existence of the pseudovacuum
state in the Hilbert space of the model and which is particularly suitable for
solving the eigenproblem for noncompact graded spin chains. We shall demonstrate
in the present paper that such an approach is offered by the method of the Baxter
$Q-$operator.

Another motivation for developing the Baxter $Q-$operator method for spin chains
with supergroup symmetries comes from two seemingly unrelated areas: the
description of the transition between plateaux in the integer quantum Hall effect
\cite{Efe93,WeiZir88,Bha99} and studies of the AdS/CFT correspondence between
supersymmetric Yang-Mills theories and strings on a nontrivial curved background
\cite{Mal97}. In both cases one has to deal with quantization of sigma models on
noncompact supergroup target spaces -- the problem that turns out to be extremely
difficult to solve. As a way out, one can try to `discretize' the sigma model and
construct a lattice spin chain of length $N$ that would flow into the former in
the continuum limit $N \to \infty$. It has been proposed \cite{Zir99,EssFraSal05}
to look for such models among integrable graded spin chains with the Hilbert
space of the type $(\mathbb{V} \otimes \bar{\mathbb{V}})^{\otimes{N/2}}$ with
$\mathbb{V}$ and $\bar{\mathbb{V}}$ being conjugated infinite-dimensional
representations of supergroups. However, the tensor product $\mathbb{V}
\otimes\bar{\mathbb{V}}$ contains in general irreducible components which have
neither the highest, nor the lowest weight vectors and, as a consequence, the
nested Bethe ansatz is not applicable. The situation here is quite similar to
that for the $SL(2)$ magnet with the spin operators being generators of the
principal series of the $SL(2;\mathbb{C})$ \cite{DerKorMan01} or lattice
sinh-Gordon model \cite{BytTes06}. In these cases, the Bethe ansatz can not be
applied by the same token as before whereas the method of the Baxter $Q-$operator
allows one to determine the exact eigenspectrum of the model.

The Baxter $Q-$operator is one of the corner-stones of quantum integrable
systems \cite{Bax72} and it has been discussed in a variety of contexts
varying from conformal field theories \cite{BazLukZam96,AntFei96} to
classical B\"acklund transformations \cite{Skl95}. Originally developed
for the six-vertex model \cite{Bax72}, the method provides a general
framework to solve the eigenproblem for transfer matrices in a variety
of integrable lattice models. Defined as a trace of the monodromy operator
over some auxiliary space $V$, the transfer matrix $T_{V}(u)$ depends on the
spectral parameter $u$ and belongs to a commutative family of operators
acting on the quantum space of the model, $[T_{V}(u), T_{V'}(u')]=0$. For
a special choice of the auxiliary space, the transfer matrix becomes a
generating function of the local Hamiltonian and conserved charges of the
model. The method of the Baxter $Q-$operator relies on the existence of
operators $ {Q}_a(u)$ (with $a=1,2,\ldots$) which act on the quantum
space of the model and commute with transfer matrices and among themselves
\be
{}[ {Q}_a(v),  {Q}_b(u) ] =  [ {Q}_a(v), T_{V}(u) ] = 0
 \, .
\ee
The number of independent $Q-$operators depends on the rank of the symmetry
group. A distinguished feature of these operators is that all transfer matrices
of the model and, as a consequence, the Hamiltonian of the model can be expressed
in terms of ${Q}_a(u)$. At present, there exists no regular procedure for
constructing $Q-$operators in a generic lattice integrable model and the number
of models for which the Baxter method has been developed is rather limited. The
latter include homogeneous Heisenberg magnets based on classical $SL(2)$ and
$SL(3)$ symmetry and their spin-offs.

In the $SL(2)$ invariant homogeneous spin chain one can explicitly construct two
Baxter operators ${Q}_\pm(u)$ \cite{Der99,Pro00}. They satisfy an operatorial
{\sl second-order} finite difference equation, the so-called TQ-relation
\be\label{TQ-SL2}
\tau_N^{(1/2)}(u) \,{Q}_\pm (u) = (u+s)^N {Q}_\pm(u+1) + (u-s)^N {Q}_\pm(u-1)\,,
\ee
and verify the Wronskian condition
\be\label{WrSL2}
{Q}_+(u){Q}_-(u+1)-{Q}_+(u+1){Q}_-(u) = \left[\frac{\Gamma(-u-s)}{\Gamma(-u+s)}
\right]^N .
\ee
Here, the half-integer spin $s$ labels irreducible representations of the $SL(2)$
group and $\tau_N^{(1/2)}(u)$ is the transfer matrix $T_{V}(u)$ with the
auxiliary space $V$ being two-dimensional spin$-1/2$ representation of the
$SL(2)$. The eigenvalues of the operator ${Q}_+(u)$ are {\sl polynomials} in $u$
that we shall denote as $P_m^{(s)}(u)$ (with nonnegative integer $m$ defining the
total $SL(2)$ spin of the model). At the same time, the eigenvalues of the second
operator are {\sl meromorphic functions} of $u$ which can be represented as
${Q}_-(u)= [\Gamma(1-u-s)]^N\times \mbox{(analytical function)}$ as far as the
order and position of poles is concerned. Another remarkable feature of the
Baxter operators is that the Hamiltonian of the $SL(2)$ spin chain can be
expressed in terms of the `polynomial' $Q-$operator as
\be\label{H-SL2}
\mathbb{H}^{\scriptscriptstyle \rm SL(2)} = \lr{\ln  {Q}_+(s)}' - \lr{\ln
{Q}_+(-s)}'\,,
\ee
where the prime denotes a derivative with respect to the spectral parameter.

For the $SL(3)$ invariant homogeneous spin chain, one already encounters three
$Q-$ope\-ra\-tors~\cite{Baz01,Hik01,DerMan06}. Similarly to \re{TQ-SL2}, three
$Q-$operators satisfy the same TQ-relation. Among them only one $Q-$operator is
polynomial in $u$ and the eigenvalues of remaining two operators are meromorphic
functions. The Baxter equation now takes the form of a finite difference equation
of the {\sl third order} and involves two transfer matrices with the auxiliary
space corresponding to two fundamental three-dimensional representations of the
$SL(3)$. The Wronskian relation involves all three $Q-$operators simultaneously
and it can be cast in a determinant form.

In the present paper we extend the method of the Baxter $Q-$operator to
integrable spin chain models with supergroup symmetry. More precisely, we present
an explicit construction of the $Q-$operators for the homogeneous Heisenberg
magnet with the quantum space in all sites corresponding to the
infinite-dimensional $SL(2|1)$ representations $[j_q,\bar j_q]$ labeled by a pair
of spins $j_q$ and $\bar j_q$. We shall argue that the model has three different
Baxter operators ${Q}_a(u)$ (with $a=1,2,3$). These operators have a number of
unusual properties as compared with models based on classical Lie symmetry.
Namely, two operators, ${Q}_1(u)$ and ${Q}_3(u)$, verify the same TQ-relation
which takes the form of a {\sl second-order} finite difference equation analogous
to \re{TQ-SL2}. For instance, in the chiral limit $\bar j_q=0$ and $j_q\neq 0$,
relevant for the $\mathcal{N}=1$ SYM theory, the TQ-relation reads (for $a=1,3$)
\ba
\label{R1}
\lefteqn{\left[\tau_N(u)\bar\tau_N(u+j_q)-(u(u+j_q))^N \right]{Q}_a(u)} &&
\\[2mm] \nonumber
&=& u^N \left[\bar\tau_N(u+j_q)- (u+j_q)^N\right]{Q}_a(u-1)+(u+j_q)^N
\left[\tau_N(u)-u^N \right]{Q}_a(u+1)
\, ,
\ea
where $\tau_N(u)$ and $\bar\tau_N(u)$ are two transfer matrices with the
auxiliary space corresponding to three-dimensional atypical representations of
the $SL(2|1)$. An important difference with \re{TQ-SL2} is that the dressing
factors themselves now depend on the $SL(2|1)$ transfer matrices. As a consequence,
there exists no Wronskian relation for the operators ${Q}_1(u)$ and ${Q}_3(u)$.
The TQ-relation for the remaining operator, ${Q}_2(u)$, is a finite-difference
equation of the {\sl first order}. In the chiral limit, it reads
\be
\left[{\bar\tau_N(u+j_q-1)-(u+j_q-1)^N}\right] {{Q}_2(u)}=
\left[{\tau_N(u)-u^N}\right]{{Q}_2(u-1)}\,.
\ee
Among the three $SL(2|1)$ Baxter operators only $\mathcal{Q}_1(u)$ is not
polynomial in $u$. Under appropriate normalization, its eigenvalues are
meromorphic functions of $u$ and their pole structure is similar to that for
eigenvalues of the $SL(2)$ operator ${Q}_-(u)$. Finally, we will demonstrate that
in analogy with the $SL(2)$ relation \re{H-SL2}, the dilatation operator in the
$\mathcal{N}=1$ SYM theory, Eqs.~\re{DO} and \re{V-super}, is given by a
logarithmic derivative of the polynomial operator ${Q}_3(u)$ in the chiral limit
$\bar j_q=0$ and $j_q=2$
\be
\mathbb{H}^{\scriptscriptstyle \rm SL(2|1)} = \lr{\ln {Q}_3(0)}' - \lr{\ln
{Q}_3(-j_q)}'\, .
\ee
Being combined with the TQ-relations \re{R1}, this leads to an exact solution to
the eigenproblem for the $SL(2|1)$ spin chain Hamiltonian \re{DO}.

The present construction of $Q-$operators for the graded $SL(2|1)$ spin chain
makes use of the approach developed in Ref.~\cite{DerMan06} in application to the
$SL(3)$ spin chain. The two main ingredients of our analysis are (i) the
factorization property of the $\mathcal{R}-$operators \cite{Der05} acting on the
tensor product of two generic, {\sl infinite-dimensional} $SL(2|1)$
representations and (ii) property of the transfer matrices with the auxiliary
space corresponding to a {\sl reducible} \cite{BazLukZam96,AntFei96,DerMan06}
(but in general indecomposable) $SL(2|1)$ representation. These properties allow
us to factorize an arbitrary transfer matrix into a product of three `elementary'
transfer matrices which we identify as Baxter $Q-$operators. In addition, they
lead to functional relations between the $SL(2|1)$ transfer matrices including
those with the auxiliary space corresponding to {\sl finite-dimensional}
representations of the $SL(2|1)$. Such representations naturally arise as
invariant components of a bigger infinite-dimensional reducible representation.
As a result, a generic finite-dimensional transfer matrix can be expressed as a
difference of two infinite-dimensional transfer matrices each given by a product
of three $Q-$operators. Remarkably, this relation can be cast into the form of
the Baxter TQ-relations. It also leads to functional relations between
finite-dimensional transfer matrices which are in agreement with similar relations
obtained in Ref.~\cite{Tsu97}. The above construction can be straightforwardly
generalized to integrable models based on supergroups of higher rank and, in
distinction to the Bethe Ansatz, it is not sensitive to the existence of the
pseudovacuum state in the quantum space of the model.

The outline of the paper is as follows. In Section 2, we introduce the $SL(2|1)$
superalgebra and present its realization most suitable for the analysis of the
model \re{DO}. Then, we define generic infinite-dimensional representations of
the $SL(2|1)$ group and describe in detail the structure of reducible
indecomposable $SL(2|1)$ representations which play a pivotal r\^ole throughout
our analysis. In Section 3, we review the formalism of factorized
$\mathcal{R}-$operators and demonstrate that a generic infinite-dimensional
transfer matrix can be factorized into the product of three mutually commuting
operators that we identify as $Q-$operators. In Section 4, we combine together
properties of reducible $SL(2|1)$ representations and factorized expressions for
the corresponding transfer matrices to obtain a representation for various
finite- and infinite-dimensional transfer matrices in terms of the Baxter
operators. In Section 5, we argue that the obtained relations yield a hierarchy
between the transfer matrices and identify one of the relations as the
TQ-equation for the Baxter operators. In Section 6, we present an exact solution
of the eigenproblem for the model \re{DO} based on the TQ-relations and establish
the correspondence with the nested Bethe ansatz solution. Section 7 contains
concluding remarks. Several appendices give detailed derivation of some results
used in the body of the paper.

\section{Noncompact $SL(2|1)$ spin chain}

As was already mentioned, noncompact $SL(2|1)$ spin chains naturally appear in
supersymmetric $\mathcal{N}=1$ Yang-Mills theory. The $SL(2|1)$ symmetry arises
as a reduction of the full superconformal symmetry group $SU(2,2|1)$ of the
four-dimensional gauge theory on the light-cone. Gauge theory leads to a
particular realization of the $SL(2|1)$ algebra on the space of functions in the
superspace $\mathcal{Z}=(z,\theta,\bar\theta)$ that we shall employ throughout
this paper. As we will argue, this representation is advantageous as far as the
construction of the Baxter operators is concerned.

A general superfield $\Phi(z,\theta,\bar\theta)$ is defined as a function of
`even' $z$ and `odd'  $\theta$ and $\bar\theta$ variables verifying the standard
anti-commutation relations $\theta^2=\bar\theta^2=0$ and $\{\theta,\bar\theta \}
=\theta\bar\theta +\bar\theta\theta=0$. A single superfield comprises four independent
functions
\be\label{Phi-phi}
\Phi(z,\theta,\bar\theta) = \chi(z) + \theta \phi(z) + \bar\theta
\bar\phi(z) + \theta\bar\theta\,\psi(z)\,,
\ee
which are assumed to be holomorphic functions of $z$ around the origin. In the
chiral limit, the superfield does not depend on $\bar\theta$ or, equivalently,
$\bar\phi(z)=\psi(z)=0$. The superfield parameterizes the quantum space in each
site of the spin chain. If its expansion in powers of $z$ is not truncated, this
space is infinite-dimensional and, therefore, the corresponding spin chain is
called noncompact. Otherwise, the superfield is a polynomial in $z$ of a finite
degree and the corresponding spin chain is compact.

\subsection{Representation of the $SL(2|1)$ superalgebra}

The superfield \re{Phi-phi} forms a representation of the $SL(2|1)$ algebra
labeled by two spins $j$ and $\bar j$. Its variation under the $SL(2|1)$
transformations is given by
\be\label{G}
\delta_G \Phi(z,\theta,\bar\theta) =  G_{j\bar j} \cdot
\Phi(z,\theta,\bar\theta)
\ee
with the operator $G_{j\bar j}$ being a linear combination of four even $L^+,L^0,
L^-, B$ and four odd $V^\pm, \bar V^\pm$ generators. Using the technique of
induced representations, they can be realized as first order differential
operators acting on the super-coordinates of the superfields
$\Phi(z,\theta,\bar\theta)$.

\begin{itemize}
\item The operators $L^-$, $V^-$ and $\bar V^-$ decrease the power in $z$,
$\theta$ and $\bar\theta$,
\ba\label{trans1}
&& L^-=-\partial_z\,,\qquad V^- = \partial_\theta + \ft12 \bar\theta
\partial_z\,,\qquad \bar V^- =\partial_{\bar\theta} + \ft12 \theta
\partial_z\,.
\ea

\item The operators $L^+$, $V^+$ and $\bar V^+$ increase the power in $z$,
$\theta$
and $\bar\theta$,
\ba\nonumber
&& V^+ = z\partial_\theta+\ft12 \bar\theta
z\partial_z+\ft12\bar\theta\theta\partial_{\theta}+\bar j\, \bar\theta \, ,
\\[2mm]
&& \label{rising}
\bar V^+ = z\partial_{\bar\theta}+\ft12 \theta
z\partial_z+\ft12\theta\bar\theta\partial_{\bar\theta}+  j\,  \theta \, ,
\\[2mm]
&& \nonumber L^+=z^2\partial_z +
z\theta\partial_{\theta}+z\bar\theta\partial_{\bar\theta}+(j+\bar
j)z+\ft12(j-\bar j)\bar\theta\theta\,.
\ea

\item The operators $J$ and $\bar J$ preserve the power in $z$, $\theta$ and
$\bar\theta$,
\ba\nonumber
J \!\!\!&=&\!\!\! L^0 + B = z\partial_z+\bar\theta\partial_{\bar\theta}+j \,,
\qquad \\ \label{trans3}
\bar{J}
\!\!\!&=&\!\!\!
L^0 -B = z\partial_z+\theta\partial_{\theta}+\bar{j}
\,.
\ea
\end{itemize}
Equations \re{trans1} -- \re{trans3} define the infinitesimal
$SL(2|1)$ transformations
of the superfields carrying the superconformal spins $j$ and $\bar
j$~\cite{BelDerKorMan04}.
In a global form the transformations generated by the Cartan generators
\re{trans3} look
like
\ba \nonumber
\e^{\lambda J} \cdot \Phi(z,\theta,\bar\theta) \!\!\!&=&\!\!\! \e^{\lambda j}
\Phi(\lambda z,\theta,\lambda\bar\theta)\,,\qquad
\\ \label{scaling}
\e^{\lambda \bar
J}\cdot \Phi(z,\theta,\bar\theta) \!\!\!&=&\!\!\! \e^{\lambda \bar j}
\Phi(\lambda z,\lambda\theta,\bar\theta) \, ,
\ea
while for other generators similar relations can be found in
Ref.~\cite{DerKarKir98}.
For our purposes it is convenient to introduce the operators $E^{AB}$ (with
${\scriptstyle A,B}=1,2,3$)
\be
\label{CartanBasisGen}
\begin{array}{lll}
E^{11}=J         & \qquad E^{12} = - V^+   &\qquad E^{13}=L^+ \\[2mm]
E^{21}=-\bar V^- & \qquad E^{22}=\bar J- J &\qquad E^{23}=-\bar V^+ \\[2mm]
E^{31}=L^-       & \qquad E^{32}=V^-       &\qquad E^{33}=-\bar J \\
\end{array}
\ee
Then, the generators of the superconformal transformations \re{trans1} --
\re{trans3} satisfy the graded $SL(2|1)$ commutation relations~\cite{Frappat96}
\ba \nonumber
{}[ E^{AB}, E^{CD} ]
\!\!\!&\equiv&\!\!\!
E^{AB} E^{CD} - (-1)^{(\bar A+\bar B)(\bar C + \bar D)} E^{CD} E^{AB} \\
&=&\!\!\!
\delta_{CB} E^{AD} - (-1)^{(\bar A+\bar B)(\bar C + \bar D)} \delta_{AD} E^{CB}
\, ,
\ea
where the indices run over ${\scriptstyle A,B,C,D} =1,2,3$ and the grading is
chosen as $\bar 1=\bar 3=0$ and $\bar 2=1$. The $SL(2|1)$ algebra has an obvious
automorphism
\be\label{auto}
J \rightleftarrows \bar J\,,\qqquad V^\pm \rightleftarrows \bar V^\pm\,,\qqquad
L^\pm \rightleftarrows L^\pm \, ,
\ee
which amounts to substituting $\theta\rightleftarrows\bar\theta$ and
$j\rightleftarrows \bar j$ in \re{trans1} -- \re{trans3}.

Following \cite{Sch76,Jar78,Marcu79,Frappat96}, one can construct the $SL(2|1)$
Casimir
operators%
\footnote{The definition of the Casimirs in
Ref.~\cite{Sch76,Jar78,Marcu79,Frappat96} involves the $GL(2|1)$ generators
$e^{AB}$. They are related to the $SL(2|1)$ generators as $E^{AB} = e^{AB}- \str
e^{AB}$ so that $\str E^{AB} = 0$.} $\mathbb{C}_p$ ($p=1,2,3,\ldots)$
\ba\nonumber
\mathbb{C}_1
\!\!\!&=&\!\!\!
\sum_A E^{AA} = 0 \, ,
\\ \label{Casimir}
\mathbb{C}_2 \!\!\!&=&\!\!\! \ft{1}{2!} \sum_{A,B} (-1)^{\bar B} E^{AB} E^{BA} =
J \bar J
+ L^+ L^- - V^+ \bar V^- - \bar V^+ V^- \, ,
\\ \nonumber
\mathbb{C}_3 \!\!\!&=&\!\!\! \ft{1}{3!}\sum_{A,B,C} (-1)^{\bar B + \bar C}
E^{AB} E^{BC}
E^{CA} = \ft{1}{2} (J-\bar J+\ft13) J \bar J + \ldots \, .
\ea
Here ellipses denote terms involving the lowering operators $V^-$, $\bar V^-$
and $L^-$ to the right and, therefore, vanishing when applied to the lowest
weight (see Eq.~\re{Omega0} below). One can verify using \re{trans1} --
\re{trans3} that the Casimir operators are diagonal for the superfield
$\Phi(z,\theta,\bar\theta)$
\ba \nonumber
\mathbb{C}_2 \cdot \Phi(z,\theta,\bar\theta ) &=& j \bar j\,
\Phi(z,\theta,\bar\theta )\,,
\\[2mm]
 \mathbb{C}_3 \cdot \Phi(z,\theta,\bar\theta) &=& \ft12 (j-\bar j+\ft13) j \bar
j\,\Phi(z,\theta,\bar\theta)\,.
\ea
Both Casimirs vanish in the (anti)chiral limit $j=0$ (or $\bar j=0$).

Let us denote by $[j,\bar j]$ the $SL(2|1)$ representation the field
$\Phi(z,\theta,\bar\theta )$ belongs to. For generic values of the spins $j$ and
$\bar j$, this representation is infinite-dimensional and irreducible. However
the representation $[j,\bar j]$ becomes reducible (but, in general,
indecomposable) for some special values of the spins. This property plays
a crucial r\^ole in our analysis and we shall describe it in details in the
next subsection.

The even generators of the $SL(2|1)$ superalgebra form the $SL(2)\otimes U(1)$
subalgebra. The operators $L^\pm$ and $L^0=\ft12(J+\bar J)$ belong to the $SL(2)$
subalgebra while $B=\ft12(J-\bar J)$ defines the $U(1)$ charge. Substituting the
superfield in \re{G} by its expression \re{Phi-phi} one finds that four functions
of $z$ entering \re{Phi-phi} form four different representations under the action
of the  $SL(2)\otimes U(1)$ generators. This corresponds to the decomposition of
a typical $SL(2|1)$ representation $[j,\bar j]$ over the $SL(2)\otimes U(1)$
multiplets~\cite{Sch76,Jar78,Marcu79,Frappat96}
\be\label{sl2-dec}
[j,\bar j\,] =  \mathcal{D}_\ell(b) \oplus \mathcal{D}_{\ell+\ft12}(b-\ft12)
\oplus \mathcal{D}_{\ell+\ft12}(b+\ft12)\oplus \mathcal{D}_{\ell+1}(b)\,,
\ee
where $\mathcal{D}_\ell(b)$ stands for the $SL(2)\otimes U(1)$ representation
labeled by the conformal $SL(2)$ spin $\ell$ and the $U(1)$ charge $b$
\be\label{l,b}
\ell = \ft12 \lr{j+\bar j}\,,\qquad b=\ft12\lr{j-\bar j}\,.
\ee

\subsection{Reducible $SL(2|1)$ representations}

By construction, the representation $[j,\bar j]$ 
is spanned by the superfields \re{Phi-phi} which are assumed to be analytical
functions of $z$ around the origin. Let us denote the corresponding
representation space as $\mathcal{V}_{j\bar j}$. The basis in the
infinite-dimensional linear graded space $\mathcal{V}_{j\bar j}$ can be chosen
as
\be\label{V-gen}
\mathcal{V}_{j\bar j} = \Span
\left\{z^k,z^k\theta,z^k\bar\theta,z^k\theta\bar\theta\, | \, k\in
\mathbb{N}\right\}\,.
\ee
One identifies among these states the lowest weight $\Omega = 1$. It is
annihilated by all lowering
$SL(2|1)$ generators%
\footnote{In virtue of $\{V^-,\bar V^-\} = - L^-$, the  relation $L^-\Omega=0$
follows from the remaining two.} and diagonalizes the Cartan generators
\be\label{Omega0}
L^-\Omega = V^-\Omega = \bar V^-\Omega = 0\,, \qquad J\, \Omega =j
\,\Omega\,,\qquad \bar J \,\Omega = \bar j\, \Omega\,.
\ee
Applying the raising operators \re{rising} to the lowest weight $\Omega=1$ one
can construct the $SL(2|1)$ invariant graded linear space
\be\label{Verma}
\mathcal{V}_\Omega= \Span \left\{ (L^+)^k \Omega\,,\ (L^+)^{k}  \bar V^+
\Omega\,, \ (L^+)^{k} V^+ \Omega \,,\  (L^+)^{k}  \bar V^+ V^+\Omega \, | \,
k\in
\mathbb{N}\right\}.
\ee
These states are given by a linear combination of the basis vectors \re{V-gen}
of the same Grassmann parity. Their explicit form can be found in
Ref.~\cite{DerKarKir98}.

For generic $j$ and $\bar j$, the two spaces are isomorphic, $\mathcal{V}_{j\bar
j}= \mathcal{V}_\Omega$. There are however special values of the spins $j$ and
$\bar j$ for which some basis vectors in \re{Verma} vanish identically. In that
case, nonvanishing states in \re{Verma} still form the $SL(2|1)$ invariant space
but it is now a subspace of $\mathcal{V}_{j,\bar j}$. In other words, the
representation $[j,\bar j]$ becomes reducible and $\mathcal{V}_\Omega$ defines
its invariant component. The corresponding values of spins are:
\begin{itemize}
\item $\bar j=0$ for $j\neq 0$ (chiral limit);
\item $j=0$ for $\bar j\neq 0$ (antichiral limit);
\item $j=\bar j=0$;
\item $j+\bar j=-n$ with $n=$ positive integer.
\end{itemize}
Let us examine the four cases one after another and decompose
$\mathcal{V}_{j,\bar j}$ over irreducible components.

\subsubsection{(Anti)chiral infinite-dimensional representations}

\begin{figure}[t]
\begin{center}
\mbox{
\begin{picture}(0,135)(230,0)
\psfrag{*}[cc][cc]{$\ast$}
\psfrag{V}[cc][cc]{$\mathbb{V}$}
\psfrag{W}[cc][cc]{$\mathcal{V}$}
\psfrag{v}[cc][cc]{$v$}
\put(0,0){\insertfig{16}{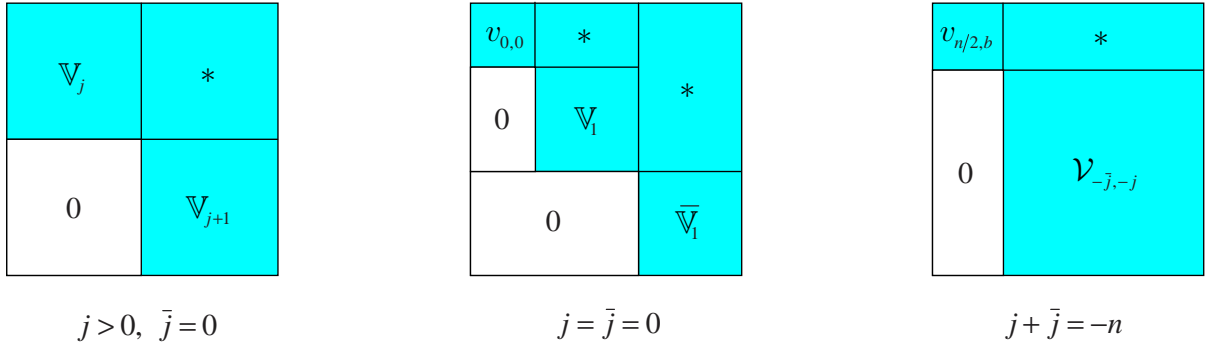}}
\end{picture}
}
\end{center}
\caption{ \label{ReducibleReps} The structure of reducible indecomposable
$SL(2|1)$ representations $[j,\bar j]$ for different values of the spins $j$ and
$\bar{j}$.}
\end{figure}

For $\bar j=0$, one finds from \re{rising} that $V^+\cdot 1 =0 $ and, therefore,
half of the basis vectors in \re{Verma} vanish identically indicating that the
corresponding $SL(2|1)$ representation $[j,0]$ becomes reducible. The nonvanishing
vectors in \re{Verma} define an infinite-dimensional $SL(2|1)$ invariant
subspace that we shall denote as $\mathbb{V}_j$. It is convenient to introduce
two supercovariant derivatives
\be\label{cov-der}
D = -\partial_{\bar\theta} + \ft12\theta\partial_z\,,\qquad
\bar D =-\partial_{\theta} + \ft12\bar\theta\partial_z
\ee
satisfying $D^2=\bar D^2=0$ and $\{ D, \bar D\}=-\partial_z$. Then, one can
verify that for $\bar j=0$ the basis vectors in \re{Verma} are annihilated by the
operator $D$ and, therefore, the space $\mathbb{V}_j$ coincides with its kernel
\be\label{chiral-irreps}
\mathbb{V}_j=\ker D=\Span \left\{1,\theta z^k,
\left(z+\ft12\bar\theta\theta\right)^{k+1}\,|\, k\in \mathbb{N} \right\}\,.
\ee
Notice that the $\bar\theta-$dependence of states in $\mathbb{V}_j$ can be removed
by a shift $z\mapsto z-\ft 12\bar\theta\theta$. The basis in the quotient space
$\mathcal{V}_{j,0}/\mathbb{V}_j $ can be constructed by imposing the {\sl
antichirality} conditions $\bar D \Phi_- =0$ and $D \Phi_- \neq 0$ for the
basis vectors,
\be\label{bas-a}
\mathcal{V}_{j,0}/\mathbb{V}_j =\Span \left\{\bar\theta,\bar\theta z^{k+1},
\left(z-\ft12\bar\theta\theta\right)^{k+1} \,|\, k\in \mathbb{N}\right\} \, .
\ee
Under the $SL(2|1)$ transformations, the states $\Phi_-\in
\mathcal{V}_{j,0}/\mathbb{V}_j$ mix with the states from the invariant subspace
$\Phi_+\in \mathbb{V}_{j}$ while the opposite is prohibited. This implies that
for $\bar j=0$ the $SL(2|1)$ generators $G_{j0}$, Eqs.~\re{trans1} --
\re{trans3}, take a block-triangular form in the basis $(\Phi_+,\Phi_-)$
\ba
\nonumber
G_{j0} \cdot \Phi_+^i &=& \Phi_+^k \, [G_{++}]^{ki}
\\[2mm] \label{G-mat}
G_{j0} \cdot \Phi_-^\alpha &=& \Phi_-^\beta \, [G_{--}]^{\beta\alpha}  +
\Phi_+^i
\, [G_{+-}]^{i\alpha}
\ea
where $G_{\pm \pm}$ are (infinite-dimensional) graded matrices. This property
can be depicted graphically as demonstrated in Fig. \ref{ReducibleReps}. The
upper diagonal block $G_{++}$ defines a (infinite-dimensional) representation of
the $SL(2|1)$ superalgebra to which we shall refer as the {\sl chiral} $SL(2|1)$
representation of spin $j$ and denote is as $[j]_+$. The corresponding
representation space is defined in \re{chiral-irreps}. The lower diagonal
block $G_{--}$ defines yet another $SL(2|1)$ representation which is
isomorphic to the chiral $SL(2|1)$ representation of spin $j+1$ (see Appendix~A
for details).

We conclude that for $\bar j=0$ the $SL(2|1)$ representation $[j,0]$ is
reducible but indecomposable. It is given by a semidirect sum of two
chiral $SL(2|1)$ representations of spins $j$ and $j+1$
\be\label{red1}
[j,0] = [j]_+\, \lsemisum  \, [j+1]_+\,.
\ee
Here `$\lsemisum$' stands for the semidirect sum with the first summand being an
invariant subspace of the whole representation space. Making use of the
automorphism of the $SL(2|1)$ superalgebra \re{auto}, one can obtain from
\re{red1} the decomposition of the $SL(2|1)$ representation $[0,\bar j]$
\be\label{red11}
[0,\bar j] = [\bar j]_- \, \lsemisum  \, [\bar j+1]_-\,,
\ee
where $[\bar j]_-$ denotes the {\sl antichiral} $SL(2|1)$ representation. It is
spanned by the states $\bar \Phi(z,\theta,\bar\theta)\in \mathbb{\bar V}_{\bar
j}$ which verify the condition $\bar D \, \bar \Phi =0$. The vector space
$\mathbb{\bar V}_{\bar j}$ can be obtained from \re{chiral-irreps} by
substituting $\theta \rightleftarrows\bar\theta$.
\be\label{antichiral-irreps}
\mathbb{\bar V}_{\bar j}=\ker \bar D=\Span\left\{1,\bar\theta z^k,
\left(z-\ft12\bar\theta\theta\right)^{k+1}\,|\, k\in \mathbb{N} \right\}\,.
\ee
So far we did not specify the value of the spin $j$ (or $\bar j$) and did not
discuss reducibility of the representation $[j]_+$ (or $[\bar j]_-$). We will
address this question in Section~2.2.3 and show that the (anti)chiral $SL(2|1)$
representations $[-n]_+$ and $[-n]_-$ are in turn reducible for $n\in \mathbb{N}$
in which case they contain a {\sl finite-dimensional} invariant component.

%
%

\subsubsection{Finite dimensional typical representations}

For $j=\bar j=0$ or $j+\bar j=-n$ (with $n\in \mathbb{Z}_+$) the representation
space $\mathcal{V}_{j,\bar j}$, Eq.~\re{V-gen}, has a finite-dimensional invariant
subspace that we shall denote as $v_{n/2\!,\,b}$ (with $b=\ft12(j-\bar{j})$).

Indeed, for these values of the spins, the state
\be\label{Omega}
\widehat\Omega =  \lr{z+\ft12 \theta\bar\theta}^{-j}\lr{z-\ft12
\theta\bar\theta}^{-\bar j}=z^{-(j+\bar j)} -\ft12 (j-\bar j) z^{-(j+\bar j)-1}
\theta\bar\theta
\ee
belongs to \re{Verma} and takes the form $\widehat\Omega \sim (L^+)^{-j-\bar
j}\,\Omega$ with $\Omega=1$. It is annihilated by all raising operators $V^+
\widehat\Omega = \bar V^+ \widehat\Omega = L^+ \widehat\Omega =0$ and, therefore,
it defines the highest weight in $\mathcal{V}_{\Omega}$. As a consequence, the
$SL(2|1)$ invariant space \re{Verma} becomes finite-dimensional. For $j=\bar{j}=0$,
it contains only one state $ v_{00}=\{1\} $, while for $j+\bar j=-n$, it has the
dimension $\dim v_{n/2\!,\,b} = 4n$ and is spanned by the states
\be\label{vn}
v_{n/2\!,\,b} = \Span\left\{1,\theta,\bar\theta, z_+^k,z_-^k,\theta
z^{k},\bar\theta z^{k}, \widehat\Omega | \, 1\le k \le n-1\right\} \, ,
\ee
where $z_\pm = z \pm \ft12 \theta\bar\theta$ and $\widehat\Omega$ is given by
\re{Omega}. The underlying $SL(2|1)$ representation is known in the literature
\cite{Sch76,Jar78,Marcu79,Frappat96} as the {\sl typical} representation and we
shall denote it as $(b,n/2)$.

Subsequent analysis goes along the same lines as in Section~2.2.1. For $j+\bar
j=-n$, the $SL(2|1)$ generators $G_{j\bar j}$, Eqs.~\re{trans1} -- \re{trans3},
take a block-triangular form in the basis $\Phi_+\in v_{n/2,\,b}$ and $\Phi_-\in
\mathcal{V}_{j\bar j}/v_{n/2,\,b}$ similarly to \re{G-mat}. The upper diagonal
block $G_{++}$ represents the $SL(2|1)$ generators of the typical representation
$(b,n/2)$ while the lower diagonal block $G_{--}$ can be mapped into generators
of the infinite-dimensional representation $[-\bar j,-j]$ described in
Section~2.1 (see Appendix~A for details). We conclude that for $j+\bar j=-n$
(with $n$ being positive integer) the $SL(2|1)$ representation $[j,\bar j]$ is
reducible and it admits the following decomposition
\be\label{red2}
[j,\bar j] = (b,n/2) \, \lsemisum  \,[-\bar j, -j]\,,
\ee
where  $b=(j-\bar j)/2$ and $(b,n/2)$ is a finite-dimensional typical $SL(2|1)$
representation \re{vn}.

For $j=\bar j=0$ the situation is more subtle. The invariant subspace $v_{00}$
contains only one state -- the lowest weight $1$, while the quotient space
$\mathcal{V}_{00}/v_{00}$ is given by a direct sum
\be\label{V00}
\mathcal{V}_{0,0}/v_{0,0}=\mathcal{V}_+ \oplus \mathcal{V}_-\,,\qquad
\mathcal{V}_\pm = \Span\{\theta_\pm ,\theta_\pm z^{k+1}, z_\pm^{k+1} \, | \,
k\in
\mathbb{N} \}\,,
\ee
where  $\theta_+=\theta$, $\theta_-=\bar\theta$ and $z_\pm = z \pm \ft12
\theta\bar\theta$. The action of the $SL(2|1)$ generators \re{trans1} --
\re{trans3} on $\mathcal{V}_\pm$ defines the infinite-dimensional chiral and
antichiral representations, $[1]_+$ and $[1]_-$, respectively (see Appendix~A for
details). As a result, for $j=\bar j=0$ the infinite-dimensional $SL(2|1)$
representation $[0,0]$ admits the following decomposition (see
Fig.~\ref{ReducibleReps})
\be
[0,0] = (0,0) \, \lsemisum  \, \lr{\, [1]_+ \oplus [1]_-}
\ee
where $(0,0)$ stands for a trivial one-dimensional $SL(2|1)$ representation.

\subsubsection{Finite-dimensional atypical representations}
\label{Reducibility3}

For $\bar j=0$ and $j=-n$ one encounters the situation when reducibility
conditions discussed in two previous subsections are satisfied simultaneously.
According to \re{red1}, the representation $[-n,0]$ decomposes into a
semi-direct sum of two infinite-dimensional chiral representations $[-n]_+$
and $[1-n]_+$. The latter representations are also reducible for positive
integer $n$.

For $j=-n$ the chiral representation $[-n]_+$ is spanned by the states
\re{chiral-irreps}. A unique feature of $\mathbb{V}_{-n}$ is that it contains
the highest weight vector $\widehat\Omega = (z+\ft12\bar\theta\theta)^n$ which is
annihilated by all raising $SL(2|1)$ generators \re{rising}. As a consequence,
the space $\mathbb{V}_{-n}$ contains a finite-dimensional invariant subspace
\be\label{v-chi}
\mathrm{v}_{n} = \Span\left\{1, \theta, z_+, \theta z, z_+^2,\ldots, \theta
z^{n-1}, z_+^n \right\}\,,
\ee
with $z_+=z+\ft12\bar\theta\theta$. It has the dimension $\dim \mathrm{v}_n = 2n
+1$ and all states in $\mathrm{v}_{n}$ are annihilated by the supercovariant
derivative $D$, Eq.~\re{cov-der}. We shall denote the corresponding $SL(2|1)$
representation as $(n)_+$. It is known in the literature
\cite{Sch76,Jar78,Marcu79,Frappat96} as the {\sl atypical} $SL(2|1)$
representation. As before, the $SL(2|1)$ generators take a block-triangular form
on $\mathrm{v}_n\oplus\mathbb{V}_{-n}/ \mathrm{v}_n$. Acting on the quotient
space $\mathbb{V}_{-n}/ \mathrm{v}_n$, they define the $SL(2|1)$ representation
$[n+1]_-$, i.e., the {\sl anti-chiral\/} infinite-dimensional representation of
spin $n+1$ (see Appendix~A for details).

We conclude that the chiral representation $[-n]_+$ (with positive integer $n$)
decomposes into a semidirect sum of atypical and antichiral representations,
\be\label{V-chi}
[-n]_+ =(n)_+ \, \lsemisum  \, [n+1]_-\,.
\ee
Making use of the automorphism of the $SL(2|1)$ superalgebra, Eq.~\re{auto},
this relation can be extended to reducible antichiral representations,
\be\label{V-chibar}
[-n]_- =(n)_- \, \lsemisum  \, [n+1]_+\,.
\ee
Here $(n)_-$ is yet another {atypical} $SL(2|1)$ representation spanned by the
states
\be\label{v-bar-chi}
\mathrm{\bar v}_{n} = \Span\{1, \bar\theta, z_-, \bar\theta z, z_-^2,\ldots,
\bar\theta z^{n-1}, z_-^n \}\,,
\ee
with $z_-=z-\ft12\bar\theta\theta$.

It is well known \cite{Sch76,Jar78,Marcu79,Frappat96} that the atypical
representations $(n)_+$ and $(n)_-$ also appear as invariant components of the
typical representation $(b,n/2)$ for $b=\pm n/2$
\ba \nonumber
(-n/2,n/2) &=& (n)_+ \, \lsemisum  \, (n-1)_+\,,
\\ \label{nn}
(n/2,n/2) &=& (n)_- \, \lsemisum  \, (n-1)_-\,.
\ea
In our analysis these relations appear as consistency conditions for the
relations \re{red2} and \re{red1} with $j=-n,\bar j=0$ and for the relations
\re{red2} and \re{red11} with $j=0,\bar j=-n$, respectively.

\subsubsection{Traces over reducible representations}

One of the fundamental objects in lattice integrable models is the transfer
matrix. It is defined as a (super)trace of the so-called monodromy operator over
a particularly chosen $SL(2|1)$ representation space $\mathcal{V}_{j,\bar j}$
(see Eq.~\re{Eq1} below). As we will argue in Section~3, a crucial r\^ole in
constructing the Baxter operator is played by transfer matrices with
$\mathcal{V}_{j,\bar j}$ being a {\sl reducible} $SL(2|1)$ representation.

Applying the results of this section, one can decompose a supertrace of an
arbitrary linear operator on $\mathcal{V}_{j,\bar j}$ into a sum of supertraces
evaluated over irreducible components of $\mathcal{V}_{j,\bar j}$. Let
\be\label{e's}
\vec e = \left\{1,\ \theta z^k,\ \bar\theta z^k,\ z_+^{k+1},\ z_-^{k+1} \, | \,
k\in \mathbb{N}\right \}
\ee
be a basis of $\mathcal{V}_{j,\bar j}$ and let us assign the grading $(-1)^{\bar
e_k}=1$ and $(-1)^{\bar e_k}=-1$ to correspondingly `even' and `odd' vectors in
this basis. Then, an arbitrary linear operator, say $O$, can be represented as
a (infinite-dimensional) graded matrix
\be\label{O-matrix}
{O} \cdot e_i = \sum_k e_k \, O_{ki}\,,
\ee
with $O_{ki}$ possessing the grading $(-1)^{\bar O_{ki}}=(-1)^{\bar
e_i+\bar{e}_k}$. The supertrace is defined then as
\be\label{str}
\str_{\mathcal{V}_{j,\bar j}} {O} = \sum_i (-1)^{\bar e_i} O_{ii}\,.
\ee
As an example relevant for further analysis, let us choose $O$ to be the
$SL(2|1)$ generators \re{CartanBasisGen} and realize them as finite dimensional
matrices on the spaces ${\rm v}_1$ and ${\rm \bar v}_1$, Eqs.~\re{v-chi} and
\re{v-bar-chi}, corresponding to the atypical representations $(1)_+$ and
$(1)_-$, respectively. It is convenient to choose the basis on ${\rm v}_1$ as
$e_1=-z_+$, $e_2=\theta$, $e_3=1$ with the grading $\bar 1=\bar 3=0$ and $\bar
2=-1$. The $SL(2|1)$ generators are given by the differential operators
\re{trans1} -- \re{trans3} with $j=-1$ and $\bar j=0$. Then, one applies
\re{O-matrix} and finds after some algebra
\be\label{e}
(E^{AB})_{kl} = (e^{AB})_{kl} - \delta_{kl} \str
e^{AB}=\delta^A_{k}\delta^B_{l}-(-1)^{\bar A}\delta^{AB}\delta_{kl}\,,
\ee
where $(e^{AB})_{kl} = \delta^A_{k}\delta^B_{l}$ are the $GL(2|1)$ generators of
the fundamental representation. Similar expressions for the generators of the
atypical representation $(1)_-$ can be obtained from \re{e} with a help of the
automorphism \re{auto}.

We have demonstrated in this section, that for reducible indecomposable
$SL(2|1)$ representations $[j,\bar j]$ the generators $G_{j,\bar j}$ take
a block-diagonal form in an appropriately chosen basis on $\mathcal{V}_{j,
\bar j}$. Obviously, the same property holds true for an arbitrary linear
operator $O$ depending on $G_{j,\bar j}$. This allows one to rewrite its
supertrace over a `big' space, $\str_{\mathcal{V}_{j,\bar j}} O $, as a sum of
supertraces over diagonal blocks corresponding to various irreducible components
of $[j,\bar j]$. In this way, one finds the following relations:
\begin{itemize}
\item From \re{red2}, for $j=-\ft12 n+b$ and $\bar j=-\ft12n-b$,
\be\label{V-expand}
\str_{{\mathcal{V}_{j,\bar j}}} O = \str_{{\mathcal{V}_{-\bar j,-j}}} O +
\str_{{v_{n/2,\,b}}} O
\, .
\ee
\item From \re{red1} and \re{red11}, for $j\neq 0$ and $\bar j\neq 0$,
respectively,
\ba\nonumber
\str_{{\mathcal{V}_{j,0}}} O &=& \str_{\mathbb{V}_{j}} O -
\str_{\mathbb{V}_{j+1}}O
\, , \\ \label{dec1}
\label{dec11} \str_{{{\mathcal{V}}_{0,\bar j}}} O &=& \str_{\mathbb{\bar
V}_{\bar
j}} O - \str_{\mathbb{\bar V}_{\bar j+1}}O
\, .
\ea
\item From \re{V-chi} and \re{V-chibar},
\ba \nonumber
\str_{\mathbb{V}_{-n}} O &=& \str_{\mathrm{v}_{n}} O-\str_{\mathbb{\bar V}_{n+1}}
O \, ,
\\ \label{mist}
\str_{\mathbb{\bar V}_{-n}} O &=& \str_{\mathrm{\bar
v}_{n}} O -\str_{\mathbb{V}_{n+1}} O\, .
\ea
\item From \re{nn},
\ba \nonumber
\str_{v_{-{n}/2,n/2}} O &=& \str_{\mathrm{v}_{n}} O - \str_{\mathrm{v}_{n-1}} O
\, , \\ \label{v-expand} \str_{v_{{n}/2,n/2\phantom{-}}} O &=& \str_{\mathrm{\bar
v}_{n}} O - \str_{\mathrm{\bar v}_{n-1}} O \, .
\ea
\end{itemize}
Notice the minus sign in the right-hand side of \re{dec1}. It comes about due to
the fact that the lowest weights in the space $\mathbb{V}_{j}$,
Eq.~\re{chiral-irreps}, and the quotient $\mathcal{V}_{j,0}/\mathbb{V}_{j}$,
Eq.~\re{bas-a}, are given by $1$ and $\bar\theta$, respectively, and have
different Grassmann parity. The minus sign in the right-hand side of \re{mist}
and \re{v-expand} has the same origin.

Later in the paper we shall heavily use the relations \re{V-expand} --
\re{v-expand} with the operator $O$ coinciding with the monodromy operator for
the $SL(2|1)$ spin chain. In that case, the supertrace of $O$ over the $SL(2|1)$
invariant space defines the transfer matrix of the model. Depending on the
choice of this (auxiliary) space, one can distinguish six different transfer
matrices summarized in Table~\ref{Table1}. There, the third column sets up
the notation for the transfer matrix and the fourth column specifies the
dimension of the corresponding auxiliary space. In what follows, we shall
refer to the transfer matrices with a (in)finite-dimensional auxiliary space
as (in)finite-dimensional ones.

\begin{table}[h]
\begin{center}
\begin{tabular}{|c|c|c|c|c|}
\hline
Representation & Vector space & Transfer matrix & Dimension & Definition \\
\hline
& & & & \\[-3.5mm]
$[j,\bar j]~$ & $\mathcal{V}_{j,\bar j}$ & $\mathcal{T}_{j,\bar j}(u)$ &
$\infty$ & \re{V-gen} \\[1mm]
$[j]_+$ & $\mathbb{V}_{j}$ & $\mathbb{T}_{j}(u)$ & $\infty$ & \re{chiral-irreps}
\\[1mm]
$[\bar j]_-$ & $\mathbb{\bar V}_{\bar j}$ & $\mathbb{\bar T}_{\bar j}(u)$ &
$\infty$ & \re{antichiral-irreps} \\[1mm]
$(b,n/2)$ & $v_{n/2,\,b}$ & $t_{n/2,\,b}(u)$ & $4n$ & \re{vn} \\[1mm]
$(n)_+$ & ${\rm v}_{n}$ & ${\rm t}_{n}(u)$ & $2n+1$ & \re{v-chi} \\[1mm]
$(n)_-$ & ${\rm \bar v}_{n}$ & ${\rm \bar t}_{n}(u)$ & $2n+1$ & \re{v-bar-chi}
\\[1mm]
\hline
\end{tabular}
\end{center}
\caption{Notations for the $SL(2|1)$ representations and the corresponding
transfer matrices used throughout the paper.}\label{Table1}
\end{table}

Equations \re{V-expand} -- \re{v-expand} allow us to establish relations between
the transfer matrices listed in Table~\ref{Table1}. A remarkable feature of these
relations, that we shall explore in Section 4, is that {\sl finite-dimensional}
transfer matrices can be expressed as a difference of {\sl infinite-dimensional}
ones. This suggests that infinite-dimensional transfer matrices should serve as
building blocks in the construction of the Baxter $Q-$operator. Indeed, we will
show in Section 3 that the $Q-$operators can be identified as the $SL(2|1)$
transfer matrices $\mathcal{T}_{j,\bar j}(u)$ for special values of the spins $j$
and $\bar j$ (see Eqs.~\re{Q=T} below).

\subsection{Invariant scalar product}

Instead of dealing with infinite-dimensional matrices \re{str}, it is more
advantageous to realize $O$ as an integral operator on the space of functions
$\Phi(z,\theta,\bar\theta) \in \mathcal{V}_{j,\bar j}$ endowed with an $SL(2|1)$
invariant scalar product. In what follows we shall assume that the spins $j$ and
$\bar j$ take real values only.

\subsubsection{General case}

For two arbitrary states belonging to an infinite-dimensional vector space
$\mathcal{V}_{j,\bar j}$ the scalar product is defined as
\be\label{scal}
\langle \Phi_2 | \Phi_1 \rangle_{{j\bar j}} = \int [\mathcal{D}
\mathcal{Z}]_{j\bar j} \, \lr{\Phi_2 \left( z, \theta, \bar\theta \right)}^*
\Phi_1 \left( z, \theta,
\bar\theta \right) ,
\ee
where $\mathcal{Z}=(z,\theta,\bar\theta)$ parameterizes the superspace and the
integration is performed over complex $z$ and four ``odd'' variables
\be\label{susy-scal}
\int [\mathcal{D} \mathcal{Z}]_{j\bar j} =\frac{j+\bar j}{j
\bar{j}} \int_{|z|\le 1} d^2 z \int d\theta d
\theta^* \int d\bar \theta d\bar\theta^* \, \mu_{j\bar j} (\mathcal{Z},
\mathcal{Z}^*)\,.
\ee
Here, `$*$' denotes the complex conjugation which acts on even and odd
coordinates according to
\be
\mathcal{Z}^* = (z^*,\theta^*,\bar\theta^*)\,,\qquad (\theta\bar\theta)^\ast =
\bar\theta^\ast\theta^\ast \,,
\ee
with $\theta$, $\bar\theta$, $\theta^*$  and $\bar\theta^*$ being mutually
independent Grassmann variables. The integration measure in \re{susy-scal} is
given by
\be\label{measure}
\mu_{j \bar{j}} (\mathcal{Z}, \mathcal{Z}^*) = \frac1{\pi}( 1 - z_+ {z}_+^* -
\theta \theta^* )^{j} ( 1 - z_- {z}_-^* - \bar\theta\bar\theta^* )^{\bar{j}}\, ,
\ee
with
$$
z_\pm = z \pm \ft12 \bar\theta \theta\,,\qquad z_\pm^* = z^* \pm \ft12
\theta^*\bar\theta^*\,.
$$
In Eq.~\re{susy-scal}, the integration goes over a unit disk in the complex
$z-$plane, $d^2 z = dz dz^*$ and the integration over the Grassmann variables
is performed according to
\be
\int d\theta d\theta^* \lr{c_0 + c_1 \theta + c_2 \theta^* + c_3 \theta
\theta^*}=c_3
\ee
with arbitrary constants $c_i$. A similar relation holds upon the substitution
$(\theta,\theta^*) \mapsto(\bar\theta,\bar \theta^*)$.

The $SL(2|1)$ scalar product \re{scal} represents a natural generalization of
the $SL(2)$ scalar product for holomorphic functions $\phi(z)$
\be\label{sl2-scal}
\vev{\phi_2|\phi_1}_s = \frac{2s-1}{\pi} \int_{|z| \leq 1}  {d^2z}\, (1-z z^*)^{2s-2}
\lr{\phi_2(z)}^* \phi_1(z)\,.
\ee
Here the (half)integer positive $s$ defines the spin of the $SL(2)$ representation
to which the states $\phi_{1,2}(z)$ belong. We recall that the states
$\Phi(z,\theta,\bar\theta)$ can be decomposed over the $SL(2)$ multiplets
carrying the spins $\ell$, $\ell +\ft12$ and $\ell+1$, Eq.~\re{sl2-dec}. Indeed,
substituting $\Phi_{1,2}(\mathcal{Z})$  in \re{scal} with its expansion
\re{sl2-dec} in powers of  $\theta$ and $\bar\theta$ and performing the
integration over Grassmann variables, one can express the $SL(2|1)$ scalar product
$\vev{\Phi_2|\Phi_1}$ as a sum of the $SL(2)$ scalar products \re{sl2-scal}
between the functions $\phi(z)$, $\chi(z)$, $\bar\chi(z)$ and $\varphi(z)$. To
save space we do not present the explicit expression.

The hermitian conjugation of the $SL(2|1)$ generators $G_{j\bar j}$ with respect
to the scalar product \re{scal} is defined conventionally as
\be\label{dagger}
\langle \Phi_2 | \, G_{j\bar j} \Phi_1 \rangle = \langle G_{j\bar j}^\dagger
\Phi_2 | \Phi_1 \rangle \, .
\ee
Replacing the generators by their explicit expressions \re{trans1} --
\re{trans3}, one integrates by parts in both sides of \re{dagger} and finds
after some algebra
\be\label{herm}
\left( L^\pm \right)^\dagger = - L^\mp \, , \qquad \left( {\bar V}^\pm
\right)^\dagger =  V^\mp \, , \qquad  J^\dagger = J \, , \qquad
\bar J^\dagger = \bar J\,.
\ee
Using these relations, one verifies that the Casimirs \re{Casimir} are hermitian
operators, $\mathbb{C}_p^\dagger = \mathbb{C}_p$, and the scalar product
\re{scal} is invariant under (complexified) $SL(2|1)$ transformations \re{G},
$\delta_G \vev{\Phi_2|\Phi_1} = 0$.

Using the scalar product \re{scal} one can realize an arbitrary $SL(2|1)$
invariant operator $O$ as an integral operator on $\mathcal{V}_{j,\bar j}$
\be\label{O-gen}
O \cdot \Phi(\mathcal{W}) = \int [\mathcal{D} \mathcal{Z}]_{j\bar j} \,
{O}(\mathcal{W},\mathcal{Z}^*) \, \Phi(\mathcal{Z})
\ee
where $\Phi(\mathcal{W})$ is an arbitrary test function and the kernel of the
operator, ${O}(\mathcal{W}, \mathcal{Z}^*)$, depends on two sets of variables
$\mathcal{W}=(w,\vartheta,\bar\vartheta)$ and
$\mathcal{Z}^*=(z^*,\theta^*,\bar\theta^*)$. In particular, the unity operator in
$\mathcal{V}_{j,\bar j}$ has the following integral representation
\be\label{unit-O}
\II\cdot \Phi(\mathcal{W})  = \int [\mathcal{D} \mathcal{Z}]_{j\bar j} \,
\mathcal{K}_{j
\bar{j}} (\mathcal{W},{\mathcal{Z}}^*) \Phi (\mathcal{Z}) =\Phi (\mathcal{W})
\, ,
\ee
with the reproducing kernel $\mathcal{K}_{j\bar{j}}
(\mathcal{W},{\mathcal{Z}}^*)$ being
\be\label{K-kernel}
\mathcal{K}_{j \bar{j}} (\mathcal{W}, {\mathcal{Z}}^*) =  \left( 1 - w_- {z}_-^*
-
\bar\vartheta \bar\theta^* \right)^{- \bar{j}} \left( 1 - w_+ {z}_+^* -
\vartheta
\theta^* \right)^{- j} \, ,
\ee
where  $w_\pm = w \pm \ft12 \bar\vartheta \vartheta$ and ${z}_\pm^* = {z}^* \pm
\ft12 \theta^* \bar\theta^*$. To verify \re{unit-O} it suffices to substitute
$\Phi(\mathcal{Z})$ with one of the basis vectors \re{e's} and perform the
integration.

Let us demonstrate that the scalar product \re{scal} is positively definite for
$j,\bar j >0$. An arbitrary state $\Phi(\mathcal{Z})\in \mathcal{V}_{j\bar j}$ can
be decomposed over the graded basis \re{e's} as
\be\label{test-state}
\Phi(z,\theta,\bar\theta) = \sum_{n\ge 0} \chi_1(n)\cdot z_+^n + \chi_2(n)\cdot
z_-^n + \phi(n) \cdot \theta z^n + \bar\phi(n)\cdot \bar\theta z^n
\, ,
\ee
with $z_\pm = z\pm \ft12 \bar\theta\theta$. Calculating the scalar product of the
basis vectors, one can express the norm of $\Phi(z,\theta,\bar\theta)$ in terms
of the expansion coefficients $\chi_{1,2}$, $\phi$ and $\bar\phi$. The basis
vectors $z_\pm^n$, $\theta z^n$ and $\bar\theta z^n$ diagonalize simultaneously
the $U(1)$ charge $B=\ft12 (J-\bar J)$, Eq.~\re{trans3}, and the $SL(2)$ Cartan
operator $L^0=\ft12 (J+\bar J)$, Eq.~\re{trans3}. By virtue of \re{herm}, the two
operators are hermitian with respect to the scalar product \re{scal} and,
therefore, the basis vectors with different values of the $U(1)$ charge and the
$SL(2)$ spin are orthogonal to each other. As a result, the norm of the state
\re{test-state} is given by
\be\label{norm}
\langle \Phi | \Phi \rangle_{{j\bar j}} = \sum_{n\ge 0} \sigma(n) \left(
\sum_{i\!,\,k=1,2} \chi_i^*(n) g^{ik}(n) \chi_k(n) + \phi^*(n) \phi(n)/j +
\bar\phi^*(n)\bar\phi(n)/\bar j \right),
\ee
where the notation was introduced for $\sigma(n) = n! \Gamma(j+\bar
j+1)/\Gamma(n+1+j+\bar j)$ and
\be
g^{ik} =
\left[
\begin{array}{cc}
(j+n)/j & 1                 \\
1       & (\bar j+n)/\bar j \\
\end{array}
\right] .
\ee
This matrix is positively definite for $j,\bar j > 0$.

\subsubsection{Reduction to the chiral representation}

We demonstrated in Section~2.2.1 that for $\bar j=0$ the representation space
$\mathcal{V}_{j,0}$ contains an invariant subspace $\mathbb{V}_{j}$. It is
spanned by the states \re{chiral-irreps}, which admit the expansion
\re{test-state} with $\chi_2(n)=\bar\phi(n)=0$. According to \re{norm}, the
states $\Phi_+ \in \mathbb{V}_{j}$ have a finite norm with respect to the scalar
product \re{scal} as $\bar j\to 0$

Let us apply \re{scal} to determine the scalar product on $\mathbb{V}_{j}$. The
states $\Phi(z,\theta,\bar\theta)\in \mathbb{V}_{j}$ verify the chirality
condition $D\,\widehat\Phi(z,\theta,\bar\theta)=0$ and, as a consequence, their
dependence on $\bar\theta$ can be eliminated by a shift in $z$
\be\label{hat}
\Phi(z,\theta,\bar\theta) =\e^{\ft12\bar\theta\theta\partial_z} \widehat
\Phi\left(z,\theta\right) \,,
\ee
with $\widehat \Phi\left(z,\theta\right)=\Phi(z,\theta,0)$. Let substitute this
relation into \re{scal} and examine the integral in the right-hand side of
\re{scal} in the limit $\bar j\to 0$. One shifts the integration variables as
$z\to z-\ft12 \bar\theta\theta$, $z^*\to z^*-\ft12 \theta^*\bar\theta^*$,
performs integration over $\bar\theta$ and $\bar\theta^*$ and, finally, obtains
the scalar product on the space of functions $\widehat \Phi(z,\theta) \in
\mathbb{V}_{j}$ as
\be\label{scal-ind}
\vev{\widehat\Phi_2|\widehat\Phi_1}_{j} = \int [\mathcal{D} Z]_{j}\,
\lr{\widehat{\Phi}_2 \left( z, \theta \right)}^* \widehat{\Phi}_1 \left( z,
\theta\right) .
\ee
Here the integration goes over the unit disk in the complex $z-$plane and two
Grassmann variables
\be\label{Z-measure}
\int [\mathcal{D} Z]_{j} = \int_{|z|\le 1} d^2 z \int d\theta^*  d \theta \,
\mu_{j} (Z,  Z^*)
\ee
with $Z=(z,\theta)$, $Z^*=(z^*,\theta^*)$ and the integration measure given by
\be\label{mu-measure}
\mu_{j} (Z,Z^*)
=\frac1{\pi} (1-z z^*-\theta\theta^*)^{j-1}\,.
\ee
By construction, the scalar product \re{scal-ind} is invariant under the
$SL(2|1)$ transformations
\be
\delta_G \widehat \Phi(z,\theta) = \widehat G_{j}\cdot \widehat \Phi(z,\theta)\,,
\ee
with the operators $\widehat G_{j}$ related to the $SL(2|1)$ generators
\re{trans1} -- \re{trans3} as $\widehat G_{j} =
\e^{-\bar\theta\theta\partial_z/2}G_{j,0} \e^{\bar\theta\theta\partial_z/2}$.
They are given by differential operators acting on $z$ and $\theta$ variables
only.

Making use of \re{scal-ind}, one defines invariant operators on $\mathbb{V}_{j}$
\be\label{O-chi}
O \cdot \widehat\Phi({W}) = \int [\mathcal{D} {Z}]_{j} \,  {O}({W},{Z}^*) \,
\widehat\Phi({Z})\,,
\ee
where the kernel ${O}({W},{Z}^*)$ depends on $W=(w,\vartheta)$ and
$Z^*=(z^*,\theta^*)$. The unity operator takes the form
\be\label{unity}
\II\cdot \widehat\Phi({W})  = \int [\mathcal{D} {Z}]_{j} \, \mathbb{K}_{j}
({W},{{Z}}^*) \widehat\Phi ({Z}) =\widehat\Phi({W}) \, ,
\ee
with the reproducing kernel expressed by
\be
\label{ReprodKernel} \mathbb{K}_{j} ({W},{{Z}}^*) = \left( 1 - w {z}^* -
\vartheta\theta^* \right)^{-j} \, .
\ee
To verify the relation \re{unity} one substitutes a test function by its general
expression $\widehat\Phi({W}) = \sum_{n\ge 0} c_1 w^n + c_2\vartheta w^n$ and
performs the integration.

The relations \re{O-gen} and \re{O-chi} allow one to manipulate operators acting
on infinite-di\-men\-sional representation spaces, $\mathcal{V}_{j,\bar j}$ and
$\mathbb{V}_{j}$, respectively. For instance, the product of operators on
$\mathcal{V}_{j,\bar j}$ corresponds to the convolution of their integral kernels
\be\label{conv}
O_1\, O_2 \cdot \Phi(\mathcal{W}) =\int [\mathcal{D} \mathcal{Z}_1]_{j\bar j}
\int [\mathcal{D} \mathcal{Z}_2]_{j\bar j} \,  {O}_1(\mathcal{W},\mathcal{Z}_1^*)
{O}_2(\mathcal{Z}_1,\mathcal{Z}_2^*) \, \Phi(\mathcal{Z}_2) \, .
\ee
The supertrace \re{str} over $\mathcal{V}_{j,\bar j}$ then reads
\be
\str_{\mathcal{V}_{j,\bar j}} O  =\int [\mathcal{D} \mathcal{Z}]_{j\bar j} \,
{O}(\mathcal{Z},\mathcal{Z}^*)\,.
\ee
Finally, the operator acting on the tensor product $\mathcal{V}_{j_1\bar
j_1}\otimes\ldots \otimes \mathcal{V}_{j_N\bar j_N}$ is represented by an
integral kernel depending on $N$ pairs of coordinates
$\{\mathcal{W},\mathcal{Z}^*\}\equiv\{\mathcal{W}_k,\mathcal{Z}^*_k\, |\, 1\le k
\le N \}$
\be\label{int-repr}
O\cdot \Phi(\mathcal{W}_1,\ldots, \mathcal{W}_N) = \int [\mathcal{D}
\mathcal{Z}_1]_{j_1\bar j_1}\ldots \int [\mathcal{D} \mathcal{Z}_N]_{j_N\bar j_N}
\,{O}(\{\mathcal{W},\mathcal{Z}^*\})\,\Phi(\mathcal{Z}_1,\ldots,
\mathcal{Z}_N)\,.
\ee
Similar relations also hold for the operators in $\mathbb{V}_{j}$. In the next
Section, we will apply them to define the Baxter ${Q}-$operator as an integral
operator on the quantum space of the model.

\section{Baxter $\mathcal{Q}-$operators as transfer matrices}

The construction of noncompact integrable $SL(2|1)$ spin chains relies on the
$\mathcal{R}-$operator which depends on a spectral parameter and acts on the
tensor product of two infinite-dimensional $SL(2|1)$ representations as
\be\label{R-def}
\mathcal{R}(u): \quad \mathcal{V}_{j_1, \bar{j}_1} \otimes
\mathcal{V}_{j_2,\bar{j}_2} \mapsto \mathcal{V}_{j_1, \bar{j}_1}\otimes
\mathcal{V}_{j_2, \bar{j}_2}\,.
\ee
In addition, it obeys the Yang-Baxter equations
\ba \nonumber
\mathcal{R}_{12}
(u - v) \mathcal{R}_{13} (u) \mathcal{R}_{23} (v) &=& \mathcal{R}_{23} (v)
\mathcal{R}_{13} (u) \mathcal{R}_{12} (u - v) \, .
\\[2mm] \label{YB}
\mathcal{R}_{12} (u - v) L_{1} (u) L_{2} (v) &=& L_{2} (v) L_{1} (u)
\mathcal{R}_{12} (u - v) \, ,
\ea
where in the first relation both sides are defined on $\mathcal{V}_{j_1,
\bar{j}_1} \otimes \mathcal{V}_{j_2,
\bar{j}_2} \otimes \mathcal{V}_{j_3, \bar{j}_3}$ and each
$\mathcal{R}_{nm}-$operator acts on $n^{\rm th}$ and $m^{\rm th}$ spaces only,
for instance, $\mathcal{R}_{12}(u) = \mathcal{R}(u) \otimes \II$. In the second
relation, $L_k(u)$ is the $SL(2|1)$ Lax operator \cite{Kul85} acting on the
tensor product $\mathcal{V}_{j_k, \bar{j}_k}\otimes {\rm v}_1$ with ${\rm v}_1$
being the (fundamental) three-dimensional atypical $SL(2|1)$ representation,
Eq.~\re{v-chi}.

For compact (typical and atypical) $SL(2|1)$ representations, solutions to the
Yang-Baxter equation are well known~\cite{Kul85,Maa94}. In the context of
noncompact spins one has to deal however with solutions to \re{YB} for the
infinite-dimensional representations \re{R-def}. The latter have been studied in
Ref.~\cite{DerKarKir98}.

\subsection{Factorized $R-$matrix}

The Lax operator $L_k(u)$ is given by a $3\times 3$ graded matrix whose entries
are differential operators representing the $SL(2|1)$ generators on the `quantum
space' $\mathcal{V}_{j_k,\bar{j}_k}$ (see Eq.~\re{Lax} below). As such, $L_k(u)$
depends on three parameters -- two spins, $j_k$ and $\bar j_k$, and the spectral
parameter $u$. Solving the second relation in \re{YB}, it is convenient to view
the Lax operators $L_1(u)$ and $L_2(v)$ as functions of the following
combinations of the above parameters
\ba
\label{u's}
\begin{array}{lll}
u_1 = u + j_1 \, , \qquad & u_2 = u + j_1 - \bar{j}_1 \, , \qquad & u_3 = u -
\bar{j}_1
\\[2mm]
v_1 = v + j_2 \, , \qquad & v_2 = v + j_2 - \bar{j}_2 \, , \qquad & v_3 = v -
\bar{j}_2 \, ,
\end{array}
\ea
so that $L_1\equiv L_1(u_1,u_2,u_3)$ and $L_2\equiv L_2(v_1,v_2,v_3)$. Then, the
second relation in \re{YB} can be rewritten as
\be\label{YBRcheck}
\check{\mathcal{R}}_{12} (u - v) L_1 (u_1,u_2,u_3) L_2 (v_1,v_2,v_3) = L_1
(v_1,v_2,v_3) L_2 (u_1,u_2,u_3) \check{\mathcal{R}}_{12} (u - v)\,,
\ee
where $\check{\mathcal{R}}_{12}(u) = {\Pi}_{12} \mathcal{R}_{12}(u)$ and the
notation was introduced for the (graded) permutation operator ${\Pi}_{12}$. For
an arbitrary state in the tensor product $\mathcal{V}_{j_1,\bar{j}_1} \otimes
\mathcal{V}_{j_2, \bar{j}_2}$ it permutes the $\mathcal{Z} = ( z, \theta,
\bar\theta)-$coordinates in the two spaces according to
\be\label{Pi}
{\Pi}_{12} \cdot \Phi ( \mathcal{Z}_1, \mathcal{Z}_2) = \Phi ( \mathcal{Z}_2,
\mathcal{Z}_1) \, .
\ee
One notices that in \re{YBRcheck} the $\check{\mathcal{R}}_{12}-$operator
interchanges the arguments of two Lax operators, $(u_1,u_2,u_3)\rightleftarrows
(v_1,v_2,v_3)$. This transformation can be split into three steps, first,
exchanging $u_3 \rightleftarrows v_3$, then $u_2 \rightleftarrows v_2$ and,
finally, $u_1 \rightleftarrows v_1$. Each step is governed by a certain
$\mathcal{R}^{(a)}-$operator (with $a=1,2,3$) leading to the following
factorized expression for the $\mathcal{R}-$matrix \cite{Der05},
\be\label{R-fact}
\mathcal{R} (u - v) = {\Pi} \, \mathcal{R}^{(1)} (u_1 - v_1) \mathcal{R}^{(2)}
(u_2 - v_2) \mathcal{R}^{(3)} (u_3 - v_3) \, .
\ee
The order in which the spectral parameters $u_j$ and $v_j$ are interchanged in
\re{YBRcheck} is not important. This allows one to write down six different
expressions for the $\mathcal{R}-$operator containing the product of operators
$\mathcal{R}^{(a)}(u_a-v_a)$ but in different order. One can show that these
expressions coincide up to an overall normalization factor.

There is the following important difference between the operators $\mathcal{R}(u)$
and  $\mathcal{R}^{(a)}(u)$. In distinction with the former, the
$\mathcal{R}^{(a)}-$operators map $\mathcal{V}_{j_1,\bar j_1}\otimes
\mathcal{V}_{j_2,\bar j_2}$ into the tensor product of two yet another $SL(2|1)$
representations:
\ba
&& \mathcal{R}^{(1)}(u): \quad \mathcal{V}_{j_1,\bar j_1}\otimes
\mathcal{V}_{j_2,\bar j_2} \mapsto  \mathcal{V}_{j_1,\bar j_1-u}\otimes
\mathcal{V}_{j_2,\bar j_2+u}
\nonumber\\[2mm]
\label{R's} && \mathcal{R}^{(2)}(u): \quad \mathcal{V}_{j_1,\bar j_1}\otimes
\mathcal{V}_{j_2,\bar j_2} \mapsto  \mathcal{V}_{j_1-u,\bar j_1+u}\otimes
\mathcal{V}_{j_2+u,\bar j_2-u}
\\[2mm]
&& \mathcal{R}^{(3)}(u): \quad \mathcal{V}_{j_1,\bar j_1}\otimes
\mathcal{V}_{j_2,\bar j_2} \mapsto  \mathcal{V}_{j_1+u,\bar j_1}\otimes
\mathcal{V}_{j_2-u,\bar j_2} \, . \nonumber
\ea
For $u\neq 0$ the operators $\mathcal{R}^{(1)}(u)$ and $\mathcal{R}^{(3)}(u)$
only modify the spins in the antichiral and chiral sectors, respectively, while
the operator $\mathcal{R}^{(1)}(u)$ changes the spins in both sectors
simultaneously
\be
\mathcal{R}^{(a)}(u) \left( G_{j_1 \bar{j}_1} + G_{j_2
\bar{j}_2} \right) = \left( G_{j_1', \bar{j}_1'} + G_{j_2', \bar{j}_2'} \right)
\mathcal{R}^{(a)}(u) \, ,
\ee
and the sum of chiral and antichiral spins is separately preserved,
$j_1+j_2=j_1'+j_2'$ and $\bar j_1+\bar j_2=\bar j_1'+\bar j_2'$. Here $G_{j,\bar
j}$ denote the $SL(2|1)$ generators, Eqs.~\re{trans1} -- \re{trans3}, and the
spins $j'_{1,2}$ and $\bar j'_{1,2}$ can be read from the right-hand side of
\re{R's}. Examining the right-hand side of \re{R-fact} on the tensor product
$\mathcal{V}_{j_1,\bar j_1}\otimes \mathcal{V}_{j_2,\bar j_2} $, one finds from
\re{R's} that the $\mathcal{R}^{(1)}-$ and $\mathcal{R}^{(2)}-$operators act on
different tensor products $ \mathcal{V}_{j_1',\bar j_1'}\otimes
\mathcal{V}_{j_2',\bar j_2'}$ with spins $j_{1,2}'$ and $\bar j_{1,2}'$ depending
on the $u-$ and $v-$parameters. One can check using \re{Pi} that the
$\mathcal{R}-$operator \re{R-fact} satisfies \re{R-def} provided that these
parameters verify the relations $(u_2 - u_3) - (v_2 - v_3) = j_1 - j_2$ and $(u_1
- u_2) - (v_1 - v_2) =
\bar{j}_1 - \bar j_2$, in agreement with \re{u's}.

The operators $\mathcal{R}^{(a)}(u)$ (with $a=1,2,3$) possess a number of
remarkable properties:
\begin{itemize}
\item For $u=0$, the $\mathcal{R}^{(a)}-$operator does not affect the
arguments of the Lax operators and, as a consequence, $\mathcal{R}^{(a)}
(u=0)$ is proportional to the identity operator. We choose the normalization
as
\be\label{R=1}
\mathcal{R}^{(a)}(u=0) = \II \qquad (a=1,2,3) \, .
\ee
\item In virtue of \re{R=1}, for special values of the spins $j_2$ and
$\bar{j}_2$, the operator $\mathcal{R}(u)$ reduces to a single
$\mathcal{R}^{(a)}-$operator
\be
\label{cases} \mathcal{R}(u) = \left \{
\begin{array}{ll}
{\Pi}\, \mathcal{R}^{(3)}(u)\,, & \mbox{for $(j_2=j_1+u\,,\
\bar j_2=\bar j_1)$} \\[2mm]
{\Pi}\,\mathcal{R}^{(2)}(-u)\,, & \mbox{for $(j_2=j_1+u\,,\
\bar j_2=\bar j_1-u)$} \\[2mm]
{\Pi}\, \mathcal{R}^{(1)}(u)\,, & \mbox{for $(j_2=j_1\,,\
\bar j_2= \bar j_1 -u)$}
\end{array}
\right.
\ee
We recall that the $\mathcal{R}-$operator acts on the tensor product
$\mathcal{V}_{j_1,\bar j_1}\otimes \mathcal{V}_{j_2,\bar j_2}$, Eq.~\re{R-def},
and, in all three cases in \re{cases}, the space $\mathcal{V}_{j_2,\bar j_2}$
depends explicitly on the spectral parameter $u$.
\item For $a>b$, the operators $\mathcal{R}_{12}^{(a)}(u)$ and
$\mathcal{R}_{23}^{(b)}(v)$
commute with each other
\be
\label{CommRuleRR} \mathcal{R}_{12}^{(a)}(u) \mathcal{R}_{23}^{(b)}(v) =
\mathcal{R}_{23}^{(b)}(v)\mathcal{R}_{12}^{(a)}(u)\,,
\ee
where the definition of the $\mathcal{R}^{(a)}_{nm}-$operators on
$\mathcal{V}_{j_1,\bar j_1} \otimes \mathcal{V}_{j_2,\bar j_2} \otimes
\mathcal{V}_{j_3,\bar j_3}$ is analogous to that of the
$\mathcal{R}_{nm}-$operators in \re{YB}.
\end{itemize}
It is straightforward to verify that the operators entering both sides of
\re{CommRuleRR} interchange the argument of three Lax operators $L_1L_2L_3$ in
the same way and, therefore, they are proportional to each other. Explicit
calculations show that the corresponding proportionality factor equals $1$ for
$a>b$ and it is different from $1$ for $a<b$.

\begin{figure}[t]
\psfrag{w1}[cc][cc]{$\mathcal{Z}_1^*$} \psfrag{w2}[cc][cc]{$\mathcal{W}_1$}
\psfrag{z1}[cc][cc]{$\mathcal{Z}_2^*$} \psfrag{z2}[cc][cc]{$\mathcal{W}_2$}
\psfrag{A}[cc][cc]{${\rm (a)}$} \psfrag{B}[cc][cc]{${\rm (b)}$}
\psfrag{j1}[cc][cc][0.8][90]{$(j_1,\bar j_1)$}
\psfrag{j2}[cc][cc][0.8][90]{$(j_2,\bar j_2)$}
\psfrag{jj1}[cc][cc][0.8][90]{$(j_1+u,\bar j_1)$}
\psfrag{jj2}[cc][cc][0.8][90]{$(j_2,\bar j_2+u)$}
\psfrag{u0}[cc][cc][0.8][45]{$(0,-u)$} \psfrag{0u}[cc][cc][0.8][-45]{$(-u,0)$}
\centerline{\includegraphics[scale=0.7]{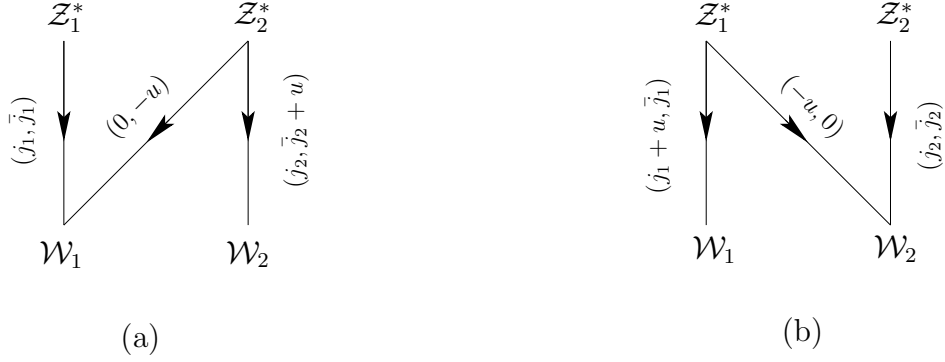}}
\caption[]{Diagrammatical representation of the integral operators
$\mathcal{R}^{(1)}(u)$ and $\mathcal{R}^{(3)}(u)$, Eq.~\re{R13-kernels}.
The arrow line with the index $(\alpha,\bar\alpha)$ and the end-points
$\mathcal{W}$ and $\mathcal{Z}^*$ represents for the kernel
$\mathcal{K}_{\alpha,\bar\alpha}(\mathcal{W},\mathcal{Z}^*)$. } \label{fig:Rk}
\end{figure}

Similarly to \re{O-gen}, the operators $\mathcal{R}^{(a)}(u)$ can be realized on
the tensor product of graded linear spaces $\mathcal{V}_{j_1,
\bar{j}_1} \otimes \mathcal{V}_{j_2, \bar{j}_2}$ as integral operators
\be\label{R-i}
\mathcal{R}^{(a)}(u)\, \Phi ( \mathcal{W}_1, \mathcal{W}_2 ) = \int
{}[\mathcal{D} \mathcal{Z}_1]_{j_1 \bar{j}_1} [\mathcal{D} \mathcal{Z}_2]_{j_2
\bar{j}_2}\, {R}^{(a)}_u (\mathcal{W}_1, \mathcal{W}_2 ; \mathcal{Z}^\ast_1,
\mathcal{Z}^\ast_2) \,\Phi (\mathcal{Z}_1, \mathcal{Z}_2) \, ,
\ee
where the integration measure $\int[\mathcal{D} \mathcal{Z}]_{j\bar{j}}$ is
defined in \re{susy-scal} and $\Phi (\mathcal{W}_1, \mathcal{W}_2)$ is a test
function on $\mathcal{V}_{j_1,\bar{j}_1} \otimes \mathcal{V}_{j_2, \bar{j}_2}$.
As in \re{conv}, the product of $\mathcal{R}^{(a)}-$operators is represented by
a convolution of the corresponding integral kernels. The integral kernels of the
operators $\mathcal{R}^{(1)}(u)$ and $\mathcal{R}^{(3)}(u)$ can be expressed in
terms of the $SL(2|1)$ reproducing kernels \re{K-kernel}
\ba\nonumber
{R}^{(1)}_u (\mathcal{W}_1, \mathcal{W}_2 ; \mathcal{Z}^\ast_1,
\mathcal{Z}^\ast_2) &=& r^{(1)}_u\,\mathcal{K}_{j_1,\bar j_1}
(\mathcal{W}_1,\mathcal{Z}_1^\ast)\mathcal{K}_{0,-u}
(\mathcal{W}_1,\mathcal{Z}_2^\ast)\mathcal{K}_{j_2,\bar j_2+u}
(\mathcal{W}_2,\mathcal{Z}_2^\ast)
\\[2mm] \label{R13-kernels}
{R}^{(3)}_u (\mathcal{W}_1, \mathcal{W}_2 ; \mathcal{Z}^\ast_1,
\mathcal{Z}^\ast_2) &=& r^{(3)}_u\,\mathcal{K}_{j_1+u,\bar j_1}
(\mathcal{W}_1,\mathcal{Z}_1^\ast)\mathcal{K}_{-u,0}
(\mathcal{W}_2,\mathcal{Z}_1^\ast)\mathcal{K}_{j_2,\bar j_2}
(\mathcal{W}_2,\mathcal{Z}_2^\ast)\,.
\ea
The factors $r^{(1)}_u$ and $r^{(3)}_u$ fix the normalization of the operators.
Let us choose them as %
\footnote{We introduced the phases $\e^{\pm i \pi u/2}$ to avoid factors
$(-1)^\#$ in expressions for the transfer matrices (see Eq.~\re{T=t+T} below).}
\begin{align} \nonumber
r^{(1)}_u&=\e^{-i\pi u/2}\,  \frac{ \bar j_2}{\bar j_2+u}\,
\frac{\Gamma(u+j_2+\bar j_2+1)}{\Gamma(j_2+\bar j_2+1)}
\,, \\[2mm] \label{r13}
r^{(3)}_u&=\e^{i\pi u/2}\,  \frac{\Gamma(u+j_1+\bar j_1+1)}{\Gamma(j_1+\bar
j_1+1)}\,.
\end{align}
It is convenient to introduce a diagrammatical representation of the kernels
\re{R13-kernels} in terms of Feynman diagrams. Let us represent the reproducing
kernel $\mathcal{K}_{j,\bar j} (\mathcal{W},\mathcal{Z}^\ast)$ as an arrow line
connecting the points $\mathcal{W}$ and $\mathcal{Z}^\ast$ and carrying the pair
of indices $j,\bar j$. Then, the kernels of the $\mathcal{R}^{(1)}-$ and
$\mathcal{R}^{(3)}-$operators, Eq.~\re{R13-kernels}, can be depicted as two
`zig-zag' diagrams shown in Fig.~\ref{fig:Rk}(a) and (b), respectively.

The remaining operator $\mathcal{R}^{(2)}(u)$ can be realized as a differential
operator acting on $\mathcal{V}_{j_1,\bar j_1} \otimes \mathcal{V}_{j_2,\bar
j_2}$
\be\label{R2}
\mathcal{R}^{(2)}(u)=\frac{j_2}{j_2+u}\left[\lr{1+u\frac{\theta_{12}
\bar D_{2}}{j_2}}\lr{1-u\frac{\bar\theta_{12}  D_{1}}{\bar j_1}} +
{u}\left(z_{1+}-z_{2-}+\theta_1\bar\theta_2\right)\frac{D_1\bar  D_2}{\bar
j_{1}j_{2}}\right],
\ee
where the notation was introduced for $z_{k\pm}=z_k\pm\ft12\bar\theta_k\theta_k$,
$\theta_{12}=\theta_1-\theta_2$ and $\bar\theta_{12}=\bar\theta_1-\bar\theta_2$.
Also, $D_1=-\partial_{\bar\theta_1}+\ft12 \theta_1\partial_{z_1}$ and $\bar
D_2=-\partial_{\theta_2}+\ft12 \bar\theta_2\partial_{z_2}$ denote the
supercovariant derivatives \re{cov-der} acting on the first and second arguments
of $\Phi(\mathcal{Z}_1,\mathcal{Z}_2)\in \mathcal{V}_{j_1,\bar j_1} \otimes
\mathcal{V}_{j_2,\bar j_2}$, respectively. The same operator can be realized as
an integral operator \re{R-i} with the kernel
\be\label{R-last}
{R}^{(2)}_u (\mathcal{W}_1, \mathcal{W}_2 ; \mathcal{Z}^\ast_1,
\mathcal{Z}^\ast_2) = \mathcal{R}^{(2)}(u)\cdot \mathcal{K}_{j_1,\bar j_1}
(\mathcal{W}_1,\mathcal{Z}_1^\ast) \mathcal{K}_{j_2,\bar j_2}
(\mathcal{W}_2,\mathcal{Z}_2^\ast),
\ee
where supercovariant derivatives $\bar D_1$ and $D_2$ entering \re{R2} act on
$\mathcal{W}_1$ and $\mathcal{W}_2$, respectively. Notice that the operator
$(u+j_2)\mathcal{R}^{(2)}(u)$ is a quadratic function of the spectral parameter
$u$ with operator-valued coefficients.

\subsection{Definition of $\mathcal{Q}-$operators}

Having the solutions to the Yang-Baxter equation \re{YB} at our disposal, we can
construct the transfer matrix for the $SL(2|1)$ spin chain of length $N$. The
Hilbert space in each site is identified with the $SL(2|1)$ representation space
$\mathcal{V}_{j_q \bar{j}_q}$. The quantum space of the model is given by the
direct product of the Hilbert spaces over the entire lattice with spins taking
the same values in all sites
\be
\label{HN}
\mathcal{H}_N
=
\underbrace{
\mathcal{V}_{j_q \bar{j}_q} \otimes \ldots \otimes \mathcal{V}_{j_q\bar j_q}
}_{N}
\, .
\ee
Let $\mathcal{V}_{j \bar{j}}$ be some reference $SL(2|1)$ representation space
(see Table~\ref{Table1}) and let us denote by $\mathcal{R}_{n0}(u)$ the
$\mathcal{R}-$operator acting on the tensor product of a quantum space in the
$n^{\rm th}$ site and the auxiliary space $\mathcal{V}_{j\bar{j}}$. By
definition~\cite{QISM}, the transfer matrix $\mathcal{T}_{j \bar{j}}(u)$ is
defined as a
supertrace of their product over all sites 
\be
\label{Eq1} \mathcal{T}_{j \bar{j}}(u) = {\str}_{\mathcal{V}_{j \bar{j}}} \left[
\mathcal{R}_{N0} (u) \ldots \mathcal{R}_{10} (u) \right] \, .
\ee
It follows from the Yang-Baxter equation \re{YB} that the transfer matrices with
different values of spins in the auxiliary space form a commutative family of
operators
\be\label{T-com}
{}[ \mathcal{T}_{j \bar{j}} (u) , \mathcal{T}_{j'\!,\, \bar{j}'} (v) ] = 0 \, ,
\ee
and, therefore, they serve as generating functionals of the Hamiltonian and of
(an infinite number of) integrals of motion.

Making use of \re{cases}, we obtain the following identities
\ba
&& \mathcal{T}_{j_q + u, \bar{j}_q} (u) 
= {\rm str}_{\mathcal{V}_{j_q + u, \bar{j}_q}} \left[ {\Pi}_{N0}
\mathcal{R}_{N0}^{(3)}(u) \ldots {\Pi}_{10} \mathcal{R}_{10}^{(3)}(u) \right]
, \nonumber\\
\label{T's}
&& \mathcal{T}_{j_q + u, \bar{j}_q - u} (u) 
={\rm str}_{\mathcal{V}_{j_q + u, \bar{j}_q - u}} \left[ {\Pi}_{N0}
\mathcal{R}_{N0}^{(2)} (-u) \ldots {\Pi}_{10} \mathcal{R}_{10}^{(2)} (-u)
\right]
, \\
&& \mathcal{T}_{j_q, \bar{j}_q - u}(u) 
={\rm
str}_{\mathcal{V}_{j_q,
\bar{j}_q - u}} \left[ {\Pi}_{N0} \mathcal{R}_{N0}^{(1)} (u) \ldots
{\Pi}_{10} \mathcal{R}_{10}^{(1)} (u) \right] , \nonumber
\ea
where $j_q$ and $\bar j_q$ are the $SL(2|1)$ spins of the quantum space \re{HN}
and $\Pi_{k0}$ is the graded permutation operator, Eq.~\re{Pi}. For $u=0$ one
applies \re{R=1} to get
\be\label{T=P}
\mathcal{T}_{j_q \bar{j}_q} (0) = \mathbb{P}\,,
\ee
where the notation was introduced for the operator of cyclic permutations on
\re{HN}
\be\label{P}
\mathbb{P} \, \Phi (\mathcal{Z}_1, \mathcal{Z}_2, {\dots}, \mathcal{Z}_N)
=
\Phi (\mathcal{Z}_2, \mathcal{Z}_3, {\dots}, \mathcal{Z}_{1}) \, .
\ee
Analogously to the $\mathcal{R}-$operator, Eq.~\re{R-fact}, a general
infinite-dimensio\-nal transfer matrix \re{Eq1} can be factorized into the
product of the three operators \re{T's}. Calculation goes along the same lines
as for the $SL(3)$ spin chain \cite{DerMan06} and details can be found in
Appendix~E. The resulting factorized expression for the transfer matrix reads
\be\label{T-factor}
\mathcal{T}_{j \bar{j}} (w) =\mathbb{P}^{-2 } \mathcal{T}_{j_q, \bar{j}_q - w_1}
(w_1) \mathcal{T}_{j_q - w_2, \bar{j}_q + w_2} (- w_2) \mathcal{T}_{j_q + w_3,
\bar{j}_q} (w_3) \, ,
\ee
with the spectral parameters  $w_1 = w - j + j_q$, $w_2 = w - j +
\bar{j} + j_q - \bar{j}_q$ and $ w_3 = w + \bar{j} - \bar{j}_q$.
In Eq.~\re{T-factor}, the dependence of the right-hand side on the spins of the
auxiliary space resides in the $w-$parameters. This suggests to introduce the
following operators
\ba
\mathcal{Q}_3 (u) \!\!\!&=&\!\!\! \mathcal{T}_{j_q -\bar j_q + u,
\bar{j}_q} (u-\bar j_q) \, , \nonumber
\\[2mm] \label{Q=T}
\mathcal{Q}_2 (u) \!\!\!&=&\!\!\! \mathcal{T}_{\bar j_q - u,  {j}_q + u} (\bar
j_q-j_q- u) \, ,
\\[2mm]
\mathcal{Q}_1(u) \!\!\!&=&\!\!\! \mathcal{T}_{j_q, \bar{j}_q-j_q - u} (u+j_q) \,
, \nonumber
\ea
and rewrite the infinite-dimensional transfer matrix \re{T-factor} as
\be
\mathcal{T}_{j \bar{j}} (u) = \mathbb{P}^{-2 } \mathcal{Q}_1 (u - j)
\mathcal{Q}_2 (u - j + \bar{j}) \mathcal{Q}_3 (u + \bar{j}) \, .
\ee
As follows from \re{T-com}, the $\mathcal{Q}-$operators defined in this way form
a commutative family of operators in the quantum space of the model \re{HN}.
Combining together \re{Q=T} and \re{T=P} one finds that the
$\mathcal{Q}-$operators coincide with the cyclic permutation operator for
special values of the spectral parameter
\be\label{Q=P}
\mathcal{Q}_1(- j_q)=\mathcal{Q}_2(\bar{j}_q - j_q) = \mathcal{Q}_3(\bar{j}_q) =
\mathbb{P}\,.
\ee
To identify the operators \re{Q=T} as Baxter operators for the $SL(2|1)$ spin
chain we have to establish the corresponding TQ-relations. This will be done in
Section~5.

\begin{figure}[t]
\psfrag{z1}[cc][cc]{$\mathcal{Z}_1^*$} \psfrag{w1}[cc][cc]{$\mathcal{W}_1$}
\psfrag{z2}[cc][cc]{$\mathcal{Z}_2^*$} \psfrag{w2}[cc][cc]{$\mathcal{W}_2$}
\psfrag{zn}[cc][cc]{$\mathcal{Z}_N^*$} \psfrag{w3}[cc][cc]{$\mathcal{W}_3$}
\psfrag{wn}[cc][cc]{$\mathcal{W}_N$}
\psfrag{a}[cc][cc]{$\boldsymbol{\alpha}_3$}
\psfrag{b}[cc][cc]{$\boldsymbol{\beta}_3$}
\centerline{\includegraphics[scale=0.6]{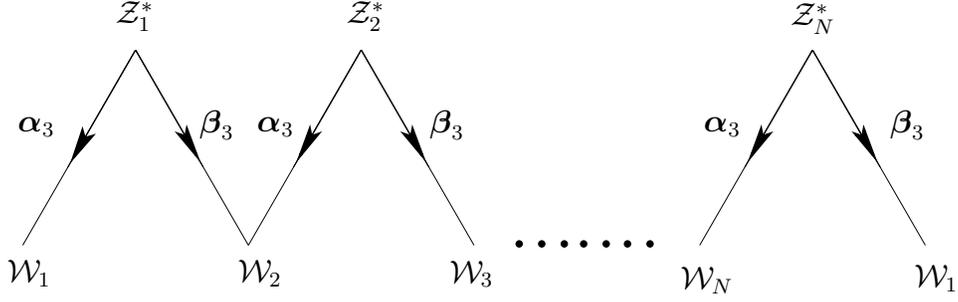}} \caption[]{Diagrammatical
representation of the operator $\mathcal{Q}_{3}(u+\bar j_q)$. The indices
specify the values of spins $\boldsymbol{\alpha}_3=(-u,0)$ and
$\boldsymbol{\beta}_3=(j_q+u,\bar j_q)$.} \label{fig:Q3}
\end{figure}

According to \re{Q=T}, the $Q-$operators are defined as infinite-dimensional
transfer matrices \re{T's} built from $\mathcal{R}^{(a)}-$operators. To determine
their explicit form, one has to evaluate supertraces over infinite-dimensional
graded spaces in the right-hand side of \re{T's}. This can be done using the
integral representation for the $\mathcal{R}^{(a)}-$operators, Eq.~\re{R-i}. In
this way, the Baxter $Q-$operators can be realized as integral operators
\re{int-repr} on the quantum space of the model \re{HN}
\be\label{Q=integral}
\mathcal{Q}_a(u)\cdot \Phi(\mathcal{W}_1,\ldots,\mathcal{W}_N) = \int
[\mathcal{D} \mathcal{Z}_1]_{j_q\bar j_q}\ldots \int [\mathcal{D}
\mathcal{Z}_N]_{j_q\bar j_q}\,
{Q}^{(a)}_u(\{\mathcal{W},\mathcal{Z}^*\})\,\Phi(\mathcal{Z}_1,\ldots,
\mathcal{Z}_N)\,,
\ee
with $\{\mathcal{W},\mathcal{Z}^*\}\equiv\{\mathcal{W}_k,\mathcal{Z}^*_k\, |\,
1\le k \le N \}$. Let us start with the operator $\mathcal{Q}_3(u)$ and examine
the operator $\Pi_{k0} \mathcal{R}_{k0}^{(3)}(u)$ entering the first relation in
\re{T's}
\ba \nonumber
\Pi_{k0} \mathcal{R}_{k0}^{(3)}(u) \Phi(\mathcal{W}_k,\mathcal{W}_0)\!\!\!\!\! &
&= \int [\mathcal{DZ}_k]_{j_q\bar j_q} \int [\mathcal{DZ}_0]_{j_0\bar
j_0}\,{R}^{(3)}_u (\mathcal{W}_0, \mathcal{W}_k ; \mathcal{Z}^\ast_k,
\mathcal{Z}^\ast_0)\Phi(\mathcal{Z}_k,\mathcal{Z}_0)
\\ \label{Pi-R}
\!\!\!\!\! & &= r^{(3)}_u\int [\mathcal{DZ}_k]_{j_q\bar j_q} \,
\mathcal{K}_{j_q+u,\bar j_q} (\mathcal{W}_0,\mathcal{Z}_k^\ast)\mathcal{K}_{-
u,0}
(\mathcal{W}_k,\mathcal{Z}_k^\ast)\Phi(\mathcal{Z}_k,\mathcal{W}_k)
\, ,
\ea
where the subscripts `0' and `$k$' refer to the auxiliary space and to the
quantum space in $k^{\rm th}$ site, respectively. Here in the second relation we
applied \re{R13-kernels} and performed $\mathcal{Z}_0-$integration with a help
of \re{unit-O}. Substituting \re{Pi-R} into the first relation in \re{T's} one
obtains the integral kernel of the operator $\mathcal{Q}_3(u)$
\be\label{Q3-kernel}
\mathcal{Q}_3(u+\bar j_q) : =
\rho_3(u) \, \prod_{k=1}^N \mathcal{K}_{-u,0}
(\mathcal{W}_k,\mathcal{Z}_k^\ast)\mathcal{K}_{j_q+u,\bar j_q}
(\mathcal{W}_{k+1},\mathcal{Z}_{k}^\ast)
\ee
where $\rho_3(u)=\left[\e^{i\pi u/2}{\Gamma(u+j_q+\bar j_q+1)}/{\Gamma(j_q+\bar
j_q+1)}\right]^N$ and periodic boundary conditions are imposed,
$\mathcal{W}_{N+1}=\mathcal{W}_{1}$. Using the diagrammatical technique
introduced earlier, the integral kernel \re{Q3-kernel} can be represented
as a zig-zag diagram shown in Fig.~\ref{fig:Q3}.

\begin{figure}[t]
\psfrag{z1}[cc][cc]{$\mathcal{Z}_N^*$} \psfrag{w1}[cc][cc]{$\mathcal{W}_1$}
\psfrag{z2}[cc][cc]{$\mathcal{Z}_1^*$} \psfrag{w2}[cc][cc]{$\mathcal{W}_2$}
\psfrag{z4}[cc][cc]{ } \psfrag{w3}[cc][cc]{$\mathcal{W}_3$}
\psfrag{wN}[cc][cc]{$\mathcal{W}_N$}
\psfrag{z3}[cc][cc]{$\mathcal{Z}_2^*$}\psfrag{w4}[cc][cc]{ }
\psfrag{zN}[cc][cc]{$\mathcal{Z}_{N-1}^*$}
\psfrag{a}[cc][cc]{$\boldsymbol{\alpha}_1$}
\psfrag{b}[cc][cc]{$\boldsymbol{\beta}_1$}
\psfrag{c1}[cc][cc]{$\boldsymbol{\alpha}_1$}
\centerline{\includegraphics[scale=0.8]{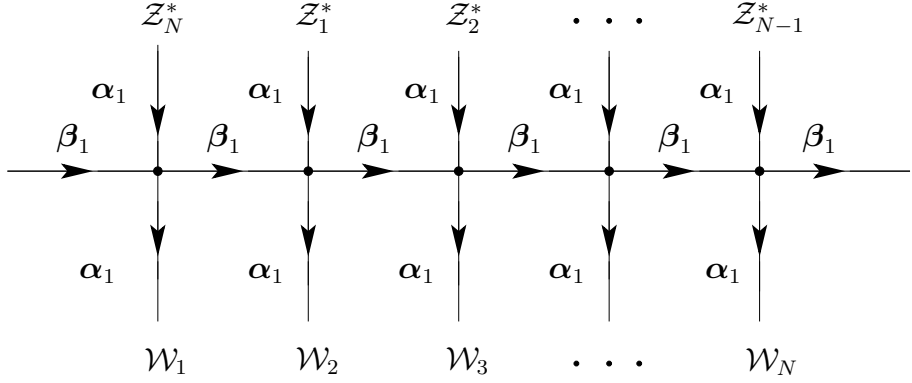}} \caption[]{Diagrammatical
representation of the operator $\mathcal{Q}_{1}(u-j_q)$. The indices specify the
values of spins $\boldsymbol{\alpha}_1=(j_q,\bar j_q)$ and
$\boldsymbol{\beta}_1=(0,-u)$. The leftmost incoming horizontal line is a
continuation of the rightmost outgoing horizontal line. An integration over
the position of `fat' vertices is implied with the measure \re{susy-scal} for $j=j_q$
and $\bar j=\bar j_q-u$.} \label{fig:Q1}
\end{figure}

Similar analysis can be performed for the operator $\mathcal{Q}_1(u)$. In that
case, $\mathcal{Z}_0-$integration in the expression for $\Pi_{k0}
\mathcal{R}_{k0}^{(1)}(u)$ can not be performed in a closed form and the
kernel of the operator $\mathcal{Q}_1(u)$ is given by an $N-$fold integral
over the auxiliary space shown diagrammatically in Fig.~\ref{fig:Q1}
\be\label{Q1-kernel}
\mathcal{Q}_1(u - j_q) : = \rho_1(u) \int \prod_{k=1}^N [\mathcal{D
Y}_{k}]_{j_q,\bar j_q-u}\, \mathcal{K}_{j_q,\bar j_q}
(\mathcal{W}_{k+1},\mathcal{Y}_{k}^\ast)\mathcal{K}_{j_q,\bar j_q}
(\mathcal{Y}_{k},\mathcal{Z}_{k}^\ast) \mathcal{K}_{0,-u}
(\mathcal{Y}_{k+1},\mathcal{Y}_{k}^\ast)\,,
\ee
with $\mathcal{Y}_{N+1}=\mathcal{Y}_1$ and $\rho_1(u)=[\e^{-i\pi u/2} (\bar
j_q-u)\Gamma(j_q+\bar j_q+1)/(\bar j_q\Gamma(j_q+\bar j_q+1-u)) ]^N$. We do not
display the explicit expression for the integral operators $\mathcal{Q}_2(u)$
since it will not be important for our purposes.

\subsection{$\mathcal{Q}-$operator for $\mathcal{N} = 1$ super-Yang-Mills
theory}

As was explained in the Introduction, infinite-dimensional $SL(2|1)$ spin chains
naturally appear in the context of the $\mathcal{N} = 1$ super-Yang-Mills theory
-- the Hamiltonian of the former coincides with the one-loop dilatation operator
of the latter, Eqs.~\re{DO} and \re{V-super}. The $\mathcal{N} = 1$ superfields
\re{Phi-dec} carry the $SL(2)$ spin $\ell=1$ and the $U(1)$ charge $b=1$, or
equivalently $j=2$ and $\bar j=0$, Eq.~\re{l,b}. As a consequence, they form an
irreducible chiral $SL(2|1)$ representation $[2]_+$ (see Table~\ref{Table1}) and
the dilatation operator \re{DO} acts on the Hilbert space
\be\label{H-N=1}
\mathcal{H}^{(\mathcal{N}=1)} = \underbrace{\mathbb{V}_2\otimes \ldots \otimes
\mathbb{V}_2}_N\,.
\ee
In this section, we will perform a reduction of the Baxter $Q-$operators
\re{Q=T}, defined on the quantum space of the model \re{HN} for arbitrary spins
$j_q$ and $\bar j_q$, to the quantum space \re{H-N=1} and establish a relation
between the dilatation operator \re{DO} and the $Q-$operators.

For $\bar j_q=0$, the linear graded space $\mathcal{V}_{j_q\!,0}$ has the
$SL(2|1)$ invariant subspace $\mathbb{V}_{j_q}$, Eq.~\re{red1}. Let us apply the
map $\mathcal{V}_{j_q\!,0}\mapsto \mathbb{V}_{j_q}$ to construct the
$Q-$operators \re{Q=T} in the quantum space $(\mathbb{V}_{j_q})^{\otimes {N}}$.
The states $\Phi\in (\mathcal{V}_{j_0,0})^{\otimes {N}} $ are projected onto
$\widehat{\Phi}\in (\mathbb{V}_{j_q})^{\otimes {N}}$ as
\be\label{Phi-hat-Phi}
\Phi(\mathcal{Z}_1,\ldots,\mathcal{Z}_N) = \prod_{k=1}^N
\e^{\ft12\bar\theta_k\theta_k\partial_{z_k}} \widehat\Phi(Z_1,\ldots,Z_N)\,,
\ee
or equivalently $ \widehat\Phi(Z_1,\ldots,Z_N) =
\Phi(\mathcal{Z}_1,\ldots,\mathcal{Z}_N)\big|_{\bar\theta_1=\ldots=\bar\theta_N=
0}$. Here, $\mathcal{Z}_k=(z_k,\theta_k,\bar\theta_k)$ and $Z_k=(z_k,\theta_k)$
(with $k=1,\ldots,N$). Going over to the Baxter $Q-$operators one finds
\be\label{Q=Q}
\mathbb{Q}_a(u) \widehat\Phi(Z_1,\ldots,Z_N)= \mathcal{Q}_a(u)
\Phi(\mathcal{Z}_1,\ldots,\mathcal{Z}_N)\big|_{\bar\theta_1=\ldots=\bar\theta_N=
0}
\,,
\ee
where $\mathcal{Q}_a(u)$ is given by \re{Q=T} and $\mathbb{Q}_a(u)$ denotes the
Baxter operator on $(\mathbb{V}_{j_q})^{\otimes {N}}$. In the right-hand side of
this relation, one applies \re{Phi-hat-Phi}, replaces $\mathcal{Q}_a(u)$ by its
integral representation \re{Q=integral}, performs the integration over $\int
d\bar\theta_k d\bar\theta_k^*$ (with $k=1,\ldots,N$) and obtains after some
algebra
\be\label{Qa}
\mathbb{Q}_a(u) \widehat\Phi(W_1,\ldots,W_N) =  \int [\mathcal{D}
{Z}_1]_{j_q}\ldots \int [\mathcal{D} {Z}_N]_{j_q}\,
\widehat{Q}^{(a)}_u(\{{W},{Z}^*\})\,\widehat\Phi({Z}_1,\ldots, {Z}_N)\,.
\ee
Here, the integration measure $\int [\mathcal{D}Z]_{j_q}$ was defined in
Eqs.~\re{Z-measure} and \re{mu-measure}. The kernel
$\widehat{Q}^{(a)}_u(\{{W},{Z}^*\})$ depends on the variables
$W_k=(w_k,\vartheta_k)$ and $Z_k^*=(z_k^*,\theta_k^*)$ (with $k=1,\ldots,N$).

Equation \re{Qa} defines the $Q-$operators acting on the quantum space
$\mathbb{V}_{j_q}\otimes \ldots \otimes \mathbb{V}_{j_q}$. Let us apply it to
construct the operator $\mathbb{Q}_3(u)$. One substitutes \re{Q3-kernel} into the
right-hand side of \re{Q=Q} and notices that for $\bar j_q\to 0$ the
normalization factor $r_u^{(3)}$ in \re{Q3-kernel} scales as $1/\bar j_q$.
Carefully examining the right-hand side of \re{Q=Q} one finds that
$\widehat{Q}^{(3)}_u(\{{W},{Z}^*\})$ approaches a finite value as $\bar j_q\to 0$
\be\label{Q-prop}
\widehat{Q}^{(3)}_u(\{{W},{Z}^*\})= \rho(u) \prod_{k=1}^N (1 - w_k z_k^* -
\vartheta_k \theta_k^*)^{ u} (1 - w_{k+1} {z}_{k}^* -
\vartheta_{k+1}\theta_{k}^\ast)^{- j_q - u}
\ee
with $\rho(u) = \left[ \e^{i\pi u/2}\Gamma(u+j_q+1)/\Gamma(j_q+1)\right]^N$.
Here, ${W}_k=(w_k,\vartheta_k)$ and ${Z}_k^*=(z_k^*,\theta_k^*)$ and periodic
boundary conditions $W_{N+1}=W_1$ are imposed. It convenient to remove the
factor $\rho(u)$ from the right-hand side of \re{Q-prop} by changing the
normalization of the operator $\mathbb{Q}_3(u)\to \mathbb{Q}_3(u)/\rho(u)$.
The resulting expression for the integral kernel can be then rewritten in
terms of the reproducing kernels \re{ReprodKernel} as
\be\label{Q=N1}
\mathbb{Q}_3(u)\big|_{(\mathbb{V}_{j_q})^{\otimes {N}}} : =  \prod_{k=1}^N
\mathbb{K}_{-u} ({W}_k,{Z}_k^\ast)\mathbb{K}_{u+j_q} ({W}_{k+1},{Z}_{k}^\ast)\,.
\ee
Analogously to \re{Q3-kernel}, this expression admits a diagrammatical
representation as a zig-zag diagram shown in Fig.~\ref{fig:Q3}. %

The operator $\mathbb{Q}_3(u)$ admits yet another integral representation which
is extremely useful for comparison with the $\mathcal{N}=1$ dilatation operator
\re{DO} and \re{V-super}. It is based on the following identity
\be
\frac{1}{A^a B^b} = \frac{\Gamma (a + b)}{\Gamma (a) \Gamma (b)} \int_0^1 d
\alpha \, \frac{\alpha^{a - 1} (1 - \alpha)^{b - 1}}{[\alpha A + (1 - \alpha)
B]^{a + b}} \, .
\ee
Applying this transformation to \re{Q-prop}, one can combine two $Z_k-$dependent
`propagators' in the right-hand side of \re{Q-prop} into a single factor $(1 -
w_{k}' {z}_{k}^* - \vartheta_{k}'\theta_{k}^\ast)^{- j_q }$ (with $w_k'=w_k\alpha
+ w_{k+1}(1-\alpha)$ and $\theta_k'=\theta_k\alpha + \theta_{k+1}(1-\alpha)$)
which coincides in its turn with the reproducing kernel $\mathbb{K}_{j_q} (
\alpha W_k + \lr{1-\alpha} W_{k+1}, Z_k^*)$, Eq.~\re{ReprodKernel}. Then, the
subsequent $Z_k-$integration is trivial making use of the property of the
reproducing kernel \re{unity}. Finally, we obtain a multiple contour integral
representation for the Baxter operator
\ba \nonumber
&& \mathbb{Q}_3 (u) \Phi ({W}_1, \dots ,{W}_N) {=} \left[\frac{\Gamma (j_q
)}{\Gamma(-u)\Gamma (u + j_q )} \right]^N \int_0^1 \prod_{k = 1}^N d \alpha_k \,
\alpha_k^{-u - 1}(1-\alpha_k)^{j_q + u-1}
\\ \label{Q=shift}
 &&\qqqquad  \times\Phi \left( \alpha_1 {W}_1 + (1-\alpha_1)
{W}_2 \, , \dots \, , \alpha_N {W}_N + (1-\alpha_N){W}_1 \right) \, ,
\ea
where ${W}_k = (w_k, \vartheta_k)$ and the notation was introduced for a linear
combination of (super-)coor\-di\-nates $\alpha W +\beta W'=(\alpha w + \beta w',
\alpha \vartheta+\beta\vartheta')$.

Let us substitute $\vartheta_1=\ldots=\vartheta_N=0$ in both sides of
\re{Q=shift}. In this limit, the superfield $\Phi(W)$, Eq.~\re{Phi-dec}, reduces
to its lowest component $\chi(w_k)$ which, in its turn, belongs to the
$SL(2)\otimes U(1)$ multiplet $\mathcal{D}_\ell(b)$ with $\ell=b=j_q/2$,
Eq.~\re{sl2-dec}. Then, the $Q-$operator in \re{Q=shift} acts on the tensor
product $(\mathcal{D}_\ell(b))^{\otimes N}$ and coincides with the known
expression for the $SL(2)$ Baxter operator $\mathbb{Q}_+
(u)$~\cite{Der99,DerKorMan03}
\be\label{Q-reduction}
\left[\mathbb{Q}_3 (u)\, \Phi ({W}_1, \dots ,{W}_N)\right]\big|_{\vartheta_k=0}
=
\mathbb{Q}^{\rm SL(2)}_+ (u+\ell)\left[\Phi ({W}_1,
\dots ,{W}_N)\right]\big|_{\vartheta_k=0}\,.
\ee
In other words, the operator \re{Q=shift} can be considered as a lift of the
$SL(2)$ Baxter $Q-$operator from the light-cone into the superspace
$W=(w,\vartheta)$. The relation \re{Q-reduction} also suggests that for
eigenstates of the $SL(2|1)$ spin chain independent of `odd'
$\vartheta-$variables, the corresponding eigenvalues of the $Q-$operator coincide
with eigenvalues of the $SL(2)$ Baxter $Q-$operator. We will return to this issue
in Section~6.3.

Let us establish the relation between the $Q-$operator \re{Q=shift} and the
dilatation operator \re{DO} and \re{V-super}. As a first step, one
examines \re{Q=shift} for $u\to 0$.  In this limit, the leading contribution
to the right-hand side of \re{Q=shift} comes from the integration in the
vicinity of $\alpha_k = 0$. Expanding the integrand in powers of $\alpha_k$
and performing the integration one gets
\be\label{Q+}
\mathbb{Q}_3(u) \Phi ({W}_1, \dots ,{W}_N)  = \mathbb{P} \left[1 + u
\,\mathbb{H}_N^+ + \mathcal{O}(u^2)\right] \Phi ({W}_1, \dots ,{W}_N)
\, ,
\ee
where $\mathbb{Q}_3(u=0)=\mathbb{P}$ is the operator of cyclic permutations
\re{P}, in agreement with the normalization condition \re{Q=P} (for $\bar
j_q=0$). The operator $\mathbb{H}_N^+$ has a structure of a nearest-neighbor
Hamiltonian, $\mathbb{H}_N^+=H_{12}^+ +\ldots+H_{N1}^+$, with the two-particle
kernel $H_{k,k+1}^+$ acting locally in $k^{\rm th}$ and $(k+1)^{\rm th}$ sites
as
\begin{eqnarray}\nonumber
\lefteqn{H_{k,k+1}^+ \, {\Phi}(...,Z_k,Z_{k+1},...) = \int_0^1
\frac{d\alpha}{\alpha} (1-\alpha)^{j_q-1}\, }&&
\\ \label{H+}
&& \qqqquad \times\bigg\{ \Phi(...,Z_k,Z_{k+1},...) -  {\Phi} (...,Z_{k}, \alpha
Z_k + (1-\alpha) Z_{k+1},...)
 \bigg\} \, .
\end{eqnarray}
Next, we examine the expansion of $\mathbb{Q}_3(u)$ around $u=-j_q$, or equivalently
$\mathbb{Q}_3(-u-j_q)$ for $u\to 0$. In this case, the leading contribution to
the right-hand side of \re{Q=shift} comes from the integration in the vicinity of
$\alpha_k=1$ and one gets
\be\label{Q-}
\mathbb{Q}_3(u-j_q) \Phi ({W}_1, \dots ,{W}_N)  =  \left[1 - u \,\mathbb{H}_N^-
+
\mathcal{O}(u^2)\right] \Phi ({W}_1, \dots ,{W}_N)
\, ,
\ee
where $\mathbb{H}_N^-=H_{12}^- +\ldots+H_{N1}^-$ and
\begin{eqnarray}\nonumber
\lefteqn{H_{k,k+1}^- \, {\Phi}(...,Z_k,Z_{k+1},...) = \int_0^1
\frac{d\alpha}{\alpha} (1-\alpha)^{j_q-1}\, }&&
\\ \label{H-}
&& \qqqquad \times\bigg\{ \Phi(...,Z_k,Z_{k+1},...) -  {\Phi} (..., (1-\alpha)
Z_{k}+\alpha Z_{k+1},Z_{k+1},...)
 \bigg\} \, .
\end{eqnarray}
We observe that the Hamiltonians $H_{k,k+1}^+$ and $H_{k,k+1}^-$ are not
invariant under the permutations of $k^{\rm th}$ and $(k+1)^{\rm th}$ particles
while their sum is.

The Hamiltonian of the $SL(2|1)$ spin chain is
defined as %
\footnote{We will demonstrate in Section~6.2.3, that this Hamiltonian naturally
arises as the first term in the expansion of the chiral transfer matrix
$\mathbb{T}_{j_q}(u)$ (see Table~\ref{Table1}) around $u=0$.}
\ba\label{H-(j+1)}
\mathbb{H}_{(\mathbb{V}_{j_q})^{\otimes N}} = \mathbb{H}_N^+ +
\mathbb{H}_N^-=H_{12} + \ldots + H_{N1}\,,
\ea
with $H_{k,k+1} = H_{k,k+1}^++H_{k,k+1}^-$. In \re{H-(j+1)}, the subscript in
the left-hand side indicates that the quantum space of the model. Making use of
\re{Q+} and \re{Q-}, the Hamiltonian and the operator of cyclic permutations can
be expressed as
\ba \nonumber
&& \mathbb{H} = \lr{\ln \mathbb{Q}_3(0)}' - \lr{\ln \mathbb{Q}_3(-j_q)}'\,,
\\[2mm] \label{H-SL21}
&& \mathbb{P}  = \mathbb{Q}_3(0)/\mathbb{Q}_3(-j_q)\,,
\ea
where prime in the first relation denotes a derivative with respect to the
spectral parameter. Written in this form, the two operators do not depend on the
normalization of the $Q-$operators. The first relation in \re{H-SL21} has a
striking similarity with a similar relation for the $SL(2)$ spin chain,
Eq.~\re{H-SL2}. Moreover, it follows from \re{Q-reduction} that the $SL(2|1)$
Hamiltonian coincides with the $SL(2)$ Hamiltonian \re{H-SL2} when projected
onto eigenstates independent of `odd' $\vartheta-$variables.

Comparing the $SL(2|1)$ Hamiltonian, Eqs.~\re{H-(j+1)}, \re{H+} and \re{H-}, with
the $\mathcal{N}=1$ dilatation operator, Eqs.~\re{DO} and \re{V-super}, one
concludes that the two operators coincide for $j_q=3-\mathcal{N}=2$ up to an
additive c-number correction. There is however the following difference between
the two models. The dilatation operator acts on the single-trace operators \re{O}
which are invariant under cyclic permutation of the superfields. This leads to
the additional constraint $\mathbb{P}=\II$ in the $\mathcal{N}=1$ theory.

\subsection{Analytical properties of the $\mathcal{Q}-$operators}

The $\mathcal{Q}-$operators satisfy finite-difference TQ-equations whose explicit
form will be established in Section~5. To determine uniquely their solutions, one
has to specify analytical properties of the $\mathcal{Q}-$operators as functions
of the spectral parameter $u$. According \re{Q=T}, the $u-$dependence enters
through the arguments of the transfer matrices and spins of the auxiliary space.
In this subsection, we will use \re{Q=T} to determine analytical properties of
the operators $\mathcal{Q}_a(u)$.

Let us start with the operator $\mathcal{Q}_2(u)$. It follows from \re{Q=T} and
\re{T's} that it is given by
\be\label{Q2=T}
\mathcal{Q}_2(u-j_q+\bar j_q) = \str_{\mathcal{V}_{j_q-u,\bar j_q+u}}
\left[\Pi_{N0} \mathcal{R}_{N0}^{(2)}(u)\ldots \Pi_{10}
\mathcal{R}_{10}^{(2)}(u)\right].
\ee
Here the $\mathcal{R}^{(2)}-$operators in all sites act on the tensor product
$\mathcal{V}_{j_q,\bar j_q}\otimes \mathcal{V}_{j_q-u,\bar j_q+u}$ and are
given by the differential operators \re{R2} with $j_1=j_q$, $\bar j_1=\bar j_q$,
$j_2=j_q-u$ and $\bar j_2=\bar j_q+u$. It is easy to see that for these values
of spins the operator $\mathcal{R}^{(2)}(u)$ is a quadratic function of $u$ with
operator-valued coefficients. In analogy with \re{O-matrix}, the operator
$\Pi_{k0}\mathcal{R}_{k0}^{(2)}(u)$ entering \re{Q2=T} can be represented in the
linear auxiliary space $\mathcal{V}_{j_q-u,\bar j_q+u}$ by (infinite-
dimensional) matrices whose entries are at most quadratic in $u$. Their explicit
form can be found in Appendix~C. Multiplying these matrices and taking their
supertrace afterwards one obtains from \re{Q2=T} that, in general, the operator
$\mathcal{Q}_2(u-j_q+\bar j_q)$ is a polynomial in $u$ of degree $2N$.

The analysis of the operators $\mathcal{Q}_1(u)$ and $\mathcal{Q}_3(u)$ goes along
the same lines. From \re{Q=T} and \re{T's} one gets
\ba \nonumber
\mathcal{Q}_3(u+\bar j_q) &=& \str_{\mathcal{V}_{j_q+u,\bar j_q}} \left[\Pi_{N0}
\mathcal{R}_{N0}^{(3)}(u)\ldots \Pi_{10} \mathcal{R}_{10}^{(3)}(u)\right],
\\ \label{inf}
\mathcal{Q}_1(u-j_q) &=& \str_{\mathcal{V}_{j_q,\bar j_q-u}} \left[\Pi_{N0}
\mathcal{R}_{N0}^{(1)}(u)\ldots \Pi_{10} \mathcal{R}_{10}^{(1)}(u)\right],
\ea
where the $\mathcal{R}^{(3)}-$ and $\mathcal{R}^{(1)}-$ operators are given by
\re{R-i} and \re{R13-kernels} with $j_1=j_q$, $\bar j_1=\bar j_q$ and spins in
the auxiliary space equal, correspondingly, to $j_2=j_q+u$, $\bar j_2=\bar j_q$
and $j_2=j_q$, $\bar j_2=\bar j_q-u$. As before, one examines analytical
properties of matrices representing the $\mathcal{R}^{(1)}-$ and
$\mathcal{R}^{(3)}-$ operators on the tensor product $\mathcal{V}_{j_q,\bar
j_q}\otimes \mathcal{V}_{j_2,\bar j_2}$ with the spins $j_2$ and $\bar j_2$
specified above. The explicit form of these matrices can be found in Appendix~C.
One finds that matrix elements of the operator $\mathcal{R}^{(1)}(u)$  are
entire functions of $u$. For the operator $\mathcal{R}^{(3)}(u)$ its matrix
elements are meromorphic functions of $u$ which admit the following representation
\be
[\mathcal{R}^{(3)}(u)]_{ik} = \e^{i\pi u/2}\Gamma(u+j_q+\bar j_q+1) p_{ik}(u)\,,
\ee
with $p_{ik}(u)$ being polynomial in $u$. This suggests that the operator
$\mathcal{Q}_1(u)$ (or, more precisely, its eigenvalues) should be entire
functions of $u$ while the operator $\mathcal{Q}_3(u)/[\e^{i\pi
u/2}\Gamma(u+j_q+1)]^N$ should be polynomial in $u$. A delicate point however is
that the supertrace in the right-hand side of \re{inf} is given by an infinite
sum over matrix elements of the $\mathcal{R}^{(a)}-$operators and it is not
obvious that analytical properties of the two are the same. This can be checked
by applying the integral operators, $\mathcal{Q}_1(u)$ and $\mathcal{Q}_3(u)$,
Eqs.~\re{Qa}, \re{Q3-kernel} and \re{Q1-kernel}, to an arbitrary test function
and examining analytical properties of the resulting Feynman integrals.

\section{Factorized transfer matrices}

We demonstrated in the previous section that the transfer matrix evaluated over
the auxiliary space $\mathcal{V}_{{\jj}\bar {\jj}}$ is factorized into a product
of three mutually commuting  $\mathcal{Q}-$operators 
\be\label{T-fact}
\mathcal{T}_{{\jj}\bar {\jj}}(u) = {\rm str}_{\mathcal{V}_{{\jj}\bar {\jj}}}
\left[\mathcal{R}_{N0}(u)\ldots \mathcal{R}_{10}(u)\right]  = \mathbb{P}^{-2
}\mathcal{Q}_1(u - {\jj})\mathcal{Q}_2(u - {\jj}+\bar {\jj})\mathcal{Q}_3(u +
\bar {\jj})\,.
\ee
Here the operators $\mathcal{T}_{{\jj}\bar {\jj}}(u)$ and $\mathcal{Q}_a(u)$ act
on the quantum space of the model \re{HN} while the operator
$\mathcal{R}_{k0}(u)$ acts on the tensor product of the quantum space in $k^{\rm
th}$ site and the auxiliary space. Notice that the dependence of the transfer
matrix $\mathcal{T}_{{\jj}\bar {\jj}}(u)$ on the spins of the auxiliary space,
${\jj}$ and $\bar {\jj}$, resides in the arguments of the $\mathcal{Q}-$operators
only.

The relation \re{T-fact} holds true for arbitrary values of the spins
${\jj}$ and $\bar {\jj}$, that is, for generic infinite-dimensional $SL(2|1)$
representation $[j,\bar j]$ (see Table~\ref{Table1}). We have seen in Section 2.2,
that for certain values of the spins, the representation $[{\jj},\bar {\jj}]$
becomes reducible and it can be decomposed into a (semidirect) sum of irreducible
components. Whenever the representation $[j,\bar j]$ becomes reducible, the
corresponding representation space can be decomposed as $ \mathcal{V}_{{\jj}
\bar{\jj}} = \mathbb{V}^+ \oplus \mathbb{V}^- $. Then, the $\mathcal{R}-$operator
acting on the tensor product $\mathcal{V}_{j_q\bar j_q}\otimes \mathcal{V}_{{\jj}
\bar{\jj}}$ has a block triangular form according to the pattern of decomposition
of space $ \mathcal{V}_{{\jj}\bar {\jj}}$ shown schematically in Fig.~\ref{ReducibleReps},
\begin{align}
\label{Rbd}
\mathcal{R}_{\mathcal{V}_{j_q \bar{j}_q} \otimes \mathcal{V}_{{\jj}\bar
{\jj}}}(u)
=
\begin{pmatrix} {\mathbb{R}}^+(u) & \star \\ 0& {\mathbb{R}}^-(u)\end{pmatrix}
\,,
\end{align}
where `$\star$' denotes an off-diagonal operator whose explicit form is not
relevant for our purposes. Here $\mathbb{R}^+(u)$ defines the $\mathcal{R}-$operator
on the invariant subspace $\mathcal{V}_{j_q\bar j_q}\otimes \mathbb{V}^+$, while
${\mathbb{R}^-(u)}$ represents the same operator on the quotient space
$\mathcal{V}_{j_q\bar j_q}\otimes \mathbb{V}^-$.%
\footnote{Additional simplifications of the $\mathcal{R}-$operator occur when
the $SL(2|1)$ representation in the quantum space $[j_q,\bar j_q]$ is reducible.}
By definition, the $\mathcal{R}-$operator satisfies the Yang-Baxter equation
\re{YB}. Substituting $\mathcal{R}_{ik}(u)$ in \re{YB} with \re{Rbd} one finds
that the operators ${\mathbb{R}}^+(u)$ and ${\mathbb{R}}^-(u)$ also satisfy the
same equation. Moreover, evaluating the transfer matrix \re{T-fact} with the
$\mathcal{R}-$operators given by \re{Rbd}, one concludes that it is given by a
sum of two (mutually commuting) transfer matrices $\mathbb{T}^\pm(u)={\rm
str}_{\mathbb{V}^\pm} \left[\mathbb{R}_{10}^\pm(u)\ldots
\mathbb{R}_{N0}^\pm(u)\right]$ evaluated over `smaller' auxiliary spaces
$\mathbb{V}^\pm$
\be\label{T-sample}
\mathcal{T}_{{\jj}\bar {\jj}}(u) = \mathbb{T}^+(u) + \mathbb{T}^-(u)\,,\qquad [
\mathbb{T}^+(u),\mathbb{T}^-(v)] =0\,.
\ee
To make this relation more precise, let us consider the values of the spins
${\jj}$ and $\bar {\jj}$ for which the representation $[{{\jj},
\bar{\jj}}]$ becomes reducible.

\subsection{Finite-dimensional transfer matrices}

We have shown in Section~2.2.2 that for $j=-n/2+b$ and $\bar j=-n/2-b$ (with $n$
positive integer) the representation $[j,\bar j]$ decomposes as \re{red2} so that
the space $\mathcal{V}_{{\jj}\bar {\jj}}$ has a finite-dimensional invariant
subspace $v_{n/2,\,b}$, Eq.~\re{vn}. Then, the $\mathcal{R}-$matrix has the form
\re{Rbd} with $\mathbb{R}^+(u)$ defined on the tensor product of the quantum
space $\mathcal{V}_{j_q\bar j_q}$ and typical $SL(2|1)$ representation space
$v_{n/2,\,b}$. The corresponding transfer matrix $\mathbb{T}^+(u)$ is just the
transfer matrix over the finite-dimensional auxiliary space $v_{n/2,\,b}$
\be
t_{n/2,\, b}(u) = {\rm str}_{v_{n/2,\,b}} \left[\mathbb{R}^+_{N0}(u)\ldots
\mathbb{R}^+_{10}(u)\right]\,,\qquad \mathbb{R}^+(u) =
\mathcal{R}_{\mathcal{V}_{j_q\bar j_q}\otimes v_{n/2,\,b}}(u)\,.
\ee
The operator $\mathbb{R}^-(u)$ defines the $\mathcal{R}-$operator on the tensor
product $\mathcal{V}_{j_q\bar j_q}\otimes \mathcal{V}_{-\bar {\jj},-{\jj}}$.
More precisely, it verifies the same Yang-Baxter equation \re{YB} as the operator
$\mathcal{R}_{\mathcal{V}_{j_q\bar j_q}\otimes \mathcal{V}_{-\bar {\jj}
-{\jj}}}(u)$ and, therefore, the two operators coincide modulo an overall
normalization factor
\be\label{c-factor}
\mathbb{R}^-(u) = c(u)\,\mathcal{R}_{\mathcal{V}_{j_q\bar j_q}\otimes
\mathcal{V}_{-\bar {\jj} -{\jj}}}(u)\,.
\ee
Notice that under our definition of the $\mathcal{R}-$operators, Eqs.~\re{R-fact}
and \re{R=1}, the normalization of operators entering this relation is uniquely
fixed. To determine $c(u)$ it is sufficient to apply both sides of \re{c-factor}
to some reference state belonging to the quotient $\mathcal{V}_{{\jj}\bar
{\jj}}/v_{n/2,\, b}$. Explicit calculations show that $c(u)=1$ (see Appendix~B
for details). By definition \re{T-sample}, the operator $\mathbb{T}^-(u)$ is the
transfer matrix built from the operators \re{c-factor}. According to \re{T-fact},
it equals $\mathcal{T}_{-\bar {\jj},-{\jj}}(u)$ and one finds from \re{T-sample}
\be\label{T=t+T}
\mathcal{T}_{{\jj}\bar {\jj}}(u) = t_{n/2,\, b}(u) + \mathcal{T}_{-\bar
{\jj},-{\jj}}(u)\,.
\ee
Then, one replaces the $\mathcal{T}-$operators by their expression \re{T-fact}
in terms of the $\mathcal{Q}-$operators and arrives at the following relation
\be\label{t-lb}
t_{n/2,\, b}(u+b) =  \mathbb{P}^{-2 }\,\mathcal{Q}_2(u - b)\left[
{\mathcal{Q}}_1(u +n/2) {\mathcal{Q}}_3(u -n/2) -  {\mathcal{Q}}_1(u -n/2)
{\mathcal{Q}}_3(u+n/2) \right].
\ee
We notice that the dependence of the transfer matrix on $b$ resides in the first
factor only \footnote{We recall that the $\mathcal{Q}-$operators only depend on
spins in the quantum space, $j_q$ and $\bar j_q$, and the spectral parameter.}
and, therefore,
\be\label{Q2-t}
\frac{t_{n/2,\, b}(u+b)}{t_{n/2,\, b'}(u+b')}
=
\frac{\mathcal{Q}_2(u - b)}{\mathcal{Q}_2(u - b')}
\ee
for arbitrary $b$ and $b'$. Choosing $b'=u+j_q-\bar j_q$ in this relation and
taking into account the relation \re{Q=P}, one finds
\be\label{t-Q2}
t_{n/2\!,\, b}(u+b) = \mathbb{P}^{-1} \mathcal{Q}_2(u - b)t_{n/2\!,\, u+j_q-\bar
j_q}(2u+j_q-\bar j_q)\,.
\ee
It follows from this relation that the operator $\mathcal{Q}_2(u)$ is given by a
ratio of two finite-dimensional transfer matrices with the auxiliary spaces
carrying the same $SL(2)$ spin $n/2$ and different values of the $U(1)$ charge.
By construction, the operator $\mathcal{Q}_2(u)$ is a polynomial in $u$ with the
operator valued coefficients (see Section~3.4). We will demonstrate in Section~6
that up to an overall c-valued normalization factor the finite-dimensional
transfer matrices $t_{n/2\!,\, b}(u)$ enjoy the same property for arbitrary $b$.
Going over to eigenvalues in both sides of \re{t-Q2}, one observes that
$\mathcal{Q}_2(u)$ divides $t_{n/2\!,\, b}(u+2b)$, that is, all roots of the
former are also roots of the latter.

For $b=\pm n/2$, the expression for the transfer matrix $t_{n/2, b}$ can be
simplified further. The reason for this is that the typical representation $(b,n/2)$
becomes reducible and it decomposes into a semidirect sum of two atypical
representations \re{nn}. Together with \re{v-expand} this suggests that for
$b=\pm n/2$ the finite-dimensional transfer matrix $t_{n/2,\pm n/2}(u)$ is given
by a sum of two atypical transfer matrices. The explicit expressions will be
given below (see Eqs.~\re{tn+n} and \re{tn-n}).

\subsection{Infinite-dimensional (anti)chiral transfer matrices}

For $\bar {\jj}=0$ the representation $[{\jj},0]$ decomposes into a semidirect
sum of two (infinite-dimensional) chiral representations of $[{\jj}]_+$ and
$[{\jj}+1]_+$, Eq.~\re{red1}. As before, the $\mathcal{R}-$operator on the
tensor product $\mathcal{V}_{j_q\bar j_q}\otimes \mathcal{V}_{{\jj} 0}$ has a
triangular form \re{Rbd}. The operators $\mathbb{R}^+(u)$ and $\mathbb{R}^-(u)$
are related to the $\mathcal{R}-$operator on the tensor products $\mathcal{V}_{j_q
\bar{j}_q}\otimes\mathbb{V}_{{\jj}}$ and $\mathcal{V}_{j_q\bar{j}_q}\otimes
\mathbb{V}_{{\jj}+1}$, respectively, as
\be\label{a-factor}
\mathbb{R}^+(u) = \mathcal{R}_{\mathcal{V}_{j\bar
j}\otimes\mathbb{V}_{{\jj}}}(u)\,,\qquad \mathbb{R}^-(u) = \alpha(u-j)
\mathcal{R}_{\mathcal{V}_{j_q\bar j_q}\otimes\mathbb{V}_{{\jj}+1}}(u)\,.
\ee
The calculation of the normalization factor $\alpha(u)$ goes along the same
lines as in \re{c-factor}. One applies the second relation to the same reference
state belonging to the quotient $\mathcal{V}_{{\jj},0}/\mathbb{V}_{{\jj}}$ and
matches its both sides. In this way one obtains (see Appendix~B for details)
\be\label{c2}
\alpha(u) = i\frac{(u+j_q)(u-\bar j_q)}{u+j_q-\bar j_q-1}\,.
\ee
Let us introduce a notation for the transfer matrix evaluated over the chiral
representation space
\be\label{T-bb}
\mathbb{T}_{{\jj}}(u)={\rm str}_{\mathbb{V}_{{\jj}}}
\left[\mathbb{R}_{N0}(u)\ldots \mathbb{R}_{10}(u)\right]\,,\qquad \mathbb{R}(u)
=\mathcal{R}_{\mathcal{V}_{j_q\bar j_q}\otimes\mathbb{V}_{{\jj}}}(u)\,.
\ee
Then, it follows from \re{T-sample} and \re{T-fact} that
\be\label{Tj0}
\mathcal{T}_{{\jj}, 0}(u) =\mathbb{T}_{{\jj}}(u) - [\alpha(u-{\jj})]^N
\mathbb{T}_{{\jj}+1}(u)=\mathbb{P}^{-2 }\mathcal{Q}_1(u - {\jj})\mathcal{Q}_2(u
-
{\jj})\mathcal{Q}_3(u)\,,
\ee
where the additional factor ``$(-1)$'' in front of the second term has the same
origin as in \re{dec1}.

It is straightforward to generalize this consideration to the anti-chiral
transfer matrices $\mathbb{\bar T}_{{\bar j}}(u)$. In this case, for ${\jj}=0$
and $\bar{\jj}\neq 0$, the representation $[0,\bar {\jj}]$ decomposes as in \re{red11}
and the relation between the transfer matrices $\mathcal{T}_{0 \bar {\jj}}(u)$
and  $\mathbb{\bar T}_{\bar {\jj}}(u)$ reads
\be\label{Tj1}
\mathcal{T}_{0,\bar {\jj}}(u) =\mathbb{\bar T}_{\bar {\jj}}(u) - [\alpha(u+\bar
{\jj}+1)]^{-N} \mathbb{\bar T}_{\bar {\jj}+1}(u)=\mathbb{P}^{-2 }\mathcal{Q}_1(u
)\mathcal{Q}_2(u +\bar {\jj})\mathcal{Q}_3(u + \bar {\jj})\,,
\ee
with the normalization factor $\alpha(u)$ defined in \re{c2}.

\subsection{Finite-dimensional (anti)chiral transfer matrices}

For ${\jj}=-n$ (with $n$ positive integer), the chiral representation $[j]_+$
decomposes into a semidirect sum of the atypical representation $(n)_+$ and the
antichiral representation $[n+1]_-$, Eq.~\re{V-chi}. As in \re{Rbd}, the
$\mathcal{R}-$operator on the tensor product ${\mathcal{V}_{j_q\bar
j_q}\otimes\mathbb{V}_{-n}}$ has a block-triangular form with the upper diagonal
block given by $\mathcal{R}_{\mathcal{V}_{j_q\bar j_q}\otimes{\rm v}_{n}}(u)$
and the lower diagonal block proportional to $\mathcal{R}_{\mathcal{V}_{j_q
\bar{j}_q}\otimes\mathbb{\bar V}_{n+1}}(u)$. The $\mathcal{R}-$operator on
${\mathcal{V}_{j_q\bar j_q}\otimes\mathbb{\bar V}_{-n}}$ admits a similar
representation. As a result, for the chiral and anti-chiral transfer matrices
one finds (see Appendix~B for details)
\ba \nonumber
&& \mathbb{T}_{-n}(u) = {\rm t}_n(u) - [\alpha(u+n+1)]^{-N} \mathbb{\bar
T}_{n+1}(u)\,,
\\[2mm] \label{tT}
&& \mathbb{\bar T}_{-n}(u) = {\rm \bar t}_n(u) - [\alpha(u-n)]^{N} \mathbb{
T}_{n+1}(u)\,.
\ea
Here, the notation was introduced for the finite-dimensional atypical transfer
matrix (see Table~\ref{Table1})
\be\label{t-def}
{\rm t}_n(u)={\rm str}_{{\rm v}_{n}} \left[\mathbb{R}_{N0}(u)\ldots
\mathbb{R}_{10}(u)\right]\,,\qquad \mathbb{R}(u)
=\mathcal{R}_{\mathcal{V}_{j\bar
j}\otimes{\rm v}_{n}}(u) \, ,
\ee
and the transfer matrix ${\rm \bar t}_n(u)$ is defined similarly, with ${\rm v}_{n}$
replaced by ${\rm\bar v}_{n}$.

Let us now establish a relation between finite-dimensional atypical transfer
matrices, ${\rm t}_{n}(u)$ and ${\rm \bar t}_{n}(u)$, and the
$\mathcal{Q}-$operators. For $b=n/2$, or equivalently ${\jj}=0$ and $\bar
{\jj}=-n$, one obtains from \re{T=t+T}
\be
t_{{n}/2, \, n/2}(u) = \mathcal{T}_{0,-n}(u)- \mathcal{T}_{n,0}(u) \,.
\ee
One applies \re{Tj0} and \re{Tj1}, takes into account \re{tT} and obtains two
different representations for $t_{{n}/2, \, n/2}(u)$ (with $n\ge 1$)
\ba \nonumber
t_{{n}/2, \, n/2}(u) &=&  {\rm \bar t}_n(u)-\left[\alpha(u-n+1)\right]^{-N} {\rm
\bar t}_{n-1}(u)
\\[2mm] \label{tn+n}
&=& \mathbb{P}^{-2}\, \mathcal{Q}_2(u -n) \left[\mathcal{Q}_1(u
)\mathcal{Q}_3(u-n) - \mathcal{Q}_1(u -n)\mathcal{Q}_3(u)\right].
\ea
In a similar manner, for $b=-n/2$, or equivalently ${\jj}=-n$ and $\bar {\jj}=0$
one gets
\ba\nonumber
t_{{n}/2,-n/2}(u) &=&  {\rm t}_n(u)-\left[\alpha(u+n)\right]^{N} {\rm t}_{n-
1}(u)
\\[2mm] \label{tn-n}
&=& \mathbb{P}^{-2 }\,\mathcal{Q}_2(u +n)\left[{\mathcal{Q}}_1(u
+n){\mathcal{Q}}_3(u)-{\mathcal{Q}}_1(u ){\mathcal{Q}}_3(u + n)\right]\,.
\ea
To obtain ${\rm t}_n(u)$ and ${\rm \bar t}_n(u)$ from these relations they have
to be supplemented by the expression for ${\rm t}_0(u)$ and ${\rm \bar t}_0(u)$.
For $n=0$ the (anti)chiral auxiliary spaces ${\rm v}_0$ (and ${\rm \bar v}_0$)
in
\re{t-def} contain only one basis vector $\{1\}$ and, as a consequence, the
transfer matrices ${\rm t}_{0}(u)={\rm \bar t}_0(u)$ reduce to a c-number. Its
value can be found as (see Appendix~B for details)
\be\label{t0}
{\rm t}_{0}(u)={\rm \bar t}_0(u)=\left[-\xi{\frac {\Gamma \left( u+1+j_q \right)
\left(u+j_q-\bar j_q \right) }{{ {\Gamma \left(-u+\bar j_q \right)}}
 }}\right]^N,
\ee
with $\xi=\e^{{-i\pi}(j_q+\bar j_q)/2}/(j_q\bar j_q)$. Combining together the
relations \re{tn+n} -- \re{t0}, one can express the transfer matrices ${\rm
t}_n(u)$ and ${\rm \bar t}_n(u)$ (with $n\ge 1$) in terms of the $\mathcal{Q}-
$operators.

\subsection{Factorization of (anti)chiral transfer matrices}

Let us demonstrate that the (anti)chiral transfer matrices
$\mathbb{T}_{{\jj}}(u)$ and $\mathbb{\bar T}_{\bar {\jj}}(u)$ can also be
expressed in terms of $\mathcal{Q}-$operators. We return to the first relation
in \re{Tj0} and observe that the difference of the two $\mathbb{T}-$operators
carrying the spins ${\jj}$ and ${\jj}+1$ is proportional to the operator
$\mathcal{Q}_3(u)$ which in its turn is ${\jj}-$independent. In the same manner,
the difference of the $\mathbb{\bar T}-$operators in \re{Tj1} is proportional to
$\mathcal{Q}_1(u)$. This suggests to write down the transfer matrices in a
factorized form
\ba \nonumber
\mathbb{T}_{{\jj}} (u) \!\!\!&=&\!\!\!   \mathbb{P}^{-2 }\,\mathcal{Q}_{3}(u)\,
\mathcal{Q}_{12}(u+1-{\jj})/(-\beta(u-j)) \,,
\\[2mm] \label{T-fact-new}
{\mathbb{\bar T}}_{\bar {\jj}} (u) \!\!\!&=&\!\!\!  \mathbb{P}^{-2
}\,\mathcal{Q}_{1}(u)\, \mathcal{Q}_{23}(u+\bar {\jj})/\beta(u+\bar j) \,,
\ea
where $\mathcal{Q}_{12}(u)$ and $\mathcal{Q}_{23}(u)$ are some operators
commuting with the $\mathcal{Q}-$operators. Here, for the later convenience, we
introduced the normalization factor satisfying the condition
\be\label{beta}
\frac{\beta(u-1)}{\beta(u)} = \left[\alpha(u)\right]^N\,.
\ee
Combining together \re{Tj0}, \re{Tj1} and \re{T-fact-new} one obtains the
relations
\ba \nonumber
\beta(u)\mathcal{Q}_{1}(u)\mathcal{Q}_{2}(u) &=& \mathcal{Q}_{12}(u) -
\mathcal{Q}_{12}(u+1)\,,
\\[2mm] \label{beta-Q}
\beta(u)\mathcal{Q}_{3}(u)\mathcal{Q}_{2}(u) &=& \mathcal{Q}_{23}(u)
-\mathcal{Q}_{23}(u+1)\,,
\ea
which establish the correspondence between the two sets of the $Q-$operators.
Here the rationale behind the subscript `$\scriptstyle 12$' is that the
$\mathcal{Q}_{12}-$operator is related to the product of $\mathcal{Q}-$operators
with the subscripts `$\scriptstyle 1$' and `$\scriptstyle 2$'.

To elucidate the origin of \re{T-fact-new} and the meaning of the new operators,
we recall that the transfer matrix $\mathbb{T}_{{\jj}} (u)$ is built from the
$\mathcal{R}-$operators \re{T-bb} acting on the tensor product
${\mathcal{V}_{j_q\bar j_q}\otimes\mathbb{V}_{{\jj}}}$. These operators appear
as an upper diagonal block in the expression for the `big' operator
$\mathcal{R}_{{\mathcal{V}_{j_q\bar j_q}\otimes\mathcal{V}_{{\jj},0}}}$,
Eq.~\re{Rbd} for $\bar {\jj}=0$. The latter operator has the factorized form
\re{R-fact} and a question arises whether $\mathcal{R}_{{\mathcal{V}_{j_q\bar
j_q}\otimes\mathbb{V}_{{\jj}}}}$ admits a similar factorized representation. To
start with, let us examine the action of the $\mathcal{R}^{(a)}-$operators
entering \re{R-fact} on the tensor product
${\mathcal{V}_{j_q\bar{j}_q}\otimes\mathcal{V}_{{\jj}0}}$. Applying \re{R's} for
$j_1=j_q$, $\bar j_1=\bar j_q$, $j_2=j$ and $\bar j_2=0$, we notice that
$\mathcal{R}^{(3)}(u)$ does not modify the zero value of the antichiral spin in
the auxiliary space, while the $\mathcal{R}^{(1)}-$ and
$\mathcal{R}^{(2)}-$operators move it away from zero. In other words, the
operator $\mathcal{R}^{(3)}(u)$ maps a reducible auxiliary space
$\mathcal{V}_{{\jj},0}$ into a reducible one  $\mathcal{V}_{{\jj}-u,0}$ and, as a
consequence, it has a block-triangular form \re{Rbd}. The same property does not
hold however for the $\mathcal{R}^{(1)}-$ and $\mathcal{R}^{(2)}-$operators
separately but it is restored in their product $\mathcal{R}^{(12)}(u)\equiv
\mathcal{R}^{(1)}(u)\mathcal{R}^{(2)}(u-\bar j_q)$ so that the
$\mathcal{R}-$operator in \re{R-fact} acquires a block-triangular form \re{Rbd}.
This suggests to use the upper diagonal block of $\mathcal{R}^{(12)}(u)$ to
construct a new transfer matrix analogous to \re{T's}. Indeed, let us denote by
$\mathcal{P}_0$ the operator that projects ${\mathcal{V}_{j_q\bar
j_q}\otimes\mathcal{V}_{{\jj}0}}$ onto its invariant component
${\mathcal{V}_{j_q\bar j_q}\otimes\mathbb{V}_{{\jj}}}$. Then, the chiral
transfer matrix \re{T-bb} can be expressed as
\be
\mathbb{T}_{{\jj}}(u)={\rm str}_{\mathcal{V}_{{\jj}0}}
\left[\mathcal{P}_0\,\mathbb{R}_{N0}(u)\ldots \mathbb{R}_{10}(u)\right]\,,\qquad
\mathbb{R}(u)
=\mathcal{R}_{\mathcal{V}_{j_q\bar{j}_q}\otimes\mathcal{V}_{{\jj}0}}(u) \, .
\ee
One substitutes the $\mathcal{R}-$operator with \re{R-fact} and repeats the same
steps that led to the factorized expression for the transfer matrix
\re{R3fromR12} to obtain the first relation in \re{T-fact-new} with
\begin{align}
\mathcal{Q}_{12}(u+1-{\jj})\sim & \str_{\mathcal{V}_{{\jj}, 0}}
\Big\{\mathcal{P}_0\, {\Pi}_{N0}\mathcal{R}^{(12)}_{N0}(u) \ldots
{\Pi}_{10}\mathcal{R}^{(12)}_{10}(u) \Big\}\,,
\end{align}
and $\mathcal{R}^{(12)}(u)=\mathcal{R}^{(1)}(u)\mathcal{R}^{(2)}(u-\bar j_q)$.
The analysis of the antichiral transfer matrix $\mathbb{\bar T}_{\bar {\jj}}(u)$
goes along the same lines with the only difference that it is now the operator
$\mathcal{R}^{(1)}(u)$ that preserves zero value of the chiral spin, ${\jj}=0$,
in the auxiliary space and the $\mathcal{Q}_{23}-$operator is built from
$\mathcal{R}^{(23)}(u)\equiv \mathcal{R}^{(2)}(u+ j_q)\mathcal{R}^{(3)}(u)$.

Substituting \re{T-fact-new} into \re{tT} one can express the finite-dimensional
atypical transfer matrices ${\rm t}_n(u)$ and ${\rm \bar t}_n(u)$ (with $n\ge
0$) in terms of the $Q-$operators
\ba\nonumber
{\rm t}_n(u)&{=}& \mathbb{P}^{-2} \left[\mathcal{Q}_1(u) \mathcal{Q}_{23}(u+n+1)
- \mathcal{Q}_3(u)\mathcal{Q}_{12}(u+n+1)\right]/\beta(u+n)\,,
\\[2mm]\label{taun}
{\rm \bar t}_n(u)&{=}& \mathbb{P}^{-2}\left[\mathcal{Q}_1(u)
\mathcal{Q}_{23}(u-n) -
\mathcal{Q}_3(u)\mathcal{Q}_{12}(u-n)\right]/\beta(u-n)\,.
\ea
Although these relations were derived for $n\ge 0$, they can be used to define
${\rm t}_n(u)$ and ${\rm \bar t}_n(u)$ for negative $n$. Then, ${\rm t}_n(u)$
and ${\rm \bar t}_n(u)$ defined in this way verify the relation
\be\label{cont}
{\rm t}_n(u) = {\rm \bar t}_{-1-n}(u)\frac{\beta(u+n+1)}{\beta(u+n)}
\ee
for an arbitrary integer $n$. For $n=0$ the two relations in \re{taun} coincide
in virtue of \re{beta-Q} leading to ${\rm t}_0(u)={\rm \bar t}_0(u)$, in
agreement with \re{t0}. They can be further simplified by excluding
$\mathcal{Q}_1$ and $\mathcal{Q}_3$ with a help of \re{beta-Q} leading to
\be\label{Q2-QQ}
{\rm t}_0(u)\, \mathcal{Q}_2(u) = \mathbb{P}^{-2}\left[ \mathcal{Q}_{12}(u)
\mathcal{Q}_{23}(u+1)-\mathcal{Q}_{23}(u)\mathcal{Q}_{12}(u+1)
\right]/[\beta(u)]^2\,,
\ee
with $\beta(u)$ given by \re{beta}.

Being combined together, Eqs.~\re{beta-Q} and \re{Q2-QQ} allow us to express
three different $\mathcal{Q}-$ope\-ra\-tors in terms of two operators
$\mathcal{Q}_{12}(u)$ and $\mathcal{Q}_{23}(u)$ only. Obviously, the same holds
true for the transfer matrices defined in \re{T-fact}, \re{T-fact-new} and
\re{taun}. To save space we do not present here their explicit expressions. One
of the consequences of this remarkable property is that a generic {\sl
infinite-dimensional} transfer matrix $\mathcal{T}_{{\jj},\bar {\jj}}(u)$ can be
expressed in terms of (anti)chiral transfer matrices%
\footnote{Another way to get this relation is to start with \re{Q2-QQ}, shift
the spectral paramater as $u\to u-{\jj}+\bar {\jj}$, multiply both sides of the
relation by $\mathcal{Q}_1(u-{\jj}) \mathcal{Q}_3(u+\bar {\jj})$ and, then,
apply \re{T-fact} and \re{T-fact-new}.}, $\mathbb{T}_{{\jj}}(u)$ and
$\mathbb{\bar T}_{\bar {\jj}}(u)$,
\begin{align}\label{gamma}
\mathcal{T}_{{\jj},\bar {\jj}}(u)=\left[\mathbb{T}_{{\jj}}(u+\bar {\jj})\,
\mathbb{\widebar T}_{\bar {\jj}}(u-{\jj})-\gamma(u-2b)\,
\mathbb{T}_{{\jj}+1}(u+\bar {\jj}) \,\mathbb{\widebar T}_{\bar
{\jj}+1}(u-{\jj})\right]/{\rm t}_0(u-2b)\,.
\end{align}
Here $b=({\jj}-\bar {\jj})/2$ and the notation was introduced for
$$
\gamma(u)=\left[ {\alpha(u)}/{\alpha(u+1)} \right]^N
$$
with $\alpha(u)$ given by \re{c2}. A similar relation also holds between
{\sl finite-dimensional} transfer matrices \re{t-lb} and \re{taun}
\be\label{t-fused}
t_{\ell\,b}(u)  = \left[ {\rm t}_n(u-{\bar n}){\rm \bar t}_{\bar n}(u+n) -
\gamma\left( u-2b \right) \,{\rm t}_{n-1}(u-{\bar n}) {\rm
\bar t}_{{\bar n}-1}(u+n) \right]/{\rm t}_0(u)\,,
\ee
where $\ell = (n+\bar n)/2$ and $b=(\bar n -n)/2$. This equation is valid for
arbitrary integer $n$ and $\bar n$ such that $\ell>0$. For nonnegative integer
$n$, ${\rm t}_n(u)$ coincides with the atypical transfer matrix \re{t-def},
while for negative integer $n$ it is related to the transfer matrix ${\rm\bar
t}_{-n-1}(u)$ through \re{cont}.

\begin{figure}[t]
\psfrag{T1}[rc][cc]{$\mathbb{T}_j,\, \mathbb{\bar T}_{\bar j} $}
\psfrag{T2}[cc][cc]{$\mathcal{T}_{j,\bar j} $} \psfrag{t1}[cc][cc]{${\rm
t}_n,\,{\rm\bar t}_n$} \psfrag{t2}[cc][cc]{$t_{\ell, b}$}
\psfrag{eq1}[cc][cc]{\re{tT}} \psfrag{eq3}[cc][cc]{\re{gamma}}
\psfrag{eq2}[cc][cc]{\re{t-fused}}  \psfrag{eq4}[cc][cc]{\re{T=t+T}}

\centerline{\includegraphics[scale=0.8]{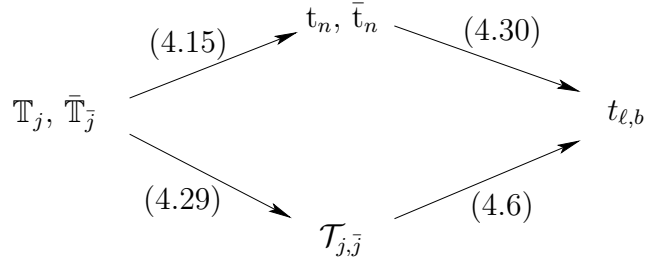}} \caption[]{
\label{HierarchyT} Hierarchy of transfer matrices (see Table~\ref{Table1}). The
number above arrowed line refers to the equation number.}
\end{figure}

\begin{figure}[t]
\psfrag{Q12}[cc][cc]{$\mathcal{Q}_{12},\, \mathcal{Q}_{23}$}
\psfrag{Q1Q2Q3}[cc][cc]{$\mathcal{Q}_{1},\, \mathcal{Q}_{2},\, \mathcal{Q}_{3}$}
\psfrag{Q13}[cc][cc]{$\mathcal{Q}_{13}$}
\psfrag{r1}[cc][cc]{\re{beta-Q}, \re{Q2-QQ}} \psfrag{r2}[cc][cc]{\re{Q13-def}}
\centerline{\includegraphics[scale=1.0]{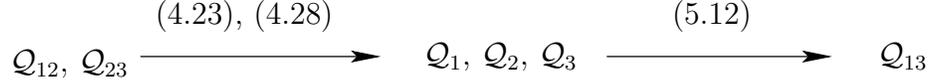}} \caption[]{
\label{HierarchyQ} Hierarchy of $Q-$operators. }
\end{figure}

The entire hierarchy of the transfer matrices is summarized in Fig.\
\ref{HierarchyT}. We observe that {\sl all} transfer matrices can be expressed in
terms of two operators only, $\mathbb{T}_j(u)$ and $\mathbb{\bar T}_{\bar j}(u)$.
Together with \re{T-fact-new}, this allows one to express the $SL(2|1)$ transfer
matrices listed in Table~\ref{Table1} in terms of Baxter $Q-$operators. So far,
we introduced two families of mutually commuting operators, $\mathcal{Q}_a(u)$
(with $a=1,2,3$) and $\mathcal{Q}_{12}(u),\, \mathcal{Q}_{23}(u)$, and one more
operator $\mathcal{Q}_{13}(u)$ will be defined later in Eq.~\re{Q13-def}. Similar
to the transfer matrices, $Q-$operators also form a hierachy shown in
Fig.~\ref{HierarchyQ}.

\section{Baxter equations}

Let us now establish the TQ-relations between the ${Q}-$operators and the
atypical transfer matrices ${\rm t}_1(u)$ and ${\rm \bar t}_1(u)$ defined over
the three-dimensional auxiliary spaces ${\rm v}_1$, Eq.~\re{v-chi} and ${\rm \bar
v}_1$, Eq.~\re{v-bar-chi}, respectively. It is well known \cite{Kul85} that these
transfer matrices are generating functions for local integrals of motion of the
$SL(2|1)$ spin chain (see Eqs.~\re{t-tbar} and \re{tau} below).

\subsection{TQ-relations}
\label{BaxterEquations}

Let us derive TQ-relations for each Baxter operator.

\subsubsection*{Operator $\mathcal{Q}_2(u)$}

We apply the relations \re{tn+n} and \re{tn-n} for $n=1$ to get
\be\label{t/t}
\frac{\mathcal{Q}_2(u)}{\mathcal{Q}_2(u-1)}
=\frac{t_{1/2,-1/2}(u-1)}{t_{1/2,1/2}(u)}\,.
\ee
{}From \re{tn+n} and \re{tn-n}, one can express the transfer matrices entering
this relation in terms of ${\rm t}_1(u)$ and ${\rm
\bar t}_1(u)$ and c-valued functions ${\rm t}_0(u)={\rm \bar t}_0(u)$,
Eq.~\re{t0}, as
\ba \nonumber
t_{1/2, 1/2}(u)
\!\!\!&=&\!\!\! {\rm \bar t}_1(u)-\Delta_-(u)\,,
\\[2mm] \label{t+-}
t_{1/2,-1/2}(u)
\!\!\!&=&\!\!\! {\rm t}_1(u)-\Delta_+(u+1)\,,
\ea
where the notation was introduced for
\ba\nonumber
&& \Delta_-(u) = {{\rm t}_0(u)}/{\alpha^N(u)}=\big[(u+j_q-\bar
j_q-1)\Delta(u-1)\big]^N\,,
\\ \label{Delta}
&& \Delta_+(u) = {\rm t}_0(u-1)\alpha^N(u)=\left[{\frac{(u+j_q)(u-\bar
j_q)}{u+j_q-\bar j_q}} 
\Delta(u-1)\right]^N
\ea
with the normalization factor
\be\label{delta}
\Delta(u)= -i \xi
\frac{\Gamma(u+j_q+1)}{\Gamma(-u+\bar j_q)}(u+j_q-\bar j_q+1) \,,
\ee
and $\xi=\e^{-i{\pi}(j_q+\bar j_q)/2}/(j_q\bar j_q)$. We recall that the
operator $\mathcal{Q}_2(u)$ is polynomial in $u$ and, therefore, the right-hand
side of \re{t/t} is a rational (operator-valued) function of $u$. This property
is not obvious in the right-hand side of \re{t/t} since the (anti)chiral
transfer matrices ${\rm t}_1(u)$ and ${\rm \bar t}_1(u)$ involve terms proportional
to the ratio of $\Gamma-$functions (see Eqs.~\re{t-tbar} and \re{delta} below). We
will demonstrate in Section 6.1, that such terms cancel against each other.

\subsubsection*{Operators $\mathcal{Q}_3(u)$ and $\mathcal{Q}_1(u)$}

For $n=\bar n=1$, or equivalently $\ell=1$ and $b=0$, the transfer matrix
$t_{\ell,\,b}$ admits two equivalent representations, Eqs.~\re{t-fused} and
\re{t-lb},
\ba \nonumber
 t_{1,\,0}(u) &=& \left[{\rm \bar t}_{1}(u+1) {\rm t}_1(u-1) -\gamma\left(
u\right) \,{\rm
\bar t}_{0}(u+1) {\rm t}_{0}(u-1)\right]/{\rm t}_0(u)
\\[2mm] \label{t10}
&=&  \mathbb{P}^{-2 }\,\mathcal{Q}_2(u)\left[ {\mathcal{Q}}_1(u +1)
{\mathcal{Q}}_3(u -1) -  {\mathcal{Q}}_1(u -1) {\mathcal{Q}}_3(u+1) \right].
\ea
Supplementing the second relation with a similar expression for the transfer
matrices
$t_{1/2, \, \pm 1/2}(u)$, Eq.~\re{t-lb},
\be \label{t1/2}
t_{1/2, \, \pm 1/2}(u\pm 1) = \pm  \mathbb{P}^{-2}\, \mathcal{Q}_2(u)
\left[\mathcal{Q}_1(u\pm 1)\mathcal{Q}_3(u) - \mathcal{Q}_1(u)\mathcal{Q}_3(u\pm
1)\right]
\ee
one finds that $\mathcal{Q}_3(u)$ satisfies the functional relation
\be \label{TQ-33}
t_{1,0}(u)\, \mathcal{Q}_3(u)=t_{1/2, \,  1/2}(u+1)\,\mathcal{Q}_3(u-1)+t_{1/2,
\, -1/2}(u-1)\,\mathcal{Q}_3(u+1)\,.
\ee
Replacing the transfer matrices by their expressions \re{t10} and \re{t+-} in
terms of atypical transfer matrices, one obtains
\be \label{TQ-3}
\boxed{\parbox[c][15mm]{150mm}{\baa \lefteqn{ \left[{\rm \bar t}_{1}(u+1) {\rm
t}_1(u-1) -\Delta_+(u) \Delta_-(u+1)\right] \mathcal{Q}_3(u)}
\\[2mm] \nonumber
&& ={\rm t}_0(u)\left[{\rm \bar
t}_1(u+1)-\Delta_-(u+1)\right]\mathcal{Q}_3(u-1)+{\rm t}_0(u)\left[{\rm
t}_1(u-1)-\Delta_+(u)\right]\mathcal{Q}_3(u+1)
\eaa}}
\ee
The operator $\mathcal{Q}_1(u)$ satisfies the same equation due to the symmetry
of \re{t10} and \re{t1/2} under exchange of the $\mathcal{Q}_1-$ and
$\mathcal{Q}_3-$operators.

\subsubsection*{Operators $\mathcal{Q}_{23}(u)$ and $\mathcal{Q}_{12}(u)$}

Let us start with the following identity
\ba \nonumber
\lefteqn{{\rm t}_1(u-1){\rm \bar t}_1(u) - {\rm t}_0(u-1){\rm \bar t}_0(u) =
\mathbb{P}^{-4}\left[\mathcal{Q}_1(u)\mathcal{Q}_3(u-1) -
\mathcal{Q}_1(u-1)\mathcal{Q}_3(u)\right]} &&
\\[2mm] \label{t11}
& & \qquad \times\left[\mathcal{Q}_{12}(u-1)
\mathcal{Q}_{23}(u+1)-\mathcal{Q}_{12}(u+1)
\mathcal{Q}_{23}(u-1)\right]/(\beta(u)\beta(u-1))\,.
\ea
It can be verified by replacing the transfer matrices in the left-hand side by
their expressions \re{taun} in terms of $Q-$operators. As follows from \re{tn+n}
and \re{tn-n}, the first factor in the right-hand side of \re{t11} is given by
$\mathbb{P}^{-2}\, t_{1/2,1/2}(u)/\mathcal{Q}_2(u-1)=\mathbb{P}^{-2}\,
t_{1/2,-1/2}(u-1)/\mathcal{Q}_2(u)$. Then, one multiplies both sides of \re{t11}
by $\mathcal{Q}_2(u)$, excludes the $\mathcal{Q}_{12}-$operator with a help of
\re{Q2-QQ} and gets the following relation for the operator $\mathcal{Q}_{23}$
\ba \nonumber
\lefteqn{ \left[{\rm t}_1(u-1){\rm \bar t}_1(u) - {\rm t}_0(u){\rm
\bar t}_0(u-1)\right]\mathcal{Q}_{23}(u)}
\\[2mm]
&& =\Delta_-(u)  t_{1/2,-1/2}(u-1)\mathcal{Q}_{23}(u-1)+\Delta_+(u)
t_{1/2,1/2}(u) \mathcal{Q}_{23}(u+1)\,,
\ea
where the `dressing factors' $\Delta_\pm(u)$ were defined in \re{Delta}. Finally,
one applies \re{t+-} and obtains
\be \label{TQ-23}
\boxed{\parbox[c][15mm]{145mm}{\baa \lefteqn{ \left[{\rm t}_1(u-1){\rm \bar
t}_1(u) - \Delta_+(u)\Delta_-(u)\right]\mathcal{Q}_{23}(u)}
\\[2mm] \nonumber
&& =\Delta_-(u)  \left[{\rm t}_1(u-1)-\Delta_+(u)\right]
\mathcal{Q}_{23}(u-1)+\Delta_+(u) \left[{\rm \bar t}_1(u)-\Delta_-(u)\right]
\mathcal{Q}_{23}(u+1)
\eaa}}
\ee
The operator $\mathcal{Q}_{12}(u)$ satisfies the same relation.

\subsubsection*{Operator $\mathcal{Q}_{13}(u)$}

As we will see in Section~5.2, it proves convenient to introduce the following
operator
\be\label{Q13-def}
\mathcal{Q}_{13}(u) = \mathbb{P} ^{-2} \left[\mathcal{Q}_1(u+1) \mathcal{Q}_3(u)
- \mathcal{Q}_1(u) \mathcal{Q}_3(u+1)\right].
\ee
The chain of ancestor relations of $Q-$operators is shown in Fig.\ \ref{HierarchyQ}.
Then, the transfer matrices \re{t1/2} can be factorized into a product of two
$Q-$operators as
\ba\nonumber
t_{1/2, \, -1/2}(u-1) &=& \mathcal{Q}_2(u) \mathcal{Q}_{13}(u-1)\,,
\\[2mm] \label{t-13}
t_{1/2, \, 1/2}(u+1) &=& \mathcal{Q}_2(u) \mathcal{Q}_{13}(u)\,.
\ea
This leads to the following relation for the operator $\mathcal{Q}_{13}(u)$
\be\label{t/t-13}
\frac{\mathcal{Q}_{13}(u-1)}{\mathcal{Q}_{13}(u)}=\frac{t_{1/2, \,
-1/2}(u-1)}{t_{1/2, \, 1/2}(u+1)}=\frac{{\rm t}_1(u-1)-\Delta_+(u)}{{\rm \bar
t}_1(u+1)-\Delta_-(u+1)}\,,
\ee
which should be compared with the analogous relation \re{t/t} for the operator
$\mathcal{Q}_2(u)$.

Notice that the TQ-relations \re{TQ-3} and \re{TQ-23} for the operators
$\mathcal{Q}_3(u)$ and $\mathcal{Q}_{23}(u)$ (as well as for  $\mathcal{Q}_1(u)$
and $\mathcal{Q}_{12}(u)$) are finite difference equations of the second-order,
while the TQ-relations \re{t/t} and \re{t/t-13} for the operators
$\mathcal{Q}_2(u)$ and $\mathcal{Q}_{13}(u)$ are of the first order only.

\subsection{Nested TQ-relations}
\label{NestedTQrel}

The TQ-relations \re{t/t}, \re{TQ-3} and \re{TQ-23} involve one
$Q-$operator and two transfer matrices, ${\rm t}_1(u)$ and ${\rm \bar t}_1(u)$.
To make a comparison with the nested Bethe Ansatz it is convenient to exclude
the antichiral transfer matrix ${\rm \bar t}_1(u)$ from the TQ-relation by
trading it for another $Q-$operator.

Let us start with \re{TQ-23} and notice that the combination of transfer
matrices in front of $\mathcal{Q}_{23}(u)$ in the left-hand side can be
rewritten with a help of \re{t+-} as ${\rm t}_1(u-1)t_{1/2,1/2}(u)+ \Delta_-(u)
t_{1/2,-1/2}(u-1)$. Then, one divides both sides of \re{TQ-23} by $t_{1/2,1/2}
(u)\mathcal{Q}_{23}(u)$ and applies \re{t/t} to get
\be\label{T-23-2}
 \boxed{ {\rm t}_1(u-1) =\Delta_+(u)
\frac{\mathcal{Q}_{23}(u+1)}{\mathcal{Q}_{23}(u)}+
\Delta_-(u)\left[\frac{\mathcal{Q}_{23}(u-1)}{\mathcal{Q}_{23}(u)}-1 \right]
\frac{\mathcal{Q}_2(u)}{\mathcal{Q}_2(u-1)} }
\ee
This relation also holds true with $\mathcal{Q}_{23}(u)$ being replaced by
$\mathcal{Q}_{12}(u)$.

The analysis of \re{TQ-3} or, equivalently, \re{TQ-33} goes along the same lines.
One uses \re{t+-} to verify that
\be
t_{1,0}(u) = \left(t_{1/2,1/2}(u+1){\rm t}_1(u-1) + [\alpha(u+1)]^{-N}
t_{1/2,-1/2}(u-1) {\rm t}_0(u+1)\right)/{\rm t}_0(u)\,.
\ee
Dividing both sides of \re{TQ-33} by $t_{1/2,1/2}(u+1) \mathcal{Q}_3(u)$, one
obtains
\be
{\rm t}_1(u-1)  ={\rm t}_0(u) \frac{\mathcal{Q}_3(u-1)}{\mathcal{Q}_3(u)}+{\rm
t}_0(u)\left[\frac{\mathcal{Q}_3(u+1)}{\mathcal{Q}_3(u)}-\frac{{\rm
t}_0(u+1)}{{\rm t}_0(u)\alpha^N(u+1)}\right]\frac{t_{1/2, \, -1/2}(u-1)}{t_{1/2,
\, 1/2}(u+1)}.
\ee
Here, in distinction with \re{T-23-2} and \re{t/t}, the ratio of the transfer
matrices cannot be expressed in terms of the $\mathcal{Q}_2-$operator. Instead,
making use of \re{t/t-13}, it can be simplified leading to yet another
expression
for the transfer matrix in terms of the operators $\mathcal{Q}_3(u)$ and
$\mathcal{Q}_{13}(u)$
\be\label{T-13-3}
\boxed{{\rm t}_1(u-1)  ={\rm t}_0(u)
\frac{\mathcal{Q}_3(u-1)}{\mathcal{Q}_3(u)}+\left[{\rm
t}_0(u)\frac{\mathcal{Q}_3(u+1)}{\mathcal{Q}_3(u)}-\Delta_-(u+1)\right] \frac{
\mathcal{Q} _{13}(u-1)}{\mathcal{Q}_{13}(u)}}
\ee
Equations \re{T-23-2} and \re{T-13-3} provide two different representations of
the transfer matrix ${\rm t}_1(u-1)$ in terms of two pairs of the $Q-$operators:
$(\mathcal{Q}_2,\mathcal{Q}_{23})$ and $(\mathcal{Q}_3,\mathcal{Q}_{13})$. One
might expect that there exist two more representations for ${\rm t}_1(u-
1)$ which involve two pairs of the operators $(\mathcal{Q}_2,\mathcal{Q}_{13})$
and $(\mathcal{Q}_3, \mathcal{Q}_{23})$.  Indeed, the former representation
follows immediately from \re{t-13} and \re{t+-}
\be\label{T-13-2}
\boxed{\phantom{\frac{Q}{Q}}{\rm t}_1(u-1)-\Delta_+(u)=\mathcal{Q}_2(u)
\mathcal{Q}_{13}(u-1)}
\ee\\[-1mm]
Lastly, one applies \re{Q13-def} and \re{taun} to verify that
\ba \nonumber
\mathcal{Q}_{13}(u) \mathcal{Q}_{23}(u+1) &=& t_0(u+1) \beta(u+1)
\mathcal{Q}_3(u) - {\rm t}_0(u)
\beta(u) \mathcal{Q}_3(u+1)
\\[2mm] \label{Q13}
&=&\beta(u)\left[\Delta_-(u+1)\mathcal{Q}_3(u) -{\rm t}_0(u) \mathcal{Q}_3(u+1)
\right].
\ea
Using this relation one excludes the operator $\mathcal{Q}_{13}(u)$ from
\re{T-13-3} and obtains the following representation for the transfer matrix
\be\label{T-23-3}
\boxed{ {\rm t}_1(u-1)-\Delta_+(u)  =\left[1-\frac{ \mathcal{Q}_{23}(u+1)}{{\rm
Q}_{23}(u)}\right] \left[{\rm t}_0(u)\frac{\mathcal{Q}_3(u-
1)}{ \mathcal{Q}_3(u)}
-\Delta_+(u) \right] }
\ee
Eqs.~\re{T-23-2}, \re{T-13-3}, \re{T-13-2} and \re{T-23-3} provide four
different expressions for the transfer matrix ${\rm t}_1(u)$ in terms of
various Baxter $Q-$operators.

We remind that the TQ-relations stay invariant under the substitution of the
operators  $\mathcal{Q}_3(u)$ and $\mathcal{Q}_{23}(u)$ with the operators
$\mathcal{Q}_1(u)$ and $\mathcal{Q}_{12}(u)$, respectively. The reason why
we wrote the TQ-relations in terms of the former operators only is that, as
we will show in the next section, their eigenvalues are given (up to an
overall normalization factor) by polynomials in $u$.

\section{Nested Bethe Ansatz}

The transfer matrix ${\rm t}_1(u)$ and the $Q-$operators commute with each other
and, therefore, can be diagonalized simultaneously. In this section, we will
apply the obtained TQ-relations to find the eigenspectrum of the transfer
matrices. The nested Bethe Ansatz provides an alternative approach to solving the
same eigenproblem. It relies on the existence of a pseudovacuum state in the
quantum space of the model and leads to expressions for the eigenvalues of the
transfer matrices in terms of two sets of Bethe roots. The $SL(2|1)$ spin chain
has in fact three different pseudovacuum states and, as a consequence, one can
construct three different nested Bethe Ansatz
solutions~\cite{Lai74,Sut75,Sch87,EssKor92}.

In this section, we will establish the correspondence between the TQ-relations
for Baxter $Q-$operators and nested Bethe Ansatz relations. In particular, we
will identify the Bethe roots as zeros of polynomial eigenvalues of certain
$Q-$operators and demonstrate that three TQ-relations, Eqs.~\re{T-23-2},
\re{T-13-3} and \re{T-23-3}, are in the one-to-one correspondence with three
different nested Bethe Ansatz solutions of the $SL(2|1)$ spin chain.

In the nested Bethe Ansatz, eigenvalues of the transfer matrices are given by
expressions similar to \re{T-23-2} with $Q-$operators replaced by polynomials
parameterized by the Bethe roots. The latter are uniquely fixed from the
condition that the transfer matrix should be polynomial in $u$. It is important
to stress that our definition of the transfer matrix ${\rm t}_1(u)$,
Eq.~\re{t-def}, differs from the conventional one $\tau_N(u)$ (see Eq.~\re{tau}
below) used in the nested Bethe Ansatz. The main difference is that the former
operator is built from the $\mathcal{R}-$matrices while the latter is
constructed from the Lax operators. We will show that the two operators differ
by an overall normalization factor and, therefore, polynomiality of ${\rm t}_1(u)$
can be restored by taking this factor out.

\subsection{Polynomial transfer matrices}

Local integrals of motion of the $SL(2|1)$ spin chain are generated by the
auxiliary transfer matrices built from the chiral and antichiral Lax operators,
$L(u)$ and $\bar L(u)$, respectively~\cite{Kul85}. By definition,
\ba
L(u)
\!\!\!&=&\!\!\!
u +  \sum_{A,B=1,2,3} (-)^{\bar{B}} e^{AB} E^{BA}
=
\lr{
\begin{array}{ccc}
u+J & -\bar V^- & L^- \\
- V^+ & u+J-\bar J &  V^- \\
L^+ & -\bar V^+ & u-\bar J \\
\end{array}}
\label{Lax}
\ea
and
\ba\label{bar-Lax}
\bar{L} (u)
\!\!\!&=&\!\!\ u +  \sum_{A,B=1,2,3} (-)^{\bar B}  \bar e^{AB} E^{BA} = \lr{
\begin{array}{ccc}
u+\bar J & -V^-& L^- \\
-\bar V^+ & u+\bar J-J & \bar V^- \\
L^+ & -V^+ & u- J \\
\end{array}}
\, ,
\ea
where $E^{BA}$ are the $SL(2|1)$ generators \re{CartanBasisGen} and the $3\times
3$ graded matrices $e^{AB}$ and $\bar e^{AB}$ represent the generators of the
three-dimensional atypical representations ${\rm v_1}$ and $\bar{\rm v}_1$,
Eqs.~\re{v-chi} and \re{v-bar-chi}, respectively. The corresponding auxiliary
transfer matrices are defined as
\ba\nonumber
\tau_N(u) \!\!\!&=&\!\!\! \str_{{\rm v}_1} \left[L_N(u) \ldots L_1(u) \right] =
u^N + i^2 q_2 \, u^{N-2} + \ldots + i^N q_N
\\[2mm] \label{tau-bar}\label{tau}
\bar\tau_N(u)
\!\!\!&=&\!\!\! \str_{\bar{\rm v}_1} \left[\bar L_N(u) \ldots \bar L_1(u)
\right]
= u^N + i^2\bar q_2 \, u^{N-2} + \ldots + i^N \bar q_N \, ,
\ea
where the supertrace is taken over three-dimensional auxiliary space. Defined in
this way, the transfer matrices $\tau_N(u)$ and $\bar\tau_N(u)$ are polynomials
in $u$ of degree $N$.  The operator valued $q-$ and $\bar q-$coefficients are
local integrals of motion of the model. It follows from \re{tau} that the
operators $q_2$ and $\bar q_2$ are proportional to the quadratic Casimir
operator
\re{Casimir}
\be\label{q2=C2}
q_2 = \bar q_2 =
- \mathbb{C}_2 + N j_q \bar j_q \,, 
\ee
with $\mathbb{C}_2=\sum_{A,B=1,2,3} (-)^{\bar B} E^{AB} E^{BA}$ and
$E^{AB}=\sum_{n=1}^N E_n^{AB}$ is given by the sum of the $SL(2|1)$ generators
over all sites. The remaining charges $q_k$ and $\bar q_k$ (with $k\ge 3$) are
given by homogeneous polynomials of degree $k$ in the $SL(2|1)$ generators. In
distinction with $q_2$, the operators $q_k\neq \bar q_k$ are not self-adjoint
with respect to the scalar product \re{scal}. Instead one finds using \re{herm}
that
\be\label{t-bar}
q_k^\dagger =\bar q_k \quad \mapsto \quad \lr{\tau_N(u)}^\dagger = (-1)^N
\bar\tau_N(-u^*)\,.
\ee
To match the auxiliary transfer matrix \re{tau} into \re{t-def}, we have to
identify the Lax operator as a special case of the $\mathcal{R}-$operator.

By definition \re{Lax}, the Lax operator $L(u)$ acts on the tensor product of
quantum space $\mathcal{V}_{j_q\bar j_q}$ and three-dimensional chiral
representation ${\rm v_1}$ and satisfies the Yang-Baxter equation. As such it
can be identified (modulo an overall normalization factor and a shift of the
spectral parameter) with one of  the $\mathcal{R}-$operators, $L(u)\sim
R_{\mathcal{V}_{j_q \bar{j}_q}\otimes {\rm v_1}}(u+c)$ or $L(u)\sim R_{{\rm v_1}
\otimes \mathcal{V}_{j_q \bar{j}_q}}(u)$. Indeed, an explicit calculation shows
(see Appendix~B) that
\ba  \label{R-L}
\mathcal{R}_{\mathcal{V}_{j_q\bar j_q}\otimes {\rm v_1}}(u) &=&
\Delta(u)\,L(u+1)\,,
\ea
where the normalization factor $\Delta(u)$ is given by \re{delta}.
According to \re{bar-Lax}, the antichiral Lax operator $\bar L(u)$ can be
obtained from the chiral operator $L(u)$, Eq.~\re{Lax}, by replacing the
auxiliary space with ${\rm \bar v}_1$.
Together with \re{R-L} this leads to $\mathcal{R}_{{\rm
\bar v_1}\otimes \mathcal{V}_{j_q\bar j_q}}(u)\sim \bar L(u+1)$.
The transfer matrix ${\rm \bar t}_1(u)$, Eq.~\re{t-def}, is built from the
operators $\mathcal{R}_{\mathcal{V}_{j_q\bar{j}_q} \otimes {\rm \bar v_1}}(u)$
rather than $\mathcal{R}_{{\rm \bar v_1}\otimes \mathcal{V}_{j_q\bar j_q}}(u)$.
It is straightforward to verify that $\bar L(u+1)$ and
$(\mathcal{R}_{\mathcal{V}_{j_q\bar j_q}\otimes {\rm \bar v_1}}(-u))^{-1}$ obey
the same Yang-Baxter equation \re{YB} and, therefore, the two operators coincide
up to an overall normalization factor. The latter can be calculated as before,
by examining the action of both operators on the same reference state leading to
\be\label{R-L-1}
\mathcal{R}_{\mathcal{V}_{j_q\bar j_q}\otimes{\rm \bar v_1}}(u+1)=
\Delta(u)(j_q+u)(\bar j_q-u)\bar L^{-1}(-u)\,.
\ee
This relation can be further simplified if one of the spins in the quantum space
vanishes, $\bar j_q j_q=0$. In particular, for $\bar j_q=0$ one has $\bar L(-u)
\bar L(u+j_q) = - u(u+j_q)$ yielding %
\footnote{The expression for $\Delta(u)$, Eq.~\re{delta}, involves a singular factor
$\xi=\e^{-i{\pi}(j_q+\bar j_q)/2}/(j_q\bar j_q)$. In what follows it is tacitly
assumed that this factor is removed in the (anti)chiral limit $j_q\bar j_q\to 0$
by changing the normalization of the $\mathcal{R}-$operators. \label{foot}}
\be
\mathcal{R}_{\mathcal{V}_{j_q0}\otimes{\rm \bar v_1}}(u+1)= \Delta(u)\bar
L(u+j_q)\,.
\ee
In a generic case, for $\bar j_q j_q\neq 0$, it is convenient to introduce the
operator $\widetilde L(u)$ related to the inverse Lax operator
\ba \nonumber
\widetilde L(u) &=& - (u+\bar j_q)(u-j_q)(u+1) \bar L^{-1}(-u+j_q-\bar j_q)
\\[2mm]
&=& (u+1)(u+ \bar L(0)) + (\bar L(0) + j_q)(\bar L(0) -\bar j_q)\,.
\ea
The Lax operators $L(u)$ and $\widetilde L(u)$ are linear and quadratic
functions in $u$, respectively. Then, one deduces from \re{R-L} and \re{R-L-1}
that, up to an overall normalization, the operators $\mathcal{R}_{\mathcal{V}_{j_q
\bar{j}_q}\otimes {\rm v_1}}(u)$ and $\mathcal{R}_{\mathcal{V}_{j_q\bar{j}_q}
\otimes{\rm\bar v_1}}(u)$ have the same property. This allows us to reveal
analytical properties of the transfer matrices ${\rm t}_1(u)$ and ${\rm \bar
t}_1(u)$, Eq.~\re{t-def}.

Substituting  \re{R-L} and \re{R-L-1} into \re{t-def} we obtain
\ba\nonumber
{\rm t}_1(u) &=&  \tau_N(u+1)\left[\Delta(u)\right]^N
\\[1mm] \label{t-tbar}
{\rm \bar t}_1(u) &=& \widetilde\tau_{2N}(u+j_q-\bar j_q-1) \left[\Delta(u-1)
/(u+j_q-\bar j_q)\right]^N,
\ea
where $\tau_N(u)$ is a polynomial of degree $N$, Eq.~\re{tau}, and
$\widetilde\tau_{2N}(u)$ is defined for general $j_q$ and $\bar j_q$ as
\ba \nonumber
\widetilde\tau_{2N}(u)&=&\str_{{\rm\bar v_1}} \left[\widetilde L_N(u) \ldots
\widetilde L_1(u) \right]
\\ \label{tau-tilde}
&=& (u(u+1))^{N} + i^2 \widetilde q_2 u^{2N-2} + i^3 \widetilde q_3 u^{2N-3} +
\ldots +i^{2N}\widetilde q_{2N}\,.
\ea
The operators $\widetilde q_k$ entering \re{tau-tilde} can be expressed in terms
of the integrals of motion $q_k$ and Casimir operators, e.g.,
\ba \nonumber
i^2\widetilde q_2 &=& \mathbb{C}_2
\\ \label{q-tilde}
i^3\widetilde q_3 &=& i^3q_3 -2 \mathbb{C}_3 +\frac13\mathbb{C}_2+
(N-1)\mathbb{C}_2 + 2(j_q-\bar j_q)\mathbb{C}_2 -N (j_q-\bar j_q) j_q\bar j_q \,
,
\ea
where $\mathbb{C}_2$ and $\mathbb{C}_3$ are given by \re{Casimir} with the
$SL(2|1)$ generators acting on the quantum space \re{HN}.

In the chiral limit $\bar j_q=0$, one has $\bar L(0)(\bar L(0) + j_q)=0$, or
equivalently $\widetilde L(u)=(u+1)\bar L(u)$, so that \re{tau-tilde} is
expressed in terms of the antichiral transfer matrix
\be\label{tau-2N}
\widetilde\tau_{2N}(u)\big|_{\bar j_q=0}=(u+1)^N\bar\tau(u)\,.
\ee
As a consequence, the second relation in \re{t-tbar} simplifies to (see
footnote~\ref{foot})
\ba \nonumber
{\rm t}_1(u)\big|_{\bar j_q=0} &=&  \tau_N(u+1)\left[\Delta(u)\right]^N
\, ,
\\[1mm] \label{t-tbar-chi}
{\rm \bar t}_1(u)\big|_{\bar j_q=0} &=&
\bar\tau_N(u+j_q-1)\left[\Delta(u-1)\right]^N
\, ,
\ea
where $\bar\tau_N(u)$ is a polynomial in $u$, Eqs.~\re{tau} and \re{t-bar}.

\subsection{Polynomial $Q-$operators}
\label{NestedABAviaQs}

The TQ-relations \re{TQ-3} and \re{TQ-23} are finite-difference equations and
they stay invariant under the multiplication of the $Q-$operators by an arbitrary
periodic function with period $1$. In order to fix this ambiguity and uniquely
determine eigenvalues of the $Q-$operators, one has to supplement the
TQ-relations with additional conditions on their solutions. These conditions were
worked out in Section~3.4. As we argued there, the operators $\mathcal{Q}_2(u)$
and $\mathcal{Q}_3(u)/[\e^{i\pi u/2}\Gamma(1+u+j_q)]^N$  are polynomial in $u$
and, therefore,
\ba \nonumber
\mathcal{Q}_2(u)\ket{\Psi_q} &=& Q_2(u)\ket{\Psi_q}\,,
\\[2mm]\label{P-2}
\mathcal{Q}_3(u)\ket{\Psi_q} &=& [\e^{i\pi u/2}\Gamma(1+u+j_q)]^N
Q_3(u)\ket{\Psi_q}\,,
\ea
where $\ket{\Psi_q}$ stands for the eigenstate of the $SL(2|1)$ spin chain
Hamiltonian and the corresponding eigenvalues $Q_2(u)$ and $Q_3(u)$ are
polynomials in $u$. The operator $\mathcal{Q}_1(u)$ verifies the same TQ-relation
\re{TQ-3} as $\mathcal{Q}_3(u)$, but in distinction with the latter its
eigenvalues are {\sl entire} functions of the spectral parameter. It is
convenient to parameterize them as
\ba
\mathcal{Q}_1(u)\ket{\Psi_q} &=& [\xi \e^{-i\pi u/2}/\Gamma(-u-j_q)]^{N}
Q_1(u)\ket{\Psi_q}\,,
\ea
where  $\xi = \e^{-i\pi(j_q+\bar j_q)/2}/(j_q\bar j_q)$ and $Q_1(u)$ is a
meromorphic function with poles located at $u+j_q\in \mathbb{N}$ of maximal
order
$N$.

Let us now examine the operator $\mathcal{Q}_{13}(u)$. This operator enters into
the factorized expression for the transfer matrix $t_{1/2, \, -1/2}(u-1)$,
Eq.~\re{t-13}. Taking into account the relations \re{t+-} -- \re{delta} one
finds
from \re{t-13}
\be\label{Q23-pol}
\mathcal{Q}_2(u)
\mathcal{Q}_{13}(u-1)=\left[\Delta(u-1)\right]^N\left[\tau_N(u)-\lr{(u+j_q)(u-
\bar
j_q)/(u+j_q-\bar j_q)}^N\right].
\ee
Since $\mathcal{Q}_2(u)$ divides the transfer matrix  $t_{1/2, \, -1/2}(u-1)$
(see \re{t-Q2}) and $\tau_N(u)$ is a polynomial in $u$, the operator
$\mathcal{Q}_{13}(u-1)/[\Delta(u-1)/(u+j_q-\bar{j}_q)]^N$ is also polynomial so
that
\be\label{P-13}
\mathcal{Q}_{13}(u)\ket{\Psi_q} = \left[\frac{\Delta(u)}{u+j_q-\bar
j_q+1}\right]^N Q_{13}(u)\ket{\Psi_q}\,,
\ee
where $\Delta(u)$ is determined in \re{delta} and $Q_{13}(u)$ is a polynomial in
$u$. We notice that  the second factor in the right-hand side of \re{Q23-pol}
simplifies in the chiral limit $\bar j_q=0$. As a result,
$\mathcal{Q}_{13}(u-1)/\Delta(u-1)$ ought to be polynomial yielding
\be\label{q13}
Q_{13}(u)\big|_{{\bar j}_q=0}= (u+j_q+1)^N\, Q_{13}^{(0)}(u)\,,
\ee
with $Q_{13}^{(0)}(u)$ being yet another polynomial.

It is instructive to compare properties of the functions ${Q}_1(u)$ and
${Q}_3(u)$ with those of the eigenvalues of the Baxter operators for the
$SL(2)$ spin chain. In that case, one encounters two operators $\mathbb{Q}_\pm
(u)$ analogous to $\mathcal{Q}_1$ and $\mathcal{Q}_3$. The eigenvalues of the
operators $\mathbb{Q}_+(u)$ and $\mathbb{Q}_-(u)$ are correspondingly polynomials
and meromorphic functions of $u$. They satisfy the second-order finite difference
TQ-relation \re{TQ-SL2} and verify the Wronskian relation \re{WrSL2}. Equation
\re{Q13-def} is an analog of the Wronskian relation for the $SL(2|1)$ spin chain.
Going over to the eigenvalues in both sides of \re{Q13-def}, one gets
\be\label{compare}
Q_1(u+1) Q_3(u) - Q_1(u) Q_3(u+1) = Q_{13}(u)\e^{2i\theta_q} \left[
\frac{\Gamma(-u-1-j_q)}{\Gamma(-u+\bar j_q)} \right]^N\,,
\ee
where $\theta_q$ is the quasimomentum, $\mathbb{P}\ket{\Psi_q} =
\e^{i\theta_q}\ket{\Psi_q}$. In distinction with \re{WrSL2}, the right-hand side
of
this relation involves the eigenvalues of the operator $\mathcal{Q}_{13}(u)$.

Let us parameterize the eigenvalues of the operators $\mathcal{Q}_{12}(u)$ and
$\mathcal{Q}_{23}(u)$ as
\ba \nonumber
\mathcal{Q}_{12}(u+1) \ket{\Psi_q} &=& \beta(u) \left[\xi\e^{-i\pi
u/2}(u+j_q-\bar j_q)/\Gamma(-u-j_q)\right]^N Q_{12}(u+1)\ket{\Psi_q}
\, ,
\\[2mm]
\label{P-23} \mathcal{Q}_{23}(u+1) \ket{\Psi_q} &=& \beta(u) \left[\e^{i\pi u/2}
(u+j_q-\bar j_q)\Gamma(u+j_q+1)\right]^N Q_{23}(u+1)\ket{\Psi_q}
\, ,
\ea
with the same normalization factor $\beta(u)$ as in \re{T-fact-new} and \re{beta}.
Substituting the second relation into \re{Q13} and \re{beta-Q}, one finds
\ba \nonumber
Q_{23}(u) Q_{13}(u-1) &=&  (u+j_q-\bar j_q)^N Q_3(u-1)-  (u+j_q)^NQ_3(u)\,,
\\[2mm]  \label{Q-relations}
Q_2(u) Q_3(u) &=& (u-\bar j_q)^N Q_{23}(u)  - (u+j_q-\bar j_q)^N Q_{23}(u+1)\,,
\ea
Due to invariance of the TQ-relations under the exchange of the $Q-$operators,
the same relation holds true upon the substitution $Q_{3}(u)$ and $Q_{23}(u)$
with $Q_1(u)$ and $Q_{12}(u)$, respectively. As before, $Q_{13}(u-1)$ divides the
right-hand side of the first relation in \re{Q-relations} and, therefore, the
functions $Q_{23}(u)$ and $Q_{12}(u)$ have the same analytical properties as
$Q_3(u)$ and $Q_1(u)$, respectively. So, $Q_{23}(u)$ is polynomial and
$Q_{12}(u)$ is a meromorphic function of $u$. One finds from \re{Q2-QQ} the two
functions satisfy a Wronskian-like condition,
\be\label{Wr}
Q_{12}(u) Q_{23}(u+1) - Q_{12}(u+1) Q_{23}(u) = Q_2(u) \e^{2i\theta_q}
\left[\frac{\Gamma(-u-j_q)}{\Gamma(1-u+\bar j_q)}\right]^N,
\ee
which should be compared with \re{compare} and \re{WrSL2}.

\subsubsection{TQ-relations}

The eigenvalues of the operators $\mathcal{Q}_2(u)$, $\mathcal{Q}_3(u)$,
$\mathcal{Q}_{13}(u)$ and $\mathcal{Q}_{23}(u)$ are parameterized by four
polynomials, Eqs.~\re{P-2}, \re{P-13} and \re{P-23}, respectively. To determine
these polynomials one has to examine the TQ-relations and replace the
transfer matrices by their eigenvalues \re{t-tbar} and \re{tau}. In this way,
one obtains three different expressions for the transfer matrix \re{tau}.

\noindent  From \re{T-23-2},
\be\label{BA1}
\tau_N(u) = - (u+j_q-\bar j_q-1)^N \frac{Q_{2}(u)}{Q_{2}(u-1)}+ (u-\bar j_q-1)^N
\frac{Q_{23}(u-1)}{Q_{23}(u)} \frac{Q_{2}(u)}{Q_{2}(u-1)}+(u+j_q)^N
\frac{Q_{23}(u+1)}{Q_{23}(u)}
.
\ee
From \re{T-13-3},
\be\label{BA2}
\tau_N(u) = {(u-\bar j_q)}^N \frac{Q_3(u-1)}{Q_3(u)}+{(u+j_q+1)}^N
\frac{Q_3(u+1)}{Q_3(u)}\frac{Q_{13}(u-1)}{Q_{13}(u)}- (u+j_q-\bar
j_q+1)^N\frac{Q_{13}(u-1)}{Q_{13}(u)}
.
\ee
From \re{T-23-3},
\be\label{BA3}
\tau_N(u) = {(u-\bar j_q)}^N \frac{Q_3(u-1)}{Q_3(u)}-{(u+j_q-\bar j_q)}^N
\frac{Q_3(u-1)}{Q_3(u)}\frac{Q_{23}(u+1)}{Q_{23}(u)} +
(u+j_q)^N\frac{Q_{23}(u+1)}{Q_{23}(u)}
.
\ee
By construction, the $Q-$functions entering these relations are polynomials in
$u$ of a finite degree. As such, they can be parameterized by their roots
\be\label{roots1}
\begin{array}{cc} \displaystyle
Q_2(u) = c_2 \prod_{k=1}^{n_2} (u-\lambda_k^{(2)})\,, &
\displaystyle \qquad {Q}_3(u)=c_3 \prod_{k=1}^{n_3} (u-\lambda_k^{(3)})\,,
\\[6mm]
\displaystyle {Q}_{13}(u)=c_{13} \prod_{k=1}^{n_{13}} (u-\lambda_k^{(13)})\,, &
\displaystyle \qquad Q_{23}(u) = c_{23} \prod_{k=1}^{n_{23}}
(u-\lambda_k^{(23)})\,,
\end{array}
\ee
where $n$'s are nonnegative integers and $c$'s are normalization constants.
Equations \re{roots1} involve four different sets of roots. The ratios of
$Q-$functions in the right-hand side of \re{BA1} -- \re{BA3} are meromorphic
functions in $u$ whereas the left-hand side involves the polynomial transfer
matrix \re{tau}. Matching analytical properties of both sides of the relations
\re{BA1} -- \re{BA3}, one equates to zero residues at all poles and arrives at
a system of coupled nested Bethe equations for the roots of $Q-$polynomials.

The polynomials \re{roots1} satisfy the additional relations \re{Q-relations}
that can be considered as consistency conditions for the system \re{BA1} --
\re{BA3}. In addition, one finds from \re{T-13-2} that
\ba \nonumber
Q_2(u) Q_{13}(u-1) &=& (u+j_q-\bar j_q)^N \tau_N(u) - [(u+j_q)(u-\bar j_q)]^N
\\[2mm] \label{Q0-relations}
&=& (Nj_q\bar j_q-q_2) u^{2N-2} + \mathcal{O}(u^{2N-3})\,,
\ea
with $\tau_N(u)$ defined in Eq.\ \re{tau}.  Substituting \re{roots1} into
\re{Q0-relations} and \re{Q-relations} and examining the asymptotic behavior of
both sides for $u\to\infty$, one finds $n_{13}=2(N-1)-n_2$ and
$n_{23}=n_2+n_3-N+1$. Thus, the four polynomials in \re{roots1} depend on two
integers only, $n_2$ and $n_3$. Since $n_{13}\ge 0$ and $n_{23}\ge 0$, they have
to satisfy the relations
\be\label{cond}
n_2 \le 2(N-1)\,,\qquad n_2+n_3 \ge N-1\,.
\ee

The nested TQ-relations \re{BA1} -- \re{BA3} involve one polynomial transfer matrix
$\tau_N(u)$ and two $Q-$polynomials. There exists another set of the TQ-relations,
Eqs.~\re{t/t}, \re{TQ-3}, \re{TQ-23} and \re{t/t-13}, which involve only one
$Q-$operator and two transfer matrices. Going over to the eigenvalues in both sides
of these relations, one can obtain the TQ-relations between the corresponding
polynomials ${Q}_2(u)$, ${Q}_3(u)$, ${Q}_{23}(u)$, ${Q}_{13}(u)$ and polynomial
transfer matrices $\tau_N(u)$ and $\tilde\tau_{2N}(u)$, Eq.~\re{t-tbar}. In the next
subsection, we  will present these relations in the chiral limit (see Eqs.~\re{TQ-}
and \re{Q23-chiral}).

\subsubsection{TQ-relations in the chiral limit}

In the chiral limit, $j_q\neq 0$ and $\bar j_q=0$, one finds from
\re{Q0-relations} and from the first relation in \re{Q-relations} that
$(u+j_q)^N$ divides the polynomial $Q_{13}(u-1)$, in agreement with \re{q13}.
One deduces from \re{roots1} and \re{q13} that $Q_{13}^{(0)}(u)$ is a
polynomial of degree $n_{13}-N$.

The TQ-relations \re{BA1}  -- \re{Q0-relations} take the following form for
$\bar
j_q=0$
\ba \nonumber
\tau_N(u) &=& \left[(u-1)^N \frac{Q_{23}(u-1)}{Q_{23}(u)}- (u+j_q-1)^N\right]
\frac{Q_{2}(u)}{Q_{2}(u-1)}+(u+j_q)^N \frac{Q_{23}(u+1)}{Q_{23}(u)}
\\ \nonumber
&=& {u}^N \frac{Q_3(u-1)}{Q_3(u)}+{(u+j_q)}^N\left[
\frac{Q_3(u+1)}{Q_3(u)}-1\right]\frac{Q_{13}^{(0)}(u-1)}{Q_{13}^{(0)}(u)}
\\ \nonumber
&=& {u}^N \frac{Q_3(u-1)}{Q_3(u)}-{(u+j_q)}^N\left[
\frac{Q_3(u-1)}{Q_3(u)}-1\right]\frac{Q_{23}(u+1)}{Q_{23}(u)}
\\[2mm] \label{NBA-chiral}
&=& u^N+Q_2(u) Q_{13}^{(0)}(u-1)
\, .
\ea
In the first three relations, the condition for the right-hand side to be
polynomial in $u$ leads to three systems of nested Bethe ansatz equations for the
roots of the $Q-$polynomials. They coincide with three different nested Bethe
ansatz solutions of the $SL(2|1)$ spin chain obtained in
Refs.~\cite{Lai74,Sut75,Sch87,EssKor92}. For $\bar{j}_q=0$ one finds from
\re{Q-relations}
\ba \nonumber
&& Q_{23}(u) Q_{13}^{(0)}(u-1) = Q_3(u-1)-Q_3(u)
\, ,
\\[2mm] \label{QQ-chiral}
&& Q_2(u) Q_3(u) = u^N Q_{23}(u)  - (u+j_q)^N Q_{23}(u+1)\,.
\ea
Finally, one takes into account the relations \re{tau-2N} and \re{t-tbar-chi}
and obtains from \re{t/t} and \re{t/t-13} in the chiral limit
\ba\label{Q2-chiral}
\frac{Q_2(u)}{Q_2(u-1)}&=&
\frac{\tau_N(u)-u^N}{\bar\tau_N(u+j_q-1)-(u+j_q-1)^N}\,,
\\
\frac{Q_{13}^{(0)}(u)}{Q_{13}^{(0)}(u-1)}&=&
\frac{\bar\tau_N(u+j_q)-(u+j_q)^N}{\tau_N(u)-u^N} \,.
\ea
From \re{TQ-3}, on finds
\ba\label{TQ-}
\lefteqn{\left[\tau_N(u)\bar\tau_N(u+j_q)-(u(u+j_q))^N \right]Q_3(u)} &&
\\[2mm] \nonumber
&=& u^N \left[\bar\tau_N(u+j_q)- (u+j_q)^N\right]Q_3(u-1) + (u+j_q)^N
\left[\tau_N(u)-u^N \right]Q_3(u+1)\,,
\ea
while Eq.\ \re{TQ-23} yields
\ba
\label{Q23-chiral} \lefteqn{\left[\tau_N(u)\bar\tau_N(u+j_q-1)-(u(u+j_q-1))^N
\right]Q_{23}(u)} &&
\\[2mm] \nonumber
&=&(u-1)^N \left[\tau_N(u)-u^N \right]Q_{23}(u-1) + (u+j_q)^N
\left[\bar\tau_N(u+j_q-1)- (u+j_q-1)^N\right]Q_{23}(u+1)
\ea
The functions $Q_1(u)$ and $Q_{12}(u)$ satisfy the same relations as
$Q_3(u)$ and $Q_{23}(u)$, respectively. The corresponding Wronskian relations
are given by \re{compare} and \re{Wr}. The relations \re{TQ-} and \re{Q23-chiral}
look similar to the TQ-relations for the $SL(2)$ magnet, Eq.~\re{TQ-SL2}. The
only difference is that the dressing factors now depend on the transfer matrices
which in their turn depend on the integrals of motion. This explains why the
Wronskian relations \re{compare} and \re{Wr} involve additional $Q-$functions
in the right-hand side.

\subsubsection{Eigenspectrum in the chiral limit}

In the chiral limit, one applies \re{H-SL21} and obtains the eigenvalues of the
Hamiltonian and the cyclic permutation operator in terms of the $Q_3-$polynomial
as
\ba \nonumber
&& {E} = \lr{\ln {Q}_3(0)}' - \lr{\ln {Q}_3(-j_q)}'\,,
\\[2mm] \label{E-SL21}
&& \e^{i\theta_q}  = {Q}_3(0)/{Q}_3(-j_q)\,.
\ea
Here the quasimomentum $\theta_q$ satisfies the relation $\e^{iN\theta_q}=1$ in
virtue of $\mathbb{P}^N=\II$.

It is well known~\cite{QISM} that in lattice integrable models the Hamiltonian
appears as a first term in the expansion of the so-called fundamental transfer
matrix in powers of the spectral parameter. The auxiliary space for this
transfer matrix coincides with the quantum space in each site. For the $SL(2|1)$
spin chain in the chiral limit, it can be identified as $\mathbb{V}_{j_q}$ and
the corresponding fundamental transfer matrix is $\mathbb{T}_{j_q}(u)$ (see
Table~\ref{Table1}). Let us demonstrate that  $\mathbb{T}_{j_q}(u)$ has the
following expansion for $u\to 0$
\be\label{T->Q}
\mathbb{T}_{j_q}(u) \sim \mathbb{P} \exp\lr{u\, \mathbb{H} + \mathcal{O}(u^2)}\,,
\ee
where $\mathbb{P}$ is the cyclic permutation operator and $\mathbb{H}$ is the
Hamiltonian of the $SL(2|1)$ spin chain. One finds from the first relation in
\re{T-fact-new} that
\be\label{Tf}
\mathbb{T}_{j_q}(u)=-\mathbb{P}^2  \left(\xi\e^{i\pi j_q/2}\right)^N
\left[u\Gamma(1+u+j_q)/ \Gamma(-u) \right]^N  Q_3(u) Q_{12}(u+1-j_q)
\ee
where we replaced the operators $\mathcal{Q}_3(u)$ and $\mathcal{Q}_{12}(u)$ by
their eigenvalues, Eqs.~\re{P-2} and \re{P-23}, respectively. We recall that the
pre-factor $\left(\xi\e^{i\pi j_q/2}\right)^N$ is singular for $\bar j_q\to 0$
(see footnote \ref{foot}) and it can be removed by changing a normalization of
the transfer matrix. To reproduce \re{T->Q} one has to exclude $Q_{12}(u+1-j_q)$
from \re{Tf}. To this end, one applies the first relation in \re{taun} for $n=0$
and obtains after some algebra in the chiral limit
\be\label{q1q23}
Q_1(u-j_q) Q_{23}(u+1-j_q) - Q_3(u-j_q) Q_{12}(u+1-j_q) = \mathbb{P}^2
\left[-\frac{\Gamma(-u)}{\Gamma(j_q-u)}\right]^N.
\ee
We observe that the right-hand side of this relation has a pole of order $N$ at
$u=0$. Since $Q_{23}(u)$ and $Q_3(u)$ are polynomials, it can only be generated
by $Q_1(u)$ and $Q_{12}(u)$ which are meromorphic functions indeed. One deduces
from \re{compare} that  $Q_{1}(u-j_q) \sim Q_{13}(u-1-j_q) \Gamma^N(-u)+
\mathcal{O}(u^0)$ but the residue at the pole $u=0$ vanishes in the chiral limit
in virtue of \re{q13}. Thus, the first term in the left-hand side of \re{q1q23}
approaches a finite value for $u\to 0$ whereas the second term scales as $1/u^N$.
Then, combining together \re{q1q23} and \re{Tf} we finally obtain
\be
\mathbb{T}_{j_q}(u)=c(u) \left[\frac{Q_3(u)}{Q_3(u-j_q)} + \mathcal{O}(u^N)
\right],
\ee
with $c(u) =\left[-u\Gamma(1+u+j_q)/ \Gamma(-u+j_q) \right]^N $. Expanding this
relation at small $u$ and taking into account \re{H-SL21} one arrives at
\re{T->Q}.

\subsection{Matching quantum numbers}

Solutions to the nested TQ-relations \re{BA1} -- \re{BA3} are parameterized
by nonnegative integers $n_2$ and $n_3$. They determine the degrees of
$Q-$polynomials in \re{roots1} and verify the condition \re{cond}. In this
subsection we demonstrate that $n_2$ and $n_3$ have a simple physical meaning --
they define the total $SL(2|1)$ spins, $J$ and $\bar J$, carried by the
eigenstates of the spin chain $\ket{\Psi_q}$.

The eigenstates of the $SL(2|1)$ spin chain belong to the quantum space \re{HN}
and they can be classified according to irreducible $SL(2|1)$ representations
entering the tensor product \re{HN}. Let us choose $\ket{\Psi_q}$ to be the
lowest weight vectors in these representations. The remaining eigenstates can be
obtained from $\ket{\Psi_q}$ by applying the raising operators. Being the lowest
weights, the eigenstates $\ket{\Psi_q}$ diagonalize the operators $J$ and $\bar
J$, Eq.~\re{trans3}, and the Casimirs \re{Casimir} acting in the quantum space
of the model \re{HN}
\be\label{C23}
\mathbb{C}_2\ket{\Psi_q} = J \bar J \ket{\Psi_q} \,,\qquad
\mathbb{C}_3\ket{\Psi_q}=\frac12 \left(J-\bar J+\frac13\right) J\bar
J\ket{\Psi_q}\,.
\ee
The total chiral and antichiral spins take the form
\be\label{total-spins}
J=m+ Nj_q\,,\qquad \bar J = \bar m+N\bar j_q\,,
\ee
with $m$ and $\bar m$ nonnegative integer. According to \re{scaling} these
integers define the transformation properties of the eigenstates under
dilatations and $U(1)$ rotations
\ba \nonumber
\Psi_q(\lambda^2 z, \lambda \theta,\lambda \bar\theta) &=& \lambda^{m+\bar m}
\Psi_q(z, \theta,\bar\theta)\,,
\\[2mm] \label{m-mbar}
\Psi_q(z, \lambda^{-1} \theta,\lambda
\bar\theta) &=& \lambda^{m-\bar m} \Psi_q(z, \theta,\bar\theta)\,,
\ea
where $(z,\theta,\bar\theta)\equiv\{z_k,\theta_k,\bar\theta_k|1\le k\le L\}$
denotes the coordinates in the quantum space \re{HN}. In other words, $(m+\bar
m)/2$ defines the scaling dimension of the eigenstates while $m-\bar m$ defines
its $U(1)$ charge.

The Casimir operators $\mathbb{C}_2$ and $\mathbb{C}_3$ enter into expressions
for the conserved charges, Eqs.~\re{q2=C2} and \re{q-tilde}, and, therefore,
determine the leading asymptotic behavior of the transfer matrices \re{tau} and
\re{tau-tilde} at large $u$. Let us consider the TQ-relation \re{t/t} and
replace the transfer matrices by their explicit expressions, Eqs.~\re{t+-} and
\re{t-tbar}. Then, one examines the asymptotic behavior of the right-hand side
of \re{t/t} for large $u$ and finds with a help of \re{tau-tilde} and
\re{q-tilde}
\ba \nonumber
\frac{ {Q}_2(u)}{ {Q}_2(u-1)} &=&
1+\lr{\frac{2\mathbb{C}_3}{\mathbb{C}_2}-\frac13 +N(j-\bar j)+N-1}\,u^{-1}+
\mathcal{O}(u^{-2})
\\ \label{Q2/Q2}
&=&1+\lr{N-1+m-\bar m}\,u^{-1}+ \mathcal{O}(u^{-2})\,.
\ea
Here we replaced the Casimir operators by their corresponding eigenvalues and
took into account \re{P-2} and \re{C23}. Since ${Q}_2(u)$ is a polynomial of
degree $n_2$, Eq.~\re{roots1}, one deduces from \re{Q2/Q2} that $n_2=N-1+m-\bar
m$. In a similar manner, one examines asymptotic behavior of both sides of
\re{BA3} for large $u$, takes into account \re{tau} and \re{q2=C2} and obtains
after some algebra
\be
-q_2 + N j_q \bar j_q  = J\bar J = (N\bar j_q + n_3) ( Nj_q + n_{23})\,.
\ee
Together with $n_{13}=2(N-1)-n_2$ and $n_{23}=n_2+n_3-N+1$ this fixes the values
of integers as
\be\label{n's}
\begin{array}{ll}
n_2=N-1+m-\bar m\,, & \qquad n_3=\bar m\,, \\[2mm]
n_{13}=N-1-m+\bar m\,, & \qquad  n_{23}=m\,.
\end{array}
\ee
In distinction with $n_3$ and $n_{23}$, the values of $n_2$ and $n_{13}$ are
bounded from above, $0\le n_2, n_{13} \le 2(N-1)$. Since $n_2$ and $n_{13}$ take
nonnegative values only, possible values of integers $m$ and $\bar m$ are
subject to the constraint $|m-\bar m| \le N-1$.  As before, simplifications
occur in the chiral limit.

\subsubsection{Chiral limit}

For $\bar j_q=0$, the quantum space in each site is given by the chiral $SL(2|1)$
representation $\mathbb{V}_{j_q}$ and the Hilbert space of the model is
$(\mathbb{V}_{j_q})^{\otimes N}$. According to \re{chiral-irreps}, the
eigenstates $\Psi_q\in (\mathbb{V}_{j_q})^{\otimes N}$ verify the chirality
condition (with $k=1,\ldots,N$)
\be
D_k\Psi_q(z,\theta,\bar\theta)  \equiv \lr{-\partial_{\bar\theta_k} +
\ft12\theta_k\partial_{z_k}}\Psi_q(z,\theta,\bar\theta) = 0
\ee
which fixes the dependence of the wave functions on $\bar\theta-$variables as $
\Psi_q = \Psi_q(z_+, \theta) $ with $z_+=z+\ft12\bar\theta\theta$. Examining the
transformation properties of $\Psi_q(z_+, \theta)$ under \re{m-mbar}, one finds
that $\bar m-m \ge 0$ in the chiral limit. Moreover, for $m-\bar m=0$ the wave
function $\Psi_q$ does not depend on $\theta$'s and it is a function of $z_+$
only. Since $\Psi_q(z_+)$ is the lowest weight, it has also to be annihilated by
the lowering operators \re{trans1}. This leads (up to an overall normalization)
to $\Psi_q=1$ or, equivalently, $m=\bar m=0$. Thus, in the chiral limit,
possible values of the integers $m$ and $\bar m$ are $m=\bar m=0$, or $ 1\le
\bar m - m \le N-1$.

Let us examine the TQ-relations \re{NBA-chiral} for different values of the
integers $m$ and $\bar m$.

\subsubsection*{$\Mybf {m=\bar m=0}$}

In this case the eigenstate has the form $\Psi_q=1$. It coincides with a
pseudovacuum state in the nested Bethe Ansatz and, therefore, it does not have
any Bethe roots associated with it. Indeed, one deduces from \re{n's} that for
$m=\bar m=0$ the polynomials $Q_3(u)$ and $Q_{23}(u)$ are reduced to a c-number,
$Q_3(u)=Q_{23}(u)=1$, and obtains from the third relation in \re{NBA-chiral} the
corresponding transfer matrix as $ \tau_N(u)=u^N$. Then, one applies
\re{QQ-chiral} to obtain (up to an overall normalization factor)
\be\label{chiral-Bax0}
Q_2(u) = u^N -(u+j_q)^N\,,\qquad Q_{13}^{(0)}(u) = 0\,.
\ee
One finds from \re{E-SL21} the corresponding energy and quasimomentum as
\be
E=0\,,\qquad \e^{i\theta}=1\,.
\ee

\subsubsection*{$\Mybf{\bar m - m=N-1}$}

In this case, the eigenstate $\Psi_q(z_+, \theta)$ has the $U(1)$ charge $N-1$
and, therefore, it is proportional to a homogeneous polynomial in $\theta$'s of
degree $(N-1)$. Requiring that $\Psi_q(z_+, \theta)$ should be annihilated by
the lowering operators \re{trans1},  one finds
\be\label{vec1}
\Psi_q(z_+, \theta)= (\theta_{1}-\theta_2)(\theta_{2}-\theta_3)\ldots
(\theta_{N-1}-\theta_N)\,\varphi_q(z_+)\sim V^- \theta_1 \ldots \theta_N
\,\phi_q(z_+)\,,
\ee
with $\phi_q(z_+)$ being a translation invariant function of $z_{k,+}$ (with
$k=1,\ldots,N$) and $V^-=\sum_k \partial_{\theta_k}+\ft12\bar\theta_k
\partial_{z_k}$ being the lowering operator in $(\mathbb{V}_{j_q})^{\otimes N}$.
As we will see in a moment, the function $\phi_q(z_+)$ coincides with the
eigenstates of the $SL(2)$ magnet of spin $s=(1+j_q)/2$.

For $\bar m-m=N-1$ one finds from \re{n's} that $n_2=0$ and, therefore,
$Q_2(u)=1$. Its substitution into \re{NBA-chiral} yields
\be\label{chiral-Bax1}
\tau_N(u)+(u+j_q-1)^N =  (u-1)^N \frac{Q_{23}(u-1)}{Q_{23}(u)} +(u+j_q)^N
\frac{Q_{23}(u+1)}{Q_{23}(u)}\,.
\ee
Shifting the spectral parameter as $u\mapsto u-\ft12(j_q-1)$ one identifies this
relation as the Baxter equation \re{TQ-SL2} for the $SL(2)$ magnet of length $N$
and spin $s=\ft12(1+j_q)=\ft12+\ell$. Denoting its polynomial solution as
$P_m^{(s)}(u)$ one finds
\be\label{chiral-Bax2}
Q_{23}(u) = P_m^{\lr{\ell+1/2}}\lr{u+\ell-\ft12}
\, ,
\ee
with $m\ge 0$. Plugging this expression into \re{chiral-Bax1} and expanding
both sides in powers of $u$ one can identify the conserved charges \re{tau}.
Making use of \re{QQ-chiral} and \re{q13} one determines the remaining
polynomials as (up to an overall normalization)
\ba \nonumber
Q_3(u) &=& u^N Q_{23}(u)  - (u+j_q)^N Q_{23}(u+1)
\, ,
\\[2mm] \label{chiral-Bax3}
Q_{13}^{(0)}(u) &=&  \tau_N(u+1) - (u+1)^N\,.
\ea
In agreement with \re{n's}, they have degree $N-1+m$ and $2(N-1)$, respectively.
Inserting \re{chiral-Bax3} and \re{chiral-Bax2} into \re{E-SL21} one finds
that the corresponding energy and quasimomentum are related to their counterparts
in the $SL(2)$ spin chain as
\ba \nonumber
E&=&\lr{\ln P_m^{\lr{\ell+1/2}}\lr{\ell+\ft12} }'-
\lr{P_m^{\lr{\ell+1/2}}\lr{-\ell-\ft12}}'=E_m^{\lr{\ell+1/2}}\,,
\\[2mm] \label{E-relation}
\e^{i\theta}&=&(-1)^{N-1}{P_m^{\lr{\ell+1/2}}\lr{\ell+\ft12}}/
{P_m^{\lr{\ell+1/2}}\lr{-\ell-\ft12}}=(-1)^{N-1}\e^{i\theta_m^{\lr{\ell+1/2}}},
\ea
where the superscript in the right-hand side refers to the spin of the $SL(2)$
magnet.

\subsubsection*{$\Mybf{\bar m - m=1}$}

In this case the eigenstate  $\Psi_q(z_+, \theta)$ has a unit $U(1)$ charge and,
therefore, it is given by a linear combination of $\theta$'s with prefactors
depending on $z_+-$variables only. The latter are fixed from the requirement
that $\Psi_q(z_+, \theta)$ has to be annihilated by the lowering generators
\re{trans1} leading to
\be\label{vec2}
\Psi_q(z_+, \theta)= \big(\sum_{k=1}^N
\theta_k\,\partial_{z_{k,+}}\big)\chi_q(z_+)=
\bar
V^- \chi_q(z_+) \,,
\ee
with $\chi_q(z_+)$ being a translation invariant function of $z_{k,+}$ (with
$k=1,\ldots,N$) and $\bar V^-=\sum_k \partial_{\bar\theta_k}+\ft12\theta_k
\partial_{z_k}$ being the lowering operator in $(\mathbb{V}_{j_q})^{\otimes N}$.
Again, we will show at the end of this subsection that $\chi_q(z_+)$ coincides
with eigenstates of the $SL(2)$ magnet of spin $s=j_q/2$.

It follows from \re{n's} and \re{q13} that $Q_{13}^{(0)}(u)$ a polynomial of
degree $n_{13}-N=\bar m -m-1$ and, therefore, for $\bar m - m=1$ it reduces to a
c-number $Q_{13}^{(0)}(u)=1$. One applies the second relation in \re{NBA-chiral}
to find
\be\label{chiral-Bax4}
\tau_N(u)+(u+j_q)^N = u^N \frac{Q_3(u-1)}{Q_3(u)}+{(u+j_q)}^N
\frac{Q_3(u+1)}{Q_3(u)}\,.
\ee
Similarly to \re{chiral-Bax1}, one substitutes $u \to u-j_q/2$ and matches the
resulting relation into the Baxter equation for the $SL(2)$ magnet of length $N$
and spin $s=j_q/2=\ell$. As a result, the polynomial solution to \re{chiral-Bax4}
reads (up to a normalization factor)
\be\label{Q3-spec}
Q_3(u) = P_{m+1}^{\lr{\ell}}\lr{u+\ell}
\ee
with $m\ge 0$. Substituting this relation into \re{QQ-chiral} one obtains the
remaining polynomials as
\ba \nonumber
Q_2(u)   &=&   \tau_N(u) - u^N
\, ,
\\[2mm]
Q_{23}(u)  &=&  Q_3(u-1)-Q_3(u)
\, .
\ea
It is straightforward to verify that the obtained expressions for the
$Q-$polynomials verify the TQ-relations \re{Q2-chiral} -- \re{Q23-chiral}. From
\re{Q3-spec} and \re{E-SL21} one finds the corresponding energy and quasimomentum as
\ba \nonumber
E&=&\lr{\ln P_{m+1}^{\lr{\ell}}\lr{\ell} }'-
\lr{P_{m+1}^{\lr{\ell}}\lr{-\ell}}'=E_{m+1}^{\lr{\ell}}
\, ,
\\[2mm]
\e^{i\theta}&=&{P_{m+1}^{\lr{\ell}}\lr{\ell}}/
{P_{m+1}^{\lr{\ell}}\lr{-\ell}}=\e^{i\theta_{m+1}^{\lr{\ell}}}\,,
\ea
where in distinction with \re{E-relation} the spin of the $SL(2)$ magnet equals
$\ell$.

\subsubsection*{$\Mybf{2 \le \bar m - m\le N-2}$}

This case is only realized for the spin chain of length $N\ge 4$. The eigenstate
$\Psi_q(z_+,\theta)$ carries the $U(1)$ charge equal to $\bar m - m$ and it is
given by a homogeneous polynomial in $\theta$'s of degree $\bar m - m$ with the
coefficient given by $z_+-$dependent functions. In distinction to \re{vec1}
and \re{vec2}, these functions are, in general, independent of each other.
Going over to $Q-$polynomials, one notices that the eigenstates \re{vec1} and
\re{vec2} corresponds to `extreme' solutions to the TQ-relations \re{NBA-chiral}
when one of the polynomials, $Q_2(u)$ or $Q_{13}^{(0)}(u)$, reduces to a c-number.
For $2 \le \bar m - m\le N-2$ the two polynomials have degrees $N-1-(\bar m-m)$
and $(\bar m -m )-1$, respectively. Their explicit form can be found from the
TQ-relations \re{NBA-chiral}.

We have demonstrated in this subsection that for $\bar m -m=1$ and $\bar m-m=N-
1$ solutions to the `chiral' TQ-relations \re{NBA-chiral} are expressed in terms
of $Q-$polynomials for the $SL(2)$ magnet. For $N=2,\, 3$ there are no other
solutions to the TQ-relations whereas for $N \ge 4$ there exist additional
solutions with $2\le \bar m -m \le N-2$. This property can be understood as
follows. We recall that in the chiral limit, $\bar j_q=0$, the $SL(2|1)$ spin
chain describes the dilatation operator in the $\mathcal{N}=1$ SYM theory. The
product of $N$ superfields, one for each site, $\Phi_N(Z)\equiv
\tr\left[\Phi(z_1,\theta_1)\ldots \Phi(z_N,\theta_N)\right]$ defines the quantum
space of the $SL(2|1)$ magnet. The chiral superfield
$\Phi(z,\theta)=\chi(z)+\theta \phi(z)$ describes a `short' $SL(2)\otimes U(1)$
multiplet built from two fields $\phi(z)$ and $\chi(z)$ carrying the $SL(2)$
spins $\ell$ and $\ell+\ft12$ (with $j_q=2\ell$) and the $U(1)$ charges $\ell$
and $\ell-\ft12$, respectively. Then, expansion of the single-trace operator in
powers of `odd' variables looks as
\be
\Phi_N(Z) =\chi_N(z) + \ldots + \phi_N(z)\,\prod_{k=1}^N \theta_k\,,
\ee
where $\chi_N(z)=\tr\left[\chi(z_1)\ldots \chi(z_N)\right]$ and
$\phi_N(z)=\tr\left[\phi(z_1)\ldots \phi(z_N)\right]$. The dilatation operator
`mixes' together different components of the sum carrying the same number of
$\theta$'s. A distinguished feature of the two components, $\chi_N(z)$ and
$\phi_N(z)$, is that the dilatation operator acts on them autonomously. For such
states, corresponding to the so-called maximal helicity
operators~\cite{BraDerMan98}, the dilatation operator can be mapped into a
Hamiltonian of the $SL(2)$ magnet of spin $\ell$ and $\ell+\ft12$, respectively.
The state $\chi_N(z)$ is a descendant of the $SL(2|1)$ lowest weight vector
\re{vec2} with $\bar m - m = 1$ while $\theta_1\ldots\theta_N \phi_N(z)$ is a
descendant of the lowest weight \re{vec1} with $\bar m - m = N-1$. In both cases,
the $SL(2|1)$ Hamiltonian effectively reduces to the Hamiltonian of the $SL(2)$
spin chain of length $N$. Thus, for $\bar m - m = 1$ and $\bar m - m = N-1$ the
solution to the $SL(2|1)$ and $SL(2)$ Baxter equations are related to each other
and the same relation holds true between the energy spectrum of two models.

We remind that the integers $m$ and $\bar m$ are related to the total spin of
the $SL(2|1)$ spin chain \re{total-spins}. A distinguished feature of this model
as compared with conventional compact spin chains is that the total spin can
now take arbitrarily large values and the energy spectrum of the model is not
restricted from above for a finite length of the spin chain. In terms of the
Baxter $Q-$operators, this property implies that the TQ-relations \re{TQ-} admit
an infinite number of polynomial solutions $Q_3(u)$ parameterized by $m$ and
$\bar m$. For small $m$ and $\bar m$ they can be worked out explicitly while
for large $m$ and/or $\bar m$ one can construct asymptotic solutions to the
TQ-relation by applying the semiclassical approach of Ref.~\cite{Kor95}.

\section{Conclusion}

In this paper, we developed an approach for systematic construction of the Baxter
$Q-$operators in integrable noncompact spin chain models with Lie supergroup
symmetry. We exposed the formalism by applying it to the generalized homogeneous
Heisenberg magnet with the $SL(2|1)$ symmetry. Apart from being the simplest case
which allows us to demonstrate all essential features of our approach, the model
has a wide spectrum of applications, ranging from superconductivity to dynamics
of four-dimensional gauge theories.

The central r\^ole in our analysis is played by noncompact transfer matrices
defined over generic infinite-dimensional $SL(2|1)$ representations in the
auxiliary space. We have demonstrated that, in comparison with conventional
compact transfer matrices, they have a number of remarkable properties. Firstly,
a generic noncompact transfer matrix is factorized into the product of three
`special' noncompact transfer matrices. We argued that the latter can be
identified as three Baxter operators $\mathcal{Q}_a(u)$ (with $a=1,2,3$).
Secondly, for certain values of spins, infinite-dimensional $SL(2|1)$
representations become reducible indecomposable and their irreducible components
define both infinite-dimensional (chiral and antichiral) and finite-dimensional
(typical and atypical) representations of the $SL(2|1)$. This property leads to
the hierarchy between the corresponding transfer matrices and allows one to
express all transfer matrices in terms of the $\mathcal{Q}-$operators.

Combining the two properties together, we derived finite difference equations
for the Baxter operators, the so-called TQ-relations. In distinction with higher
rank classical Lie groups, the TQ-relations for the $SL(2|1)$ group are of the
second order at most. Two out of the three Baxter operators, $\mathcal{Q}_1(u)$
and $\mathcal{Q}_3(u)$, satisfy the same second order finite-difference equation
while the operator $\mathcal{Q}_2(u)$ obeys a finite-difference equation of the
first order. The former equation is quite similar in structure to the
TQ-relation for the $SL(2)$ magnet. Important difference being however that its
c-number dressing factors are now replaced by (operator valued) compact transfer
matrices.

The TQ-relations are invariant under the multiplication of the $Q-$operators by
an arbitrary periodic function with period $1$ and, therefore, they have to be
supplemented by additional conditions on their solutions. To deduce these
conditions one needs an explicit expression for the $Q-$operators. We
demonstrated that for the $SL(2|1)$ magnet under consideration the eigenvalues
of the operators $\mathcal{Q}_3(u)$ and $\mathcal{Q}_2(u)$ are polynomials in $u$
and the eigenvalues of the operator $\mathcal{Q}_1(u)$ are meromorphic functions
of $u$. This property is intimately related to the fact that the quantum space
of the model contains a pseudovacuum state which is annihilated by the lowering
$SL(2|1)$ generators in all sites. It also allows one to demonstrate the
equivalence of the Baxter $Q-$operator method and the nested Bethe ansatz
approach. Parameterizing the polynomial eigenvalues of the operators
$\mathcal{Q}_3(u)$ and $\mathcal{Q}_2(u)$ by their roots, we showed that the
TQ-relations for the Baxter operators lead to a system of coupled equations for
the roots which coincide with similar relations in the nested Bethe ansatz
solution for the $SL(2|1)$ magnet.

The advantage of the Baxter $Q-$operator method, though, is that it does not rely
on the existence of the pseudovacuum and, therefore, it can be applied to the
models which do not possess such a state. Indeed, the derivation of the
TQ-relations is based on the decomposition of infinite-dimensional $SL(2|1)$
representations over irreducible components and it is not sensitive to a detailed
structure of the representation space. The latter information is encoded in
analytical properties of the $Q-$operators. Polynomiality of the $Q-$operators is
in one-to-one correspondence with the existence of the pseudovacuum state. If the
quantum space of the model does not have it, the nested Bethe ansatz is not
applicable and the eigenvalues of the $Q-$operators are, in general, meromorphic
functions of $u$. To identify their analytical properties (position and order of
poles, asymptotic behavior at infinity) one has to explicitly construct the
corresponding $Q-$operators. Integrable $SL(2|1)$ spin chain with the quantum
space of the form $(\mathbb{V} \otimes \mathbb{\bar V})^{\otimes N/2}$ mentioned
in the Introduction provides an example of the model to which the nested Bethe
ansatz is not applicable. It would be interesting to apply our approach to this
model and work out its exact solution using the method of the Baxter
$Q-$operator.

In this paper, we have only focused on the eigenvalue problem for the $SL(2|1)$
magnet, relating its energy spectrum to that of the Baxter $Q-$operators. Another
advantage of these operators is that they can be also used to construct the
eigenfunctions of the $SL(2|1)$ magnet in the representation of separated
variables (SoV)~\cite{Skl85}. In spite of the fact that the SoV method has been
formulated awhile ago, the number of models for which it has been successfully
implemented is limited. In the SoV representation, the eigenfunction factorizes
into a product of functions depending on a single separated variable. In the case
of the $SL(2)$ spin chain, going over through an explicit construction of the SoV
representation, one can show that the latter functions coincide with polynomial
eigenvalues of the Baxter $Q-$operator~\cite{DerKorMan03}. It is expected that
this property is rather general and it should also hold for spin chains with Lie
(super)symmetry of high rank including the $SL(2|1)$ group. This question
deserves additional study.

The construction of $Q-$operators can be extended to spin chains with the
$SL(2|\mathcal{N})$ symmetry \cite{Ess92,Sal99,Arn03}. These models have recently
attracted attention in light of the gauge/string duality, most notably for the
maximally supersymmetric $\mathcal{N} = 4$ gauge theory. We expect that a generic
noncompact $SL(2|\mathcal{N})$ transfer matrices can be factorized into products
of $(\mathcal{N} + 2)$ distinct $\mathcal{Q}-$operators. Two of them,
$\mathcal{Q}_1(u)$ and $\mathcal{Q}_{\mathcal{N} + 2}(u)$, are analogous to the
$SL(2)$ Baxter $Q-$operators while the remaining operators $\mathcal{Q}_a$ (with
$a = 2, \dots, \mathcal{N} + 1$) should reflect a nontrivial $SU(\mathcal{N})$
group structure of the model. We demonstrated in this paper that for
$\mathcal{N}=1$, the operators $\mathcal{Q}_1(u)$ and
$\mathcal{Q}_{\mathcal{N}+1}(u)$ for the $SL(2|1)$ magnet can be obtained from
the $SL(2)$ operators by a lift from the light-cone to the $\mathcal{N}=1$
superspace $z\mapsto \mathcal{Z}=(z,\theta,\bar\theta)$. Going over from
$\mathcal{N}=1$ to higher $\mathcal{N}$, one should simply enlarge the number of
`odd' dimensions in the superspace $\mathcal{Z}=(z,\theta^A,\bar\theta_A)$ with
${\scriptstyle A}=1,\ldots, \mathcal{N}$. In other words, for arbitrary
$\mathcal{N}$, the Baxter operators $\mathcal{Q}_1(u)$ and
$\mathcal{Q}_{\mathcal{N}+2}(u)$ can be represented by the same Feynman diagrams
as shown in Figs.~\ref{fig:Q3} and \ref{fig:Q1}. The only difference with the
$\mathcal{N}=1$ expressions is that the reproducing kernel and the integration
measure should be modified to take into account the contribution from extra
$(\mathcal{N}-1)$ `odd' coordinates in the superspace. Moreover, in the chiral
limit, the operator $\mathcal{Q}_{\mathcal{N}+2}(u)$ has the form identical to
\re{Q=shift}. The only change is that for arbitrary $\mathcal{N}$ the Baxter
operator in \re{Q=shift} acts in the space of functions defined in the
$(\mathcal{N} + 1)-$dimensional superspace $W=(w,\theta^A)$. Remarkably enough,
substitution of the operator $\mathcal{Q}_{\mathcal{N}+2}(u)$ into \re{H-SL21}
yields the Hamiltonian which coincides with the one-loop dilatation operator in
the $\mathcal{N}-$extended SYM theory, Eqs.~\re{DO} and~\re{V-super}. A detailed
study of the $SL(2|\mathcal{N})$ Baxter $Q-$operators will be presented
elsewhere.

\section*{Acknowledgements}

We are grateful to D. Karakhanyan for collaboration at an early stage  of the
project. Three of us (A.B., A.M.\ and S.D.) would like to thank Laboratoire de
Physique Th\'eorique (Orsay) for hospitality. G.K. is most grateful to A. Tsvelik
for discussions. The work was supported in part by the U.S.\ National Science
Foundation under grant no.\ PHY-0456520 (A.B.), by the RFFI grant 05-01-00922 and
the DFG grant 436 Rus 17/9/06 (S.D.), by the ANR under grant BLAN06-3\_143793
(G.K.) and by the Helmholtz Association under contract VH-NG-004 (A.M.).

\appendix

\setcounter{section}{0} \setcounter{equation}{0}
\renewcommand{\theequation}{\Alph{section}.\arabic{equation}}

\section{Reducible representations of the $SL(2|1)$}

The $SL(2|1)$ representation $[j,\bar j]$ is reducible for values of spins $j$
and $\bar j$ specified in Section~2.2.

\subsubsection*{$\Mybf {\bar j=0}$}

In this case, the $SL(2|1)$ generators defined in Eqs.~\re{trans1} --
\re{trans3}
depend on the
spin $j$.%
\footnote{The $j-$dependence resides in three generators $J$, $\bar V^+$ and
$L^+$ only.} Denoting them as $G_j$ one finds that the superconformal derivative
$D$, \re{cov-der}, intertwines the $SL(2|1)$ generators of spin $j$ and $j+1$
\be\label{rel-S}
D\,  G_j = (-1)^{\bar G} G_{j+1} \, D
\ee
with the grading $\bar G=0$ and $\bar G=1$ for even and odd generators,
respectively. Let us consider the state $\Phi_+(z,\theta,\bar\theta)\in
\mathcal{V}_{j,0}$ which satisfies the chirality condition $
D\,\Phi_+(z,\theta,\bar\theta)=0$. Applying both sides of \re{rel-S} to
$\Phi_+(z,\theta,\bar\theta)$ one finds that the state
$G_j\Phi_+(z,\theta,\bar\theta)$ is also annihilated by the supercovariant
derivative, $D\, G_j\Phi_+(z,\theta,\bar\theta) =0$. Therefore, the zero modes
of the operator $D$ form the $SL(2|1)$ invariant subspace $\mathbb{V}_j$,
Eq.~\re{chiral-irreps}. From
$D=-\e^{\bar\theta\theta\partial/2}\partial_{\bar\theta}
\e^{-\bar\theta\theta\partial/2}$ one finds that the states
$\Phi_+(z,\theta,\bar\theta)\in \mathbb{V}_j$ have the following form
\be\label{ch-V}
D \,\Phi_+=0 \quad \mapsto \quad  \Phi_+(z,\theta,\bar\theta) = \chi(z_+) +
\theta \phi(z_+)
\ee
with $z_+=z+\ft 12 \bar\theta\theta$ and $\chi(z)$ and $\phi(z)$ being analytical
inside the unit disk $|z|<1$. Expanding the right-hand side of \re{ch-V} in
powers of $z$ one finds that the basis vectors in the graded linear space
$\mathbb{V}_j$ are given by homogeneous polynomials in $\theta$ and
$z+\ft12\bar\theta\theta$ only, Eq.~\re{chiral-irreps}. The states \re{ch-V} form
the chiral $SL(2|1)$ representation $[j]_+$. Similar to \re{sl2-dec}, it can be
decomposed over the $SL(2)\otimes U(1)$
multiplets~\cite{Sch76,Jar78,Marcu79,Frappat96}
\be
[j]_+ =  \mathcal{D}_\ell(\ell) \oplus \mathcal{D}_{\ell+1/2}(\ell-\ft12) \,,
\ee
with $\chi(z_+) \in  \mathcal{D}_\ell(\ell)$ and $\phi(z_+)\in
\mathcal{D}_{\ell+1/2}(\ell-\ft12)$.

In Eq.~\re{G-mat}, the diagonal blocks $G_{++}$ and $G_{--}$ define two
different representations of the $SL(2|1)$ generators on the invariant subspace,
$\mathbb{V}_j$, and the quotient, $\mathcal{V}_{j,0}/\mathbb{V}_j$, respectively.
One applies both sides of \re{rel-S} to $\Phi_-\in
\mathcal{V}_{j,0}/\mathbb{V}_j$ and makes use of \re{G-mat} together with $D \,
\Phi_+=0$ to obtain
\be\label{GD}
G_{j+1} \, D \, \Phi_-^\alpha  = (-1)^{\bar G} D \, G_j \cdot \Phi_-^\alpha
=(-1)^{\bar G} D \, \Phi_-^\beta\, [G_{--}]^{\beta\alpha}\,.
\ee
It follows from this relation that (infinite-dimensional) graded matrix $G_{--}$
represents of the $SL(2|1)$ generators of spin $j+1$ on the space spanned by the
states $ D\, \Phi_-$ with $\Phi_-\in \mathcal{V}_{j,0}/\mathbb{V}_j$. Since $D
\cdot D \Phi_- =0$, this space is isomorphic to the chiral $SL(2|1)$ irreps
\be\label{iso-chir}
D \Phi_- \in \mathbb{V}_{j+1}= \Span\left\{1,\theta z^k,
\left(z+\ft12\bar\theta\theta\right)^{k+1} |\, k \in \mathbb{N} \right\}\,.
\ee
The $SL(2|1)$ generators on $\mathbb{V}_{j+1}$ are given by the same expressions
as before, Eqs.~\re{trans1} -- \re{trans3}, with $j$ replaced by $j+1$ and $\bar
j=0$.

\subsubsection*{$\Mybf{j+\bar j=-n}$}

Let us introduce the following operator
\be\label{W}
\mathcal{I} =  j \,(D \bar D)^{n} - \bar j \,(\bar D D)^{n} = (-\partial_z)^{n-1}
\left[j \,D \bar D - \bar j \,\bar D D\right]
\ee
where $D$ and $\bar D$ are supercovariant derivatives \re{cov-der}. One can
verify that the $SL(2|1)$ generators $G_{j,\bar j}$ defined in \re{trans1} -
\re{trans3} satisfy the relation
\be\label{W-int}
\mathcal{I}  \,G_{j,\bar j} = G_{-\bar j, -j}\, \mathcal{I}   \,,
\ee
and, therefore, $\mathcal{I} $ intertwines the corresponding $SL(2|1)$
representations $[j,\bar j]$ and $[-\bar j,-j]$. This relation is analogous to
\re{rel-S} and, as before, it implies that zero modes of the operator
$\mathcal{I} $ belong to the $SL(2|1)$ invariant subspace $v_{n/2,\,b}= \ker
\mathcal{I} $, Eq.~\re{vn},
\be
\mathcal{I} \, \Phi_+(z,\theta,\bar\theta) =0 \quad \mapsto \quad
\Phi_+(z,\theta,\bar\theta) \in v_{n/2,\,b}\,.
\ee
The basis in the quotient space $\mathcal{V}_{j,\bar j}/v_{n/2,\,b}$  can be
constructed by applying the raising operators $V^+$, $\bar V^+$ and $L^+$ to a
given reference state $\Omega'$ which does not belong to \re{vn} and has a
smallest possible degree ($=n$) in $z$. One possible choice could be
\be\label{Omega'}
\Omega' = z^{n} -\ft12\alpha  z^{n-1} \theta\bar\theta\,,
\ee
with $\alpha \neq j-\bar j$ so that $\Omega'$ is different from the highest
weight \re{Omega}. Notice that this state is not the lowest weight in
$\mathcal{V}_{j,\bar j}$. However, the states $V^-\Omega'$, $\bar V^- \Omega'$
and $L^-\Omega'$ belong to $v_{n/2,\,b}$ and, therefore, equal zero in the
quotient $\mathcal{V}_{j,\bar j}/v_{n/2,\,b}$. Similar to \re{GD}, one applies
both sides of \re{W-int} to $\Phi_-\in\mathcal{V}_{j,\bar j}/v_{n/2,\,b}$ and
obtains that the states $\mathcal{I} \Phi_-$ form the $SL(2|1)$ representation
$[-\bar j, -j]$
\be
\mathcal{I} \,\Phi_-(z,\theta,\bar\theta) \in \mathcal{V}_{-\bar j, -j}
\ee
with $\mathcal{I} \, \Omega' \sim 1$.

\subsubsection*{$\Mybf{j=\bar j=0}$}

The invariant subspace $v_{00}$ contains only one state $1$ and the quotient
space $\mathcal{V}_{0,0}/v_{00}$ takes the form \re{V00}. The spaces
$\mathcal{V}_+$ and $\mathcal{V}_-$ contain the states $\Omega_+=\theta$ and
$\Omega_-=\bar\theta$, respectively, such that the vectors $V^-\Omega_\pm$,
$\bar V^-\Omega_\pm$ and $L^-\Omega_\pm$ either vanish or have zero projection
onto $\mathcal{V}_{0,0}/v_{00}$. This allows one to realize $\mathcal{V}_+$ and
$\mathcal{V}_-$ as lowest weight representations built over these two states.
The reference states $\Omega_\pm$ verify the (anti)chirality conditions
\be
D \,\Omega_+ = \bar D\, \Omega_- =0
\ee
and the same condition is fulfilled for all states inside $\mathcal{V}_\pm$. One
applies both sides of \re{rel-S} to $\Phi_-(z,\theta,\bar\theta)\in
\mathcal{V}_-$
and finds that for $j=0$ the states $ D\,\Phi_-(z,\theta,\bar\theta)$ form the
$SL(2|1)$ invariant chiral representation space $[1]_+$, Eq.~\re{chiral-irreps}.
Notice that the value of the chiral spin is shifted from $j=0$ to $j=1$ as a
consequence of \re{rel-S}. The same result can also be obtained by examining the
eigenvalues of the Cartan operators $J$ and $\bar J$ for the reference state
$\Omega_-$, $J\Omega_-=\Omega_-$ and $\bar J \Omega_-=0$. In a similar manner,
the chiral representation $\mathcal{V}_+$ can be mapped into the antichiral
representation $[1]_-$.

\subsubsection*{$\Mybf{j=-n,\ \bar j=0}$}

According to \re{red1} the representation $[-n,0]$ decomposes into a semidirect
sum of two chiral representations $[-n]_+$ and $[-n+1]_+$ which are both
indecomposable reducible for $n\ge 1$. Let us consider the representation
$[-n]_+$ and introduce the operator
\be
\mathcal{I} =-\bar D (D \bar D)^n= - \bar D(-\partial_z)^n\,.
\ee
It annihilates the invariant subspace ${\rm v}_n$ defined in \re{v-chi},
$\mathcal{I} \,\Phi_+=0$ for $\Phi_+\in\mathrm{v}_n$, and intertwines the
$SL(2|1)$ representations $[-n]_+$ and $[n+1]_-$
\be\label{Wn}
\mathcal{I}  G_{-n, 0}  = (-1)^{\bar G}  G_{0,1+n}\mathcal{I} \,.
\ee
The quotient space $\mathbb{V}_{-n}/ \mathrm{v}_n$ is spanned by the states
\be\label{V-n}
\mathbb{V}_{-n}/ \mathrm{v}_n= \Span\left\{\theta z^{n}, z_+^{n+1},\theta
z^{n+1}, z_+^{n+2}, \ldots \right\}\,.
\ee
As before, one applies both sides of \re{Wn} to $\Phi_-\in \mathbb{V}_{-n}/
\mathrm{v}_n$ and finds that $\mathcal{I} \, \Phi_-$ form the antichiral
$SL(2|1)$ representation $[n+1]_-$
\be
\mathcal{I} \, \Phi_- \in \mathbb{\bar V}_{n+1} =\Span\left\{1,\bar\theta z^k,
z_-^{k+1}\,|\, k\in \mathbb{N} \right\}\,,
\ee
with $z_\pm = z \pm \ft12 \bar\theta\theta$. One concludes that $[-n]_+$
decomposes into a semidirect sum of $\mathrm{v}_n$ and $\mathbb{\bar V}_{n+1}$.

\section{Calculation of the normalization factors}

In this Appendix we calculate the normalization factors entering the expressions
for the $\mathcal{R}-$mat\-rix, Eqs.~\re{c-factor} and \re{a-factor}. In all
cases, the calculation goes through the same main steps. Let $[j,\bar j]$ be a
reducible indecomposable representation of the $SL(2|1)$ and let $\mathbb{V}^+$
be invariant subspace of $\mathcal{V}_{j,\bar j}$. The $\mathcal{R}-$operator
acting on the tensor product $\mathcal{V}_{j_q,\bar j_q}\otimes \mathcal{V}_{j,
\bar j}$ has a block-triangular form \re{Rbd}. The upper diagonal block
$\mathbb{R}^+(u)$ defines the $\mathcal{R}-$operator on $\mathcal{V}_{j_q,\bar j_q}
\otimes \mathbb{V}^+$. It can be defined by applying both sides of \re{Rbd} to
the same test vector $\Phi^+\in \mathbb{V}^+$
\be
\mathcal{R}_{\mathcal{V}_{j_q,\bar j_q}\otimes \mathbb{V}^+}(u) \Phi^+ =
\mathcal{R}_{\mathcal{V}_{j_q,\bar j_q}\otimes \mathcal{V}_{j,\bar j}}(u)
\Phi^+\,.
\ee
The expression in the right-hand side of this relation can be evaluated using
explicit expression for the operator $\mathcal{R}_{\mathcal{V}_{j_q,\bar
j_q}\otimes \mathcal{V}_{j,\bar j}}(u)$ from Ref.~\cite{DerKarKir98}. The
invariant subspace can be defined as a kernel of a certain operator $\mathcal{I}
$, so that $\mathcal{I} \Phi^+=0$ for $\Phi^+\in \mathbb{V}^+$. The same operator
maps the quotient space $\mathcal{V}_{j,\bar j}/\mathbb{V}^+$ into yet another
representation, say $\mathcal{V}_{j',\bar j'}$ leading to
\be\label{WR}
\mathcal{I}\, \mathcal{R}_{\mathcal{V}_{j_q,\bar j_q}\otimes \mathcal{V}_{j,\bar
j}}(u) \Phi^- = c(u) \mathcal{R}_{\mathcal{V}_{j_q,\bar j_q}\otimes
\mathcal{V}_{j',\bar j'}}(u) \mathcal{I}\, \Phi^-\,,
\ee
where $\Phi^-$ is an arbitrary test vector in $\mathcal{V}_{j,\bar
j}/\mathbb{V}^+$. Using explicit expressions for the $\mathcal{R}-$ and
$\mathcal{I}-$operators, one can evaluate both sides of this relation and, then,
determine the normalization factor $c(u)$. The calculation can be simplified by
choosing $\Phi_-$ to be the lowest weight in the tensor product
$\mathcal{V}_{j_q,\bar j_q}\otimes \mathcal{V}_{j,\bar j}$.  Then, the vector
$\mathcal{I} \Phi^-$ is automatically the lowest weight in $\mathcal{V}_{j_q,\bar
j_q}\otimes \mathcal{V}_{j',\bar j'}$ and the two $\mathcal{R}-$operators
entering \re{WR} are diagonalized simultaneously. As a result, $c(u)$ is given by
the ratio of the corresponding eigenvalues.

A complete classification of the lowest weights in the tensor product of two
generic $SL(2|1)$ representations  $\mathcal{V}_{j_q,\bar j_q}\otimes
\mathcal{V}_{j,\bar j}$ can be found in Ref.~\cite{DerKarKir98}. One finds among
them the states  $1$ and $(\bar\theta_1-\bar\theta_{2}) (z_1-z_{2})^k$ (with $k
\in \mathbb{N}$) that we shall use as reference states $\Phi^-$ in \re{WR}. The
action of the $\mathcal{R}-$operator on these states looks
like~\cite{DerKarKir98}
\ba \nonumber
 \mathcal{R}_{\mathcal{V}_{j_q,\bar
j_q}\otimes \mathcal{V}_{j,\bar j}}(u)\cdot 1 &=& r(j,\bar j)\,,
\\[2mm] \label{RR}
\mathcal{R}_{\mathcal{V}_{j_q,\bar j_q}\otimes \mathcal{V}_{j,\bar j}}(u) \cdot
\bar\theta_{12} z_{12}^k &=& r_k(j,\bar j)\,\bar\theta_{12}
z_{12}^k\,,
\ea
where $\bar\theta_{12}=\bar\theta_1-\bar\theta_2$, $z_{12}=z_1-z_2$ and the
notation was introduced for
\ba\nonumber
r_k(j,\bar j) &=&  \xi {{ \left( -1 \right) ^{k}\e^{i\pi(j+\bar j)/2} \left(
u+j_{{q}}-j+ \bar j \right) \left( u-{{\bar j}}_{{q}}-j+ \bar j \right) }}
\frac{\Gamma \left( k+u+j_{{q}}+{{\bar j}}+1 \right) }{\Gamma \left( k-u+{{\bar
j}}_{q}+j+1 \right)}
\\ \label{rr}
r(j,\bar j) &=& \frac {u+j_{{q}}-j-{\bar j}_{q}}{u+j_{{q}}-j+{\bar j}} r_0(j,\bar
j)\,,
\ea
with  $\xi=\e^{{-i\pi}(j_q+\bar j_q)/2}/(j_q\bar j_q)$.

\subsubsection*{$\Mybf{j+\bar j=-n}$}

Let us choose a reference state in \re{WR} as
\be
\Phi^-(\mathcal{Z}_1,\mathcal{Z}_2) = (\bar\theta_1-\bar\theta_2)
(z_1-z_2)^n\,,\qquad \mathcal{I} \,\Phi^-(\mathcal{Z}_1,\mathcal{Z}_2) =
(\bar\theta_1-\bar\theta_2)\,,
\ee
where the operator $\mathcal{I}$ is given by the differential operator \re{W}
acting on $\mathcal{Z}_2-$coordi\-nates. In Eq.~\re{WR}, this operator
intertwines the $SL(2|1)$ representations $\mathcal{V}_{j,\bar j}$ and
$\mathcal{V}_{-\bar j,-j}$ leading to $ c(u) =  {r_n(j,\bar j)}/{r_0(-\bar j,
-j)} = 1\,, $ in agreement with \re{c-factor}.

\subsubsection*{$\Mybf {\bar j=0}$}

According to \re{rel-S} and \re{iso-chir}, the operator $\mathcal{I}=D$
intertwines the $SL(2|1)$ representations $V_{j,0}$ and $\mathbb{V}_{j+1}$. One
chooses the reference state in \re{WR} as $\Phi^+=\bar\theta_1-\bar\theta_2$ and
obtains
\be
D_2\, \mathcal{R}_{\mathcal{V}_{j_q,\bar j_q}\otimes \mathcal{V}_{j,0}}(u)\cdot
\bar\theta_{12}
= c(u) \mathcal{R}_{\mathcal{V}_{j_q,\bar j_q}\otimes\mathbb{V}_{j+1}}(u)\cdot 1
= c(u) \mathcal{R}_{\mathcal{V}_{j_q,\bar
j_q}\otimes\mathcal{V}_{j+1,0}}(u)\cdot
1\,,
\ee
where in the last relation we took into account that $\mathbb{V}_{j+1}$ is an
invariant subspace of $\mathcal{V}_{j+1,0}$. One applies \re{RR} and \re{rr} to
get $ c(u) = {r_0(j,0)}/{r(j+1,0)}=\alpha(u-j) $ in agreement with \re{c2}.

\subsubsection*{$\Mybf{j=\bar j=0}$}

The space $\mathcal{V}_{0,0}$ has invariant subspace ${\rm v}_0=\{1\}$. For
$j=\bar j=0$ one finds from \re{RR} and \re{t-def}
\be
\mathcal{R}_{\mathcal{V}_{j_q,\bar j_q}\otimes \mathcal{V}_{0,0}}(u)\cdot 1 =
\mathcal{R}_{\mathcal{V}_{j_q,\bar j_q}\otimes {\rm v}_0}(u)\cdot 1 = r(0,0)
\ee
and, therefore,
$
{\rm t}_0(u) \cdot 1 =\left[r(0,0)\right]^N,
$
in agreement with \re{t0}.

\subsubsection*{$\Mybf{j=0,\ \bar j=-n}$}

The operator $\mathcal{I}=(-D) (-\partial_z)^n$ intertwines the representations
$\mathbb{\bar V}_{-n}$ and $\mathbb{V}_{n+1}$. Choosing
$\Phi^-=\bar\theta_{12}z_{12}^n$ in \re{WR} one finds
\be
(-D_2) (-\partial_{z_2})^n \, \mathcal{R}_{\mathcal{V}_{j_q,\bar j_q}\otimes
\mathbb{\bar V}_{-n}}(u) \,\bar\theta_{12}z_{12}^n = c(u) n!\,
\mathcal{R}_{\mathcal{V}_{j_q,\bar j_q}\otimes \mathbb{V}_{n+1}}(u)\cdot 1
\ee
and, therefore, $ c(u) = {r_n(0,-n)}/{r(n+1,0)}=\alpha(u-n) $ leading to \re{tT}.

\subsubsection*{Relation between Lax and $\mathcal{R}-$operators}

Let us examine the relation \re{R-L-1}. The operators entering both sides of
\re{R-L-1} act in the tensor product $ \mathcal{V}_{j_q,\bar j_q}\otimes
{\rm\bar
v}_1$. The basis in the three-dimensional space ${\rm\bar v}_1$ can be chosen as
$\{1,\bar\theta_2,z_2-\ft12\bar\theta_2\theta_2\}$, Eq.~\re{v-bar-chi}, while
the
basis in $\mathcal{V}_{j_q,\bar j_q}$ can be defined as in \re{V-gen}. One
verifies that the states $1$ and $\bar\theta_{12}$ belong to
$\mathcal{V}_{j_q,\bar j_q}\otimes {\rm\bar v}_1$ and define there the lowest
weight vectors. One uses the explicit expression for the Lax operator
\re{bar-Lax} to obtain
\be\label{L1}
\bar L(-u+1) \cdot 1 = \left( 1-u-j_q \right) \,,\qquad
\bar L(-u+1) \cdot \bar\theta_{12} = \left( -u+ \bar j_q-
j_q\right) \bar\theta_{12}\,.
\ee
The operator $\mathcal{R}_{ \mathcal{V}_{j_q,\bar j_q}\otimes {\rm\bar v}_1}(u)$
appears as an upper diagonal block of the operator
$\mathcal{R}_{\mathcal{V}_{j_q,\bar j_q}\otimes \mathcal{V}_{0,-1}}(u)$ and,
therefore, one gets from \re{RR}
\be\label{RR1}
\mathcal{R}_{ \mathcal{V}_{j_q,\bar j_q}\otimes \mathcal{V}_{0,-1}}(u)\cdot 1 =
r(0,-1)\,,
\qquad \mathcal{R}_{ \mathcal{V}_{j_q,\bar j_q}\otimes \mathcal{V}_{0,-1}}(u)
\cdot
\bar\theta_{12} = r_0(0,-1)\bar\theta_{12}\,.
\ee
One uses \re{L1} and \re{RR1} to verify with a help of \re{rr} that
\be
\Delta(u-1)=\frac{r(0,-1)}{u-\bar j_q-1}= \frac{r_0(0,-1)(u+j_q-\bar
j_q)}{(u+j_q-1)(u-\bar j_q-1)}\,,
\ee
in agreement with \re{R-L-1} and \re{delta}. In the similar manner, one uses the
relations $\mathcal{R}_{ \mathcal{V}_{j_q,\bar j_q}\otimes {\rm v}_1}(u)\cdot 1 =
r(-1,0)$ and $L(u+1) \cdot 1 = (u+1-\bar j_q)$ to verify \re{R-L} with $\Delta(u)
= r(-1,0)/(u-\bar j_q+1)$.

\section{Matrix representation of the $\mathcal{R}-$operators}

The operators $\mathcal{R}^{(a)}(u)$ (with $a=1,2,3$) are defined in
Eqs.~\re{R-i} -- \re{R-last}. According to \re{R's}, they map the tensor product
of two infinite-dimensional $SL(2|1)$ representations $\mathcal{V}_{j_1,\bar
j_1}\otimes \mathcal{V}_{j_2,\bar j_2}$ into tensor product of two yet another
representations $\mathcal{V}_{j_1',\bar j_1'}\otimes \mathcal{V}_{j_2',\bar
j_2'}$ with $j_1+j_2=j_1'+j_2'$ and $\bar j_1+\bar j_2=\bar j_1'+\bar j_2'$.
Both tensor products can be decomposed over irreducible components
as~\cite{Marcu79,Frappat96,DerKarKir98}
\be\label{tensor}
[j_1,\bar j_1] \otimes [j_2,\bar j_2] = [j_{12},\bar j_{12}]+\sum_{n\ge 1} 2
[j_{12}+n,\bar j_{12}+n]+ [j_{12}+n-1,\bar j_{12}+n] + [j_{12}+n,\bar j_{12}+n-1]
\, ,
\ee
where $j_{12}=j_1+j_2$, $\bar j_{12}=\bar j_1+\bar j_2$ and the factor $2$
inside the sum takes into account multiplicity. The  $\mathcal{R}^{(a)}-$operators
map each irreducible component in the right-hand side of \re{tensor} into similar
component in the tensor product $[j_1',\bar j_1'] \otimes [j_2',\bar j_2']$
carrying the same spins. In particular, they transform the lowest weight vectors
in $\mathcal{V}_{j_1,\bar j_1}\otimes \mathcal{V}_{j_2,\bar j_2}$ into those in
$\mathcal{V}_{j_1',\bar j_1'}\otimes \mathcal{V}_{j_2',\bar j_2'}$. In the
right-hand side of \re{tensor}, the lowest weight of the $[j_{12},\bar j_{12}]$
representation is $1$ and the $\mathcal{R}^{(a)}-$operators act as
\be\label{r(a)}
\mathcal{R}^{(a)}(u) \cdot 1 = r^{(a)}_u\,.
\ee
For $n\ge 1$ the lowest weights in four $SL(2|1)$ components in \re{tensor}
are~\cite{DerKarKir98}
\be
e_i = \big\{(Z_{12}+\ft12 \theta_{12}\bar\theta_{12})^n,\ (Z_{12}-\ft12
\theta_{12}\bar\theta_{12})^n,\ \bar\theta_{12} Z_{12}^n,\ \theta_{12}
Z_{12}^n\,\big\}\,,
\ee
where $Z_{12} = z_{1}-z_{2} + \ft 12 \theta_1 \bar\theta_2 + \ft12 \bar\theta_1
\theta_2$ and $1\le i\le 4$. Then, the $\mathcal{R}^{(a)}-$operators can be
represented by a graded $4\times 4$ matrix
\be
\mathcal{R}^{(a)}(u) \cdot e_i = \sum_{k=1}^4 e_k\, [\mathcal{R}^{(a)}(u)]_{ki}
\,,\qquad  [\mathcal{R}^{(a)}(u)]_{ki} = \lr{\
\begin{array}{cccc}
\star & \star & 0 & 0 \\
\star & \star & 0 & 0 \\
0 & 0 & \star & 0 \\
0 & 0 & 0 & \star \\
\end{array} }
\, ,
\ee
where `$\star$' denote entries that may take nonvanishing values.

The calculation of $r_u^{(a)}$ and the matrices $[\mathcal{R}^{(a)}(u)]_{ki}$ is
straightforward with a help of \re{R-i} -- \re{R-last}. The expressions for
$r_u^{(1)}$ and $r_u^{(3)}$ are given in \re{r13}, while $r_u^{(2)}$ can be
easily found from \re{R2} as $ r_u^{(2)} =  {j_2}/{(j_2+u)}$. In what follows, we
present explicit expressions for the matrices $[\mathcal{R}^{(a)}(u)]_{ki}$
evaluated for $n\ge 1$ and for the spins $j_1$, $\bar j_1$, $j_2$ and $\bar j_2$
taking the same values as in the definition of the $\mathcal{Q}-$operators,
Eqs.~\re{T's} and \re{Q=T}. Namely, $j_1=j_q$, $\bar j_1=\bar j_q$ and the
remaining spins are fixed as follows:
\begin{itemize}
\item $j_2=j_q$, $\bar j_2=\bar j_q-j_q-u$ :
\be\label{R1-me}
\mathcal{R}^{(1)}(u+j_q) := \e^{-i\pi(u+j_q)/2} \frac{\Gamma
 \left( n+1+j_q+{ {\bar j}}_{q} \right)}{\Gamma
 \left( n+1-u+{{\bar j}}_{q} \right)} \left[ \begin {array}{cccc} -{\frac
{u+j_{q}-{\it
{\bar j}}_{q}}{{\it {\bar j}}_{q}}}&-{\frac { \left( u+j_{q} \right) j_{q}}{
\left( n+j_{{q}}+{\it {\bar j}}_{q} \right) {\it {\bar
j}}_{q}}}&0&0\\\noalign{\medskip}0&- {\frac {-n+u-{\it {\bar
j}}_{q}}{n+j_{q}+{\it {\bar j}}_{q}}}&0&0
\\\noalign{\medskip}0&0&1&0\\\noalign{\medskip}0&0&0&-{\frac {u+j_{{q}
}-{{\bar j}}_{q}}{{ {\bar j}}_{q}}}\end {array} \right];
\ee

\item $j_2=\bar j_q-u$, $\bar j_2=j_q+u$ :
\be
\mathcal{R}^{(2)}(u+j_q-\bar j_q):=\left[ \begin {array}{cccc} {\frac
{u+j_{q}}{{\it {\bar j}}_{q}}}&0&0&0\\\noalign{\medskip}-{\frac { \left(
n+j_{q}+{\it {\bar j}}_{q} \right)
 \left( u+j_{q}-{\it {\bar j}}_{q} \right) }{j_{q}{\it {\bar j}}_{q}}}&-{
\frac {u-{\it {\bar j}}_{q}}{j_{q}}}&0&0\\\noalign{\medskip}0&0&-{\frac { \left(
u-{\it {\bar j}}_{q} \right)  \left( u+j_{q} \right) }{j_{q}{ \it {\bar
j}}_{q}}}&0\\\noalign{\medskip}0&0&0&1\end {array} \right];
\ee

\item $j_2=j_q-\bar j_q+u$, $\bar j_2=\bar j_q$ :
\be\label{R3-me}
\mathcal{R}^{(3)}(u-\bar j_q) :=\e^{i\pi(u-\bar j_q)/2} {\frac {\Gamma  \left(
n+1+u+j_{q} \right) }{\Gamma \left( n+1+j_{q}+ {\it {\bar j}}_{q} \right)
}}\left[ \begin {array}{cccc} 1&-{\frac { \left( u-{\it {\bar j}}_{q} \right)
{\it {\bar j}}_{q}}{j_{q} \left( n+u+j_{q} \right) }}&0&0
\\\noalign{\medskip}0&{\frac { \left( n+j_{q}+{\it {\bar j}}_{q}
 \right)  \left( u+j_{q}-{\it {\bar j}}_{q} \right) }{j_{q} \left( n+u
+j_{q} \right) }}&0&0\\\noalign{\medskip}0&0&1&0\\\noalign{\medskip}0
&0&0&{\frac
{u+j_{q}-{\it {\bar j}}_{q}}{j_{q}}}\end {array} \right].
\ee

\end{itemize}
For $j_2=j_q$ and $\bar j_2=\bar j_q$ these matrices reduce to unity matrix in
agreement with \re{R=1}.

The relations \re{R1-me} -- \re{R3-me} are in agreement with expressions for the
corresponding matrix elements from the second paper in~\cite{Der05}. Notice that
the operators  $\mathcal{R}_{3}(u_1,u_2,u_3|v_1)$,
$\mathcal{R}_{2}(u_1,u_2|v_2,v_3)$ and $\mathcal{R}_{1}(u_1,|v_1,v_2,v_3)$
introduced there coincide with the operators $\mathcal{R}^{(a)}(u)$ (with
$a=1,2,3$) up to the normalization factors
\ba \nonumber
\mathcal{R}^{(3)}(u_3-v_3) &=& \e^{i\pi(u_3-v_3)/2}\frac{ u_3-v_3}{u_2-u_3}
\,\mathcal{R}_{3}(u_1,u_2,u_3|v_1)\,,
\\[2mm] \nonumber
\mathcal{R}^{(2)}(u_2-v_2) &=& \frac{u_2-v_2}{(u_1-u_2)(u_2-v_3)}
\,\mathcal{R}_{2}(u_1,u_2|v_2,v_3)\,,
\\[2mm]
\mathcal{R}^{(1)}(u_1-v_1) &=& \e^{-i\pi(u_1-v_1)/2}\frac{u_1-v_1}{u_1-v_2}
\,\mathcal{R}_{1}(u_1,|v_1,v_2,v_3)\,,
\ea
with the $v-$ and $u-$parameters given by \re{u's}.

\section{Integral representation of  $\mathcal{R}^{(3)}-$operator}
\label{IntegralKernelsR}

In this Appendix, we demonstrate that the operator $\mathcal{R}^{(3)}(u)$
entering the factorized expression for the $\mathcal{R}-$operator \re{R-fact}
can be realized in the space of functions $\Phi (\mathcal{W}_1, \mathcal{W}_2)
\in \mathcal{V}_{j_1, \bar{j}_1} \otimes \mathcal{V}_{j_2, \bar{j}_2}$ as an
integral operator \re{R13-kernels}. The calculations for
$\mathcal{R}^{(1)}-$operator are identical with minor modifications.

The operator $\mathcal{R}^{(3)}(u)$ acts in the tensor product $
\mathcal{V}_{j_1,\bar{j}_1} \otimes \mathcal{V}_{j_2, \bar{j}_2}$ and changes
the spins of the vector spaces as in \re{R's}. This translates into the
following relation
\be
\label{CommR3generator} \mathcal{R}^{(3)}(u) \left( G_{j_1 \bar{j}_1} + G_{j_2
\bar{j}_2} \right) = \left( G_{j_1 + u, \bar{j}_1} + G_{j_2
- u, \bar{j}_2} \right) \mathcal{R}^{(3)}(u) \, ,
\ee
where $G_{j,\bar j}$ denote the $SL(2|1)$ generators in
$\mathcal{V}_{j,\bar{j}}$, Eqs.~\re{trans1} -- \re{trans3}. By definition, the
operator $\mathcal{R}^{(3)}(u_3-v_3)$ interchanges the spectral parameters $u_3
\rightleftarrows v_3$ in the product of two Lax operators \re{YBRcheck} and
leaves the remaining parameters intact. One can show that this property leads to
the following system of relations \cite{Der05}
\be
\label{CommR3coordinate} [\mathcal{R}^{(3)}(u), w_2] = [\mathcal{R}^{(3)}(u),
\vartheta_2] = [\mathcal{R}^{(3)}(u), \bar\vartheta_2] = 0 \, ,
\ee
where $\mathcal{W}_2=(w_2,\vartheta_2,\bar\vartheta_2)$ denote the coordinates
in the graded spaces $\mathcal{V}_{j_2, \bar{j}_2}$ and  $\mathcal{V}_{j_2-u,
\bar{j}_2}$. Equation \re{CommR3coordinate} implies that $\mathcal{R}^{(3)}(u)$
acts nontrivially only on the $\mathcal{W}_1-$coordinates of a test function
\be\label{R3-K}
\mathcal{R}^{(3)}(u) \Phi (\mathcal{W}_1, \mathcal{W}_2) = \int {}[\mathcal{D}
\mathcal{Z}_1]_{j_1 \bar{j}_1} \,{R}_u^{(3)} (\mathcal{W}_1 ;
\mathcal{Z}^\ast_1)
\Phi (\mathcal{Z}_1, \mathcal{W}_2) \, .
\ee
The same relation can be rewritten in terms of the reproducing kernel
\re{K-kernel} as
\be
\label{R3kernel} {R}_u^{(3)} (\mathcal{W}_1 ; \mathcal{Z}^\ast_1) =
\mathcal{R}^{(3)}(u) \,\mathcal{K}_{j_1 \bar{j}_1} (\mathcal{W}_1 ;
\mathcal{Z}^\ast_1) \, .
\ee
Indeed, substituting this relation into \re{R3-K} one performs
$\mathcal{Z}_1-$integration with a help of \re{unit-O} and arrives at the
identity. The reproducing kernel $\mathcal{K}_{j \bar{j}} (\mathcal{W} ;
\mathcal{Z}^\ast)$, Eq.~\re{K-kernel}, can be rewritten as a sequence of finite
$SL(2|1)$ transformations applied to the lowest weight vector in
$\mathcal{V}_{j,\bar j}$, namely,
\be
\label{ReproducingKernelFiniteGroup} \mathcal{K}_{j \bar{j}} (\mathcal{W} ;
\mathcal{Z}^\ast) =
{\rm e}^{ - \bar\theta^\ast V^+_{ j \bar{j} } } {\rm e}^{ - \theta^\ast
\bar{V}^+_{j \bar{j}} } {\rm e}^{ z_-^\ast L^+_{j \bar{j}} } \cdot 1 \, ,
\ee
where $\mathcal{Z}^\ast=(z^*,\theta^*,\bar\theta^*)$ plays the r\^ole of the
transformation parameters and the raising generators $V^+_{j \bar{j}}$,
$\bar{V}^+_{j \bar{j}}$ and $L^+_{j \bar{j}}$ are given by the differential
operators \re{rising} acting on functions depending on
$\mathcal{W}=(w,\vartheta,\bar\vartheta)$. This fact combined together with the
commutation relations \re{CommR3generator} and \re{CommR3coordinate} allows one
to find the integral kernel ${R}^{(3)}(\mathcal{W}_1 ; \mathcal{Z}^\ast_1)$
algebraically.

At the first step, one applies \re{ReproducingKernelFiniteGroup}  to rewrite the
product of two reproducing kernels as
\ba\label{KK}
\mathcal{K}_{j_1 \bar{j}_1} (\mathcal{W}_1 ; \mathcal{Z}^\ast_1)
\mathcal{K}_{j_2
\bar{j}_2} (\mathcal{W}_2 ; \mathcal{Z}^\ast_1) = {\rm e}^{ - \bar\theta^\ast_1
\left( V^+_{j_1 \bar{j}_1} + V^+_{j_2 \bar{j}_2} \right) } {\rm e}^{ -
\theta^\ast_1 \left( \bar{V}^+_{j_1 \bar{j}_1} + \bar{V}^+_{j_2 \bar{j}_2}
\right) } {\rm e}^{ z_{1,-}^\ast \left( L^+_{j_1 \bar{j}_1} + L^+_{j_2
\bar{j}_2} \right) } \cdot 1\,.
\ea
Let us apply the operator $\mathcal{R}^{(3)}(u)$ to both sides of this relation.
In the left-hand side, it only acts on the $\mathcal{W}_1-$dependent kernel,
while in the right-hand side it can be moved to the right across the exponent by
virtue of \re{CommR3generator}  so that
\ba\label{KKR}
\lefteqn{\mathcal{K}_{j_2
\bar{j}_2} (\mathcal{W}_2 ; \mathcal{Z}^\ast_1)
\left[\mathcal{R}^{(3)}(u)\mathcal{K}_{j_1 \bar{j}_1} (\mathcal{W}_1 ;
\mathcal{Z}^\ast_1)\right]}&&
\\[2mm]
&& = {\rm e}^{ - \bar\theta^\ast_1 \left( V^+_{j_1 + u, \bar{j}_1} + V^+_{j_2 -
u, \bar{j}_2} \right) } {\rm e}^{ - \theta^\ast_1 \left( \bar{V}^+_{j_1 + u,
\bar{j}_1} + \bar{V}^+_{j_2 - u, \bar{j}_2} \right) } {\rm e}^{ z_{1,-}^\ast
\left( L^+_{j_1 + u, \bar{j}_1} + L^+_{j_2 - u, \bar{j}_2} \right) } \,
\mathcal{R}^{(3)}(u) \cdot 1 \, . \nonumber
\ea
Finally, taking into account the relation \re{r(a)}, $\mathcal{R}^{(3)}(u)
\cdot 1 = r^{(3)}_u$, one finds from \re{KKR}, \re{R3kernel} and \re{KK}
\ba
{R}_u^{(3)} (\mathcal{W}_1 ; \mathcal{Z}^\ast_1) &=& r^{(3)}_u \,
\mathcal{K}_{j_1+u
\bar{j}_1} (\mathcal{W}_1 ; \mathcal{Z}^\ast_1) \mathcal{K}_{j_2-u
\bar{j}_2} (\mathcal{W}_2 ; \mathcal{Z}^\ast_1)/\mathcal{K}_{j_2
\bar{j}_2} (\mathcal{W}_2 ; \mathcal{Z}^\ast_1)
\nonumber
\\[2mm] \label{R-zig}
&=& r^{(3)}_u \, \mathcal{K}_{j_1+u
\bar{j}_1} (\mathcal{W}_1 ; \mathcal{Z}^\ast_1) \mathcal{K}_{-u,
0} (\mathcal{W}_2 ; \mathcal{Z}^\ast_1)\,.
\ea
Matching the relations \re{R3-K} and \re{R-zig} into \re{R-i} one finds the
integral kernel of the operator $\mathcal{R}^{(3)}(u)$
\be
{R}_u^{(3)} (\mathcal{W}_1, \mathcal{W}_2 ; \mathcal{Z}^\ast_1,
\mathcal{Z}^\ast_2) ={R}_u^{(3)} (\mathcal{W}_1 ; \mathcal{Z}^\ast_1)
\mathcal{K}_{j_2,\bar j_2} (\mathcal{W}_2,\mathcal{Z}_2^\ast)\,,
\ee
which coincides with \re{R13-kernels}.

\section{Factorization of the transfer matrix}

\begin{figure}[t]
\begin{center}
\psfrag{PR1}[cc][lc]{$\Pi_{k0}\mathcal{R}_{k0}^{(a)}=$}
\psfrag{R1}[cc][cc]{$\mathcal{R}_{k0}^{(a)}=$} \psfrag{R}[cc][cc]{$\scriptstyle
\mathcal{R}^{(a)}$}
\psfrag{0}[cc][cc]{$\scriptstyle 0^\ast$} \psfrag{k}[cc][cc]{$\scriptstyle
k^\ast$}\psfrag{0'}[cc][cc]{$\scriptstyle 0$} \psfrag{k'}[cc][cc]{$\scriptstyle
k$}
\insertfig{9}{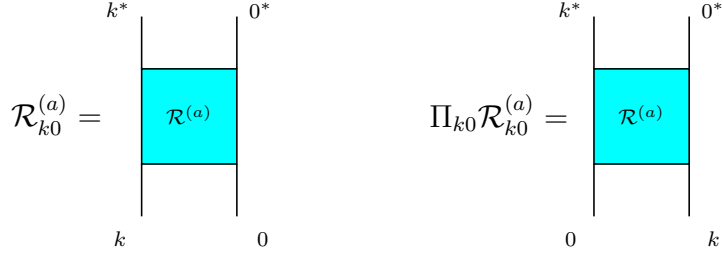}
\end{center}
\caption{ \label{R-MatrixGraph} Graphical representation of the kernel
${R}^{(a)}_u(\mathcal{W}_k, \mathcal{W}_0 ; \mathcal{Z}^\ast_k,
\mathcal{Z}^\ast_0)$. }
\end{figure}%

\begin{figure}[t]
\begin{center}
\psfrag{Rb}[cc][cc]{$\scriptstyle \mathcal{R}^{(b)}$}
\psfrag{Ra}[cc][cc]{$\scriptstyle \mathcal{R}^{(a)}$}
\psfrag{=}[cc][cc]{$\boldsymbol{=}$}
\psfrag{0}[cc][cc]{$\scriptstyle 0^\ast$} \psfrag{k}[cc][cc]{$\scriptstyle
k^\ast$}\psfrag{0'}[cc][cc]{$\scriptstyle 0$} \psfrag{k'}[cc][cc]{$\scriptstyle
k$}
\psfrag{1'}[cc][cc]{$\scriptstyle 1^\ast$} %
\psfrag{2'}[cc][cc]{$\scriptstyle 2^\ast$} %
\psfrag{3'}[cc][cc]{$\scriptstyle 3^\ast$} %
\psfrag{1}[cc][cc]{$\scriptstyle 1$} %
\psfrag{2}[cc][cc]{$\scriptstyle 2$} %
\psfrag{3}[cc][cc]{$\scriptstyle 3$} %
\insertfig{8}{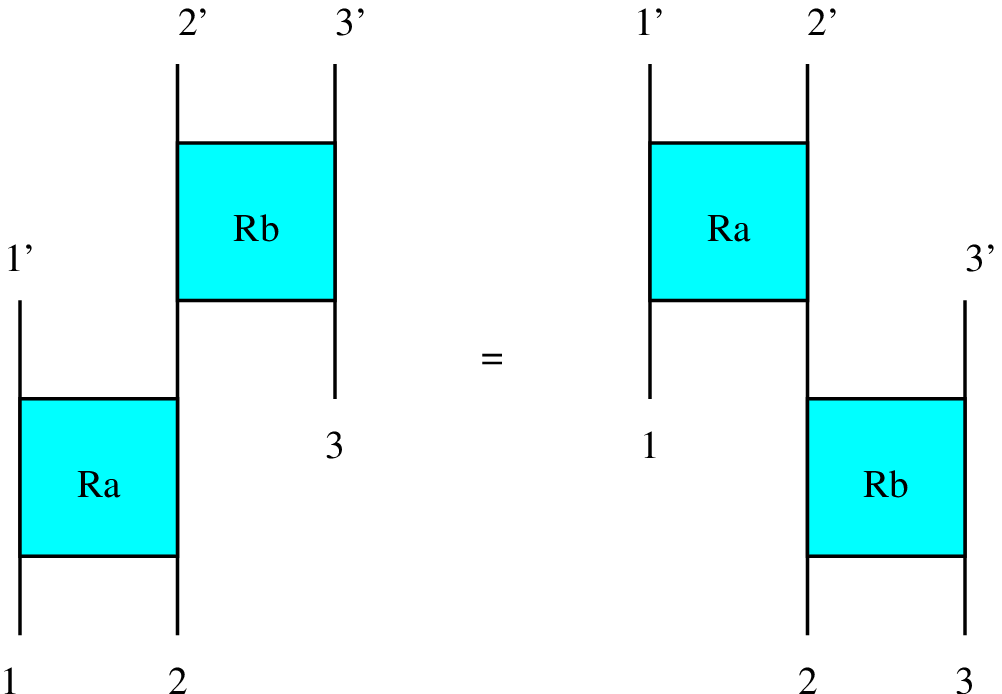}
\end{center}
\caption{ \label{CommutationGraph} Graphical representation of Eq.\
\re{CommRuleRR}.}
\end{figure}%

%
\begin{figure} 
\begin{center}
\psfrag{R3}[cc][cc]{$\scriptstyle \mathcal{R}^{(3)}$}
\psfrag{R12}[cc][cc]{$\scriptstyle \mathcal{R}^{(12)}$}
\psfrag{=}[cc][cc]{$\boldsymbol{=}$}
\psfrag{1p}[cc][cc]{$\scriptstyle 1$} \psfrag{1}[cc][cc]{$\scriptstyle
1^\ast$}\psfrag{2p}[cc][cc]{$\scriptstyle 2$} \psfrag{2}[cc][cc]{$\scriptstyle
2^\ast$}
{\insertfig{17}{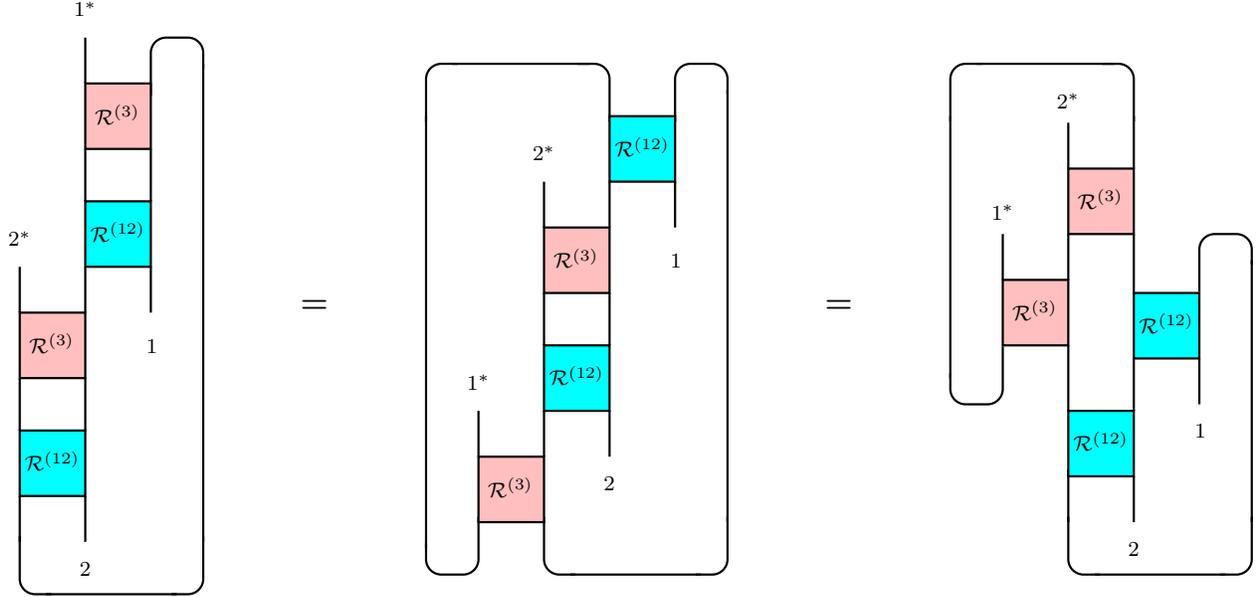}}
\end{center}
\caption{\label{Steps} Steps in factorization of the transfer matrix
$\mathcal{T}_{j \bar{j}} (w)$ for spin chain of length $N=2$.}
\end{figure}%

Derivation of the factorized expression for the transfer matrix \re{T-factor}
relies on the commutativity property of the $\mathcal{R}^{(a)}-$operators,
Eq.~\re{CommRuleRR}. To simplify manipulations with integral operators
${R}^{(a)}_{k0}(u)$, it is convenient to introduce  a diagrammatic representation
of their integral kernel ${R}^{(a)}_u (\mathcal{W}_k, \mathcal{W}_0 ;
\mathcal{Z}^\ast_k, \mathcal{Z}^\ast_0)$ as shown in Fig.\ \ref{R-MatrixGraph}.
The arguments of the kernel define the coordinates of four end-points of the
diagram. The product of operators in the left-hand side of \re{CommRuleRR} is
represented by the integral kernel
\be\label{example}
\mathcal{R}_{12}^{(a)}(u) \mathcal{R}_{23}^{(b)}(v):=\int {}[\mathcal{D}
\mathcal{Z}_2']_{j_2' \bar{j}_2'}  {R}^{(a)}_u (\mathcal{W}_1, \mathcal{W}_2 ;
\mathcal{Z}^\ast_1, \mathcal{Z'}^\ast_2) {R}^{(b)}_v (\mathcal{Z}_2',
\mathcal{W}_3 ; \mathcal{Z}^\ast_2, \mathcal{Z}^\ast_3)
\ee
where the $SL(2|1)$ representation $[j_2',\bar j_2']$ in the `intermediate'
space
is $[j_2,\bar j_2-v]$ for $b=1$, $[j_2-v,\bar j_2+v]$ for $b=2$ and $[j_2+v,\bar
j_2]$ for $b=3$. Convolution of two integral kernels in \re{example} can be
represented as the diagram shown in the left panel of Fig.\
\ref{CommutationGraph}. In the similar manner, the right panel of Fig.\
\ref{CommutationGraph} corresponds to convolution of the kernels in the
right-hand side of \re{CommRuleRR}.

Let us start with a general expression for the transfer matrix \re{Eq1} and
substitute the $\mathcal{R}-$operators by the factorized expression \re{R-fact}
(see left panel in Fig.~\ref{Steps})
\be
\mathcal{R}_{k0} (w) = {\Pi}_{k0} \mathcal{R}^{(12)}_{k0} (w_1, w_2)
\mathcal{R}^{(3)}_{k0}(w_3)\,,
\ee
where $\mathcal{R}^{(12)}_{k0}(w_1, w_2) = \mathcal{R}^{(1)}_{k0}(w_1)
\mathcal{R}^{(2)}_{k0}(w_2)$, the permutation operator ${\Pi}_{k0}$ was defined
in \re{Pi} and the parameters $w_a=u_a-v_a$ are given by \re{u's} with $w=u-v$.
Next, one makes use of cyclicity of the supertrace in \re{Eq1} to move the
right-most operator $\mathcal{R}^{(3)}_{10} (w_3)$ in front of
$\mathcal{R}^{(12)}_{{N}0}(w_1, w_2)$ (see central panel in Fig.~\ref{Steps})
and, then, applies the relation \re{CommRuleRR} shown in
Fig.~\ref{CommutationGraph} to interchange $\mathcal{R}^{(3)}-$ and
$\mathcal{R}^{(12)}-$operators. The result is displayed in Fig.\ \ref{Steps}, to
the right and reads in the symbolic form
\be
\label{R3fromR12} \mathcal{T}_{j \bar{j}} (w) = \str \left[ {\Pi}_{N0}
\mathcal{R}^{(12)}_{N0} (w_1, w_2) \ldots {\Pi}_{10}\mathcal{R}^{(12)}_{10}(w_1,
w_2) \right] \mathbb{P}^{-1} \str \left[ {\Pi}_{N0}\mathcal{R}^{(3)}_{N0}(w_3)
\ldots {\Pi}_{10}\mathcal{R}^{(3)}_{10}(w_3) \right] \, ,
\ee
where $\mathbb{P}$ is the operator of cyclic permutation \re{P}. The last factor
in this relation can be identified as the transfer matrix $\mathcal{T}_{j_q + w,
\bar{j}_q} (w_3)$, Eq.~\re{T's}. Repeating the above consideration for the first
factor in Eq.\ \re{R3fromR12}, we eventually arrive at
\ba \nonumber
\mathcal{T}_{j \bar{j}} (w) \!\!\!&=&\!\!\! \str \left[ {\Pi}_{N0}
\mathcal{R}^{(1)}_{N0}(w_1) \ldots {\Pi}_{10}\mathcal{R}^{(1)}_{10}(w_1) \right]
\mathbb{P}^{-1}
\\ \nonumber
&\times&\!\!\! \str \left[ {\Pi}_{N0}\mathcal{R}^{(2)}_{N0}(w_2) \ldots
{\Pi}_{10}\mathcal{R}^{(2)}_{10}(w_2) \right] \mathbb{P}^{-1} \str \left[
\mathbb{P}_{10}\mathcal{R}^{(3)}_{10}(w_3) \ldots
\mathbb{P}_{N0}\mathcal{R}^{(3)}_{N0}(w_3) \right] \, .
\ea
This relation can be rewritten in terms of the transfer matrices \re{T's} as
follows
\ba
\nonumber \mathcal{T}_{j \bar{j}} (w) \!\!\!&=&\!\!\! \mathcal{T}_{j_q,
\bar{j}_q
- w_1} (w_1) \mathbb{P}^{-1} \mathcal{T}_{j_q - w_2, \bar{j}_q + w_2} (- w_2)
\mathbb{P}^{-1} \mathcal{T}_{j_q + w_3, \bar{j}_q} (w_3)
\\[2mm]
&=&\!\!\! \mathbb{P}^{-2 } \mathcal{T}_{j_q, \bar{j}_q - w_1} (w_1)
\mathcal{T}_{j_q - w_2, \bar{j}_q + w_2} (- w_2) \mathcal{T}_{j_q + w_3,
\bar{j}_q} (w_3) \, ,
\ea
with the spectral parameters $w_1 = w - j + j_q$, $w_2 = w - j +
\bar{j} + j_q - \bar{j}_q$ and $w_3 = w + \bar{j} - \bar{j}_q$.



\begin{thebibliography}{99}
%
\bibitem{And90}
P.W. Anderson, Phys. Rev. Lett. 65 (1990) 2306.
%
\bibitem{FoeKar92}
A. Foerster, M. Karowski, Nucl. Phys. B 396 (1993) 611.
%
\bibitem{EssKor92}
F.H.L. E{\ss}ler, V.E. Korepin, Phys. Rev. B 46 (1992) 9147.
%
\bibitem{BelDerKorMan04}
A.V. Belitsky, S.E. Derkachov, G.P. Korchemsky, A.N. Manashov,
Phys. Lett. B 594 (2004) 385;
%
Nucl. Phys. B 708 (2005) 115.
%
\bibitem{Soh85}
S.J. Gates, M.T. Grisaru, M. Rocek, W. Siegel,
Front. Phys. 58 (1983) 1; \\
%
M.F. Sohnius,
Phys. Rept. 128 (1985) 39.
%
\bibitem{Efe93}
K.B. Efetov, {\sl Supersymmetry And Theory Of Disordered Metals}, Adv. Phys. 32
(1983) 53.
%
\bibitem{WeiZir88}
H.A. Weidenm\"uller, M. Zirnbauer, Nucl. Phys. B 305 (1988) 339.
%
\bibitem{Bha99}
M.J. Bhaseen, I.I. Kogan, O.A. Solovev, N. Tanigichi, A.M. Tsvelik,
Nucl. Phys. B 580 (2000) 688.
%
\bibitem{Mal97}
J.M. Maldacena,
Adv. Theor. Math. Phys. 2 (1998) 231; \\
%
S.S. Gubser, I.R. Klebanov, A.M. Polyakov,
Phys. Lett. B 428 (1998) 105; \\
%
E. Witten,
Adv. Theor. Math. Phys. 2 (1998) 253.
%
\bibitem{Zir99}
M.R. Zirnbauer,
hep-th/9905054.
%
\bibitem{EssFraSal05}
F.H.L. E{\ss}ler, H. Frahm, H. Saleur,
Nucl. Phys. B 712 (2005) 513.
%
\bibitem{DerKorMan01}
S.E. Derkachov, G.P. Korchemsky, A.N. Manashov,
Nucl. Phys. B 617 (2001) 375;
%
Nucl. Phys. B 645 (2002) 237.
%
\bibitem{BytTes06}
A.G. Bytsko, J. Teschner,
J. Phys. A 39 (2006) 12927.
%
\bibitem{Bax72}
R.J. Baxter,
Annals Phys. 70 (1972) 193;
%
{\sl Exactly Solved Models in Statistical Mechanics},
Academic Press (London, 1982).
%
\bibitem{BazLukZam96}
V.V. Bazhanov, S.L. Lukyanov, A.B. Zamolodchikov, Comm. Math. Phys. 190 (1997) 247.
%
\bibitem{AntFei96}
A. Antonov, B. Feigin,
Phys. Lett. B 392 (1997) 115.
%
\bibitem{Skl95}
M. Gaudin, V. Pasquier,
J. Phys. A 25 (1992) 5243; \\
%
V.B. Kuznetsov, E.K. Sklyanin, J. Phys. A 31 (1998) 2241; \\
V.B. Kuznetsov, M. Salerno, E.K. Sklyanin,  J.Phys. A33 (2000) 171; \\
E.K. Sklyanin, nlin.SI/0009009.
%
\bibitem{Der99}
S.E. Derkachov,
J. Phys. A 32 (1999) 5299; \\
%
S.E. Derkachov, A.N. Manashov,
J. Phys. A 39 (2006) 4147.
%
\bibitem{Pro00}
G.P. Pronko, Commun.Math.Phys. 212 (2000) 687.
%
\bibitem{Baz01}
V.V. Bazhanov, A.N. Hibberd, S.M. Khoroshkin,
Nucl. Phys. B 622 (2002) 475.
%
\bibitem{Hik01}
K. Hikami,
Nucl. Phys. B 604 (2001) 580.
%
\bibitem{DerMan06}
S.E. Derkachov, A.N. Manashov,
J. Phys. A 39 (2006) 13171.
%
\bibitem{Der05}
S.E.~Derkachov,~{\sl Zapiski~nauchnuch~seminarov~POMI}, 335 (2006) 134, 164;
math.qa/0503396, math.qa/0503410.
%
\bibitem{Tsu97}
Z. Tsuboi, J. Phys. A 30 (1997) 7975;
%
Physica A 252 (1998) 565;
%
J. Phys. A 31 (1998) 5485.
%
\bibitem{DerKarKir98}
S.E. Derkachov, D. Karakhanyan, R. Kirschner,
Nucl. Phys. B 583 (2000) 691.
%
\bibitem{Frappat96}
L. Frappat, P. Sorba, A. Sciarrino,
{\sl Dictionary on Lie algebras and superalgebras}, Academic Press (London,
2000);
hep-th/9607161.
%
\bibitem{Sch76}
M. Scheunert, W. Nahm, V. Rittenberg,
J. Math. Phys. 18 (1977) 155.
%
\bibitem{Jar78}
P.D. Jarvis, H.S. Green,
J. Math. Phys.  20 (1979) 2115.
%
\bibitem{Marcu79}
M. Marcu,
J. Math. Phys. 21 (1980) 1277;
%
J. Math. Phys. 21 (1980) 1284.
%
\bibitem{Kul85}
P.P. Kulish, J. Sov. Math. 35 (1986) 2648;
%
Lett. Math. Phys.  {10} (1985) 87.
%
\bibitem{Maa94}
Z. Maassarani, J. Phys. A {28} (1995) 1305; \\
%
P.B. Ramos, M.J. Martins, Nucl. Phys. B {474} (1996) 678; \\
%
M.P. Pfannmuller, H. Frahm, Nucl. Phys. B {479} (1996) 575.
%
\bibitem{DerKorMan03}
S.E. Derkachov, G.P. Korchemsky, A.N. Manashov,
J. High Ener. Phys. 0310 (2003) 053.
%
\bibitem{Lai74}
C.K. Lai,
J. Math. Phys. 15 (1974) 167.
%
\bibitem{Sut75}
B. Sutherland,
Phys. Rev. B 12 (1975) 3795.
%
\bibitem{Sch87}
P. Schlottmann, Phys. Rev. B 36 (1987) 5177.
%
\bibitem{QISM}
L.A. Takhtajan, L.D. Faddeev,
Russ. Math. Survey 34 (1979) 11; \\
%
E.K. Sklyanin, L.A. Takhtajan, L.D. Faddeev,
Theor. Math. Phys. 40 (1980) 688; \\
%
V.E. Korepin, N.M. Bogoliubov, A.G. Izergin, {\sl Quantum inverse scattering
method and correlation functions}, Cambridge Univ. Press (Cambridge, 1993).
%
\bibitem{BraDerMan98}
V.M. Braun, S.E. Derkachov, A.N. Manashov,
Phys. Rev. Lett. 81 (1998) 2020; \\
%
V.M. Braun, S.E. Derkachov, G.P. Korchemsky, A.N. Manashov,
Nucl. Phys. B 553 (1999) 355; \\
%
A.V. Belitsky,
Phys. Lett. B 453 (1999) 59;
%
Nucl. Phys. B 574 (2000) 407.
%
\bibitem{Kor95}
G.P. Korchemsky,
Nucl. Phys. B 462 (1996) 333; \\
%
S.E. Derkachov, G.P. Korchemsky, A.N. Manashov,
Nucl. Phys. B 566 (2000) 203;
%
Nucl. Phys. B 661 (2003) 533.
%
\bibitem{Skl85}
E.K. Sklyanin,
Lect. Notes Phys. 226 (1985) 196;
%
in ``Quantum Group and Quantum Integrable Systems,'' ed.\ Mo-Lin Ge,
World Scientific, (Singapore, 1992) pp.\ 63--97, hep-th/9211111;
%
Progr. Theor. Phys. Suppl. 118 (1995) 35.
%
\bibitem{Ess92}
F.H.L. E{\ss}ler, V. Korepin, K. Schoutens,
Phys. Rev. Lett. 68 (1992) 2960.
%
\bibitem{Sal99}
H. Saleur,
Nucl. Phys. B 578 (2000) 552.
%
\bibitem{Arn03}
D. Arnaudon, J. Avan, N. Crampe, A. Doikou, L. Frappat, E. Ragoucy,
Nucl. Phys. B 687 (2004) 257.
%
\end{thebibliography}
\end{document}